\documentclass[twocolumn,linenumbers,trackchanges]{aastex631}
\usepackage{lineno}
\nolinenumbers
\usepackage{silence} 
\usepackage{natbib}
\usepackage{amsmath}
\usepackage{float}
\usepackage{times}
\usepackage{enumitem}
\usepackage{etoolbox}
\usepackage{graphicx}
\usepackage{amsmath}
\usepackage{bm}
\usepackage{graphics}
\usepackage{url}
\usepackage{epsfig}
\usepackage{wrapfig}
\usepackage{ulem}
\usepackage{verbatim}
\usepackage{float}
\usepackage{lastpage}
\usepackage{CJK}
\usepackage[whole]{bxcjkjatype} 
\usepackage{CJKutf8}
\usepackage[caption=false]{subfig}

\def\lsim{\hbox{\rlap{\raise 0.425ex\hbox{$<$}}\lower 0.65ex\hbox{$\sim$}}}
\def\gsim{\hbox{\rlap{\raise 0.425ex\hbox{$>$}}\lower 0.65ex\hbox{$\sim$}}}

\def\arcdeg{\mbox{$^\circ$}}

\newcommand\N[1]{{\color{blue} \bf #1}} 

\shorttitle{SN\,2023ixf from Shock Breakout to Nebular}
\shortauthors{Zheng et al.}

\begin{document}

\title{SN\,2023ixf in the Pinwheel Galaxy M101: From Shock Breakout to the Nebular Phase}

\correspondingauthor{WeiKang Zheng, Luc Dessart, 
Alexei V. Filippenko; Yi Yang; Thomas de Jaeger.}
\author[0000-0002-2636-6508]{WeiKang Zheng},
\email{weikang@berkeley.edu, dessart@iap.fr, afilippenko@berkeley.edu; yi\_yang@mail.tsinghua.edu.cn; dejaeger.thomas@gmail.com}
\affiliation{Department of Astronomy, University of California, Berkeley, CA 94720-3411, USA}

\author[0000-0003-0599-8407]{Luc Dessart},
\affiliation{Institut d'Astrophysique de Paris, CNRS-Sorbonne Universit\'e, 98 bis boulevard Arago, F-75014 Paris, France}

\author[0000-0003-3460-0103]{Alexei~V.~Filippenko}
\affiliation{Department of Astronomy, University of California, Berkeley, CA 94720-3411, USA}

\author[0000-0002-6535-8500]{Yi Yang
\begin{CJK}{UTF8}{gbsn}
(杨轶)
\end{CJK}}
\affiliation{Physics Department, Tsinghua University, Beijing, 100084, China}
\affiliation{Department of Astronomy, University of California, Berkeley, CA 94720-3411, USA}

\author[0000-0001-5955-2502]{Thomas~G.~Brink}
\affiliation{Department of Astronomy, University of California, Berkeley, CA 94720-3411, USA}

\author[0000-0001-6069-1139]{Thomas de Jaeger},\affiliation{LPNHE, (CNRS/IN2P3, Sorbonne Universit\'{e}, Universit\'{e} Paris Cit\'{e}), Laboratoire de Physique Nucl\'{e}aire et de Hautes \'{E}nergies, 75005, Paris, France}

\author[0000-0002-4951-8762]{Sergiy~S.Vasylyev}
\affiliation{Department of Astronomy, University of California, Berkeley, CA 94720-3411, USA}

\author[0000-0001-9038-9950]{Schuyler D.~Van Dyk}
\affiliation{Caltech/IPAC, Mail Code 100-22, Pasadena, CA 91125, USA}

\author[0000-0002-1092-6806]{Kishore C. Patra}
\affiliation{Department of Astronomy, University of California, Berkeley, CA 94720-3411, USA}

\author[0000-0002-3934-2644]{Wynn V. Jacobson-Gal\'an}
\affiliation{Department of Astronomy, University of California, Berkeley, CA 94720-3411, USA}
\affiliation{Department of Astronomy and Astrophysics, California Institute of Technology, Pasadena, CA 91125, USA}

\author[0009-0000-2503-140X]{Gabrielle E. Stewart}
\affiliation{Department of Astronomy, University of California, Berkeley, CA 94720-3411, USA}

\author[0009-0005-8159-8490]{Efrain Alvarado III}
\affiliation{Department of Astronomy, University of California, Berkeley, CA 94720-3411, USA}
\affiliation{Department of Physics and Astronomy, San Francisco State University, 1600 Holloway Avenue, San Francisco, CA 94132, USA}

\author{Veda Arikatla}
\affiliation{Department of Astronomy, University of California, Berkeley, CA 94720-3411, USA}

\author{Pallas Beddow}
\affiliation{Department of Astronomy, University of California, Berkeley, CA 94720-3411, USA}

\author{Andreas Betz}
\affiliation{Department of Astronomy, University of California, Berkeley, CA 94720-3411, USA}

\author{Emma Born}
\affiliation{Department of Astronomy, University of California, Berkeley, CA 94720-3411, USA}

\author{Kate Bostow}
\affiliation{Department of Astronomy, University of California, Berkeley, CA 94720-3411, USA}

\author[0000-0002-6523-9536]{Adam~J.~Burgasser}
\affiliation{Department of Astronomy \& Astrophysics, University of California at San Diego, La Jolla, CA 92093, USA}

\author{Osmin Caceres}
\affiliation{Department of Physics and Astronomy, University of California, Los Angeles, CA 90095-1547, USA}

\author[0009-0005-5223-1606]{Evan M. Carrasco}
\affiliation{Department of Astronomy \& Astrophysics, University of California, Santa Cruz, CA 95064, USA}

\author[0000-0001-9984-5131]{Elma Chuang}
\affiliation{Department of Astronomy, University of California, Berkeley, CA 94720-3411, USA}

\author[0009-0001-2794-8278]{Asia DeGraw}
\affiliation{Department of Astronomy, University of California, Berkeley, CA 94720-3411, USA}

\author[0000-0002-3739-0423]{Elinor~L.~Gates}
\affiliation{University of California Observatories/Lick Observatory, Mount Hamilton, CA 95140}

\author{Eli Gendreau-Distler}
\affiliation{Department of Astronomy, University of California, Berkeley, CA 94720-3411, USA}

\author{Cooper Jacobus}
\affiliation{Department of Astronomy, University of California, Berkeley, CA 94720-3411, USA}

\author{Connor Jennings} 
\affiliation{Department of Astronomy, University of California, Berkeley, CA 94720-3411, USA}

\author[0000-0002-1480-9041]{Preethi~R.~Karpoor}
\affiliation{Department of Astronomy \& Astrophysics, University of California at San Diego, La Jolla, CA 92093, USA}

\author{Paul Lynam}
\affiliation{University of California Observatories/Lick Observatory, Mount Hamilton, CA 95140}

\author{Ann Mina}
\affiliation{Department of Astronomy, University of California, Berkeley, CA 94720-3411, USA}

\author{Katherine Mora}
\affiliation{Department of Astronomy, University of California, Berkeley, CA 94720-3411, USA}

\author[0009-0009-7665-6827]{Neil Pichay}
\affiliation{Department of Astronomy, University of California, Berkeley, CA 94720-3411, USA}

\author{Jyotsna Ravi}
\affiliation{Department of Astronomy, University of California, Berkeley, CA 94720-3411, USA}

\author[0000-0002-5376-3883]{Jon Rees}
\affiliation{University of California Observatories/Lick Observatory, Mount Hamilton, CA 95140}

\author[0000-0003-0427-8387]{R. Michael Rich}
\affiliation{Department of Physics and Astronomy, University of California, Los Angeles, CA 90095-1547, USA}

\author{Sophia Risin}
\affiliation{Department of Astronomy, University of California, Berkeley, CA 94720-3411, USA}

\author[0000-0002-7393-3595]{Nathan R. Sandford}
\affiliation{Department of Astronomy and Astrophysics, University of Toronto, 50 St. George Street, Toronto ON, M5S 3H4, Canada}

\author[0000-0002-1445-4877]{Alessandro Savino}
\affiliation{Department of Astronomy, University of California, Berkeley, CA 94720-3411, USA}

\author[0000-0002-1420-1837]{Emma~Softich}
\affiliation{Department of Astronomy \& Astrophysics, University of California at San Diego, La Jolla, CA 92093, USA}

\author[0000-0002-9807-5435]{Christopher~A.~Theissen}
\affiliation{Department of Astronomy \& Astrophysics, University of California at San Diego, La Jolla, CA 92093, USA}

\author[0009-0002-2209-4813]{Edgar P. Vidal}
\affiliation{Department of Physics and Astronomy, Tufts University, Medford, MA 02155, USA}
\affiliation{Department of Astronomy, University of California, Berkeley, CA 94720-3411, USA}

\author[0009-0000-0753-4345]{William Wu}
\affiliation{Department of Astronomy, University of California, Berkeley, CA 94720-3411, USA}

\author{Yoomee Zeng}
\affiliation{Department of Astronomy, University of California, Berkeley, CA 94720-3411, USA}

\begin{abstract}
We present photometric and spectroscopic observations of SN\,2023ixf covering from day one to 442 days after explosion. SN\,2023ixf reached a peak $V$-band absolute magnitude of $-18.2 \pm 0.07$, and light curves show that it is in the fast-decliner (IIL) subclass with a relatively short ``plateau'' phase (fewer than $\sim 70$ days). Early-time spectra of SN\,2023ixf exhibit strong, very narrow emission lines from  ionized circumstellar matter (CSM), possibly indicating a Type IIn classification. But these flash/shock-ionization emission features faded after the first week and the spectrum evolved in a manner similar to that of typical Type II SNe, unlike the case of most genuine SNe~IIn in which the ejecta interact with CSM for an extended period of time and develop intermediate-width emission lines.
We compare observed spectra of SN\,2023ixf with various model spectra to understand the physics behind SN\,2023ixf.
Our nebular spectra (between 200-400 d) match best with the model spectra from a 15\,$\rm M_{\odot}$ progenitor which  experienced  enhanced mass loss a few years before explosion. A last-stage mass-loss rate of $\dot{M} = 0.01\, \rm M_{\odot}\,yr^{-1}$ from the r1w6 model matches best with the early-time spectra, higher than $\dot{M} \approx 2.4 \times 10^{-3}\, \rm M_{\odot}\,yr^{-1}$ derived from the ionized H${\alpha}$ luminosity at 1.58\,d. 
We also use SN\,2023ixf as a distance indicator and fit the light curves to derive the Hubble constant by adding SN\,2023ixf to the existing sample; we obtain H$_{0}=73.1^{+3.68}_{-3.50}$\,km\,s$^{-1}$\,Mpc$^{-1}$, consistent with the results from SNe~Ia and many other independent methods.

\end{abstract}

\keywords{supernovae: individual (SN\,2023ixf) --- techniques: spectroscopic}


\section{Introduction}\label{sec:intro}

Type II supernovae (hereafter SNe II) are characterized by strong hydrogen Balmer lines in their optical spectra \citep[e.g.,][]{filippenko_optical_1997}. Thanks to progenitor detections \citep[e.g.,][]{vandyk03,smartt09,smartt15,
VanDyk2017}, it is commonly accepted that SNe~II are massive-star explosions at the end of their lives ($\gtrsim 8\,{\rm M}_{\odot}$). While historically, SNe~II were subgrouped based on their light-curve mythologies as SNe~IIP with a long-duration plateau and SNe~IIL with faster linear decline rates in magnitude, recent studies suggest that the light curves actually form a more continuous distribution \citep[e.g.,][]{anderson_characterizing_2014, sanders_toward_2015, valenti_diversity_2016,galbany16,dejaeger_berkeley_2019}.

However, even if SNe~II constitute a uniform physical family, a great variety of objects is seen in terms of their photometric and spectroscopic properties. This diversity results from differences among progenitors and explosion mechanisms (e.g., mass of hydrogen envelope, radius, metallicity). For example, SNe~II whose optical spectra exhibit long-lasting, intermediate-width emission lines that indicate expansion speeds of a few hundred to $\sim 1000$ km~s$^{-1}$, but little or no P~Cygni absorption component, have been dubbed SNe~IIn by \cite{Schlegel90}; the ``n" indicates the presence of the relatively ``narrow" component. Examples were studied by \cite{Filippenko89,Filippenko91a,Filippenko91b}, \cite{Stathakis91}, and others thereafter. The SN ejecta in these objects are likely interacting with circumstellar matter (CSM). However, \cite{Ransome21} examine a wide variety of objects proposed to be SNe~IIn, concluding that many of them do not even actually correspond to the terminal explosions of stars and are thus ``SN impostors," a term coined by \cite{VanDyk00} and subsequently used by many other authors.

In some cases, SNe~IIn also show very narrow emission lines from gas having speeds $< 100$\,km\,s$^{-1}$; these are probably produced by flash or
photoionization of CSM soon after shock breakout. 
Well-known examples include (among many others) SN\,1998S \citep{Leonard00}, SN\,2013fs \citep{Yaron17}, SN\,2013cu \citep{Gal-Yam14}, SN\,2014G \citep{Terreran16}, SN\,2017ahn \citep{Tartaglia21},  SN\,2020pni \citep{Terreran22}, and SN\,2020tlf \citep{Jacobson22}. These very narrow emission lines often disappear within a few days or a week, and the SN thereafter evolves in a manner more similar to that of typical Type II SNe \citep{Bruch23,Jacobson24}.  
Such objects, with only fleeting narrow emission lines and no long-lasting intermediate-width component, do not necessarily qualify as genuine SNe~IIn, although opinions regarding this differ among authors.

Turning now to one aspect of their cosmological significance, SNe~II have been established as useful independent extragalactic distance indicators \citep[e.g.,][]{hamuy02,dejaeger20}, especially for determining the Hubble-Lema{\^i}tre constant (see \citealt{review} for a review) and addressing the current tension between early- and late-Universe measurements \citep{riess_comprehensive_2022}.

Owing to its exceptionally early detection in a nearby galaxy, SN\,2023ixf provides a unique opportunity to explore and understand the zoo of SNe~II; photometry and spectra were acquired starting $< 1$\,day after the explosion. Moreover, SN\,2023ixf allows us to test the accuracy of different extragalactic distance methods found in the literature. 

SN\,2023ixf was discovered by Koichi Itagaki on 2023-05-19 at 17:27:15 (UTC dates are used throughout this paper) 
at a {\it Clear}-band magnitude of 14.9 \citep{2023TNSTR1158....1I}, in the nearby galaxy Messier\,101 (M101, also known as NGC 5457 and informally as the ``Pinwheel Galaxy").
Since M101 is a famous face-on spiral galaxy, it is regularly imaged by amateur astronomers throughout the world. After Itagaki reported the discovery of SN\,2023ixf, many groups including amateurs and professionals reported prediscovery detections \citep{Perley2023, filippenko_filippenko_2023,Fulton2023,Zhang2023,Hamann2023,Limeburner2023,mao_onset_2023, Yaron2023,Koltenbah2023,Chufarin2023}; see, in particular, the early-time unfiltered light curve of \citet{Sgro23} obtained with Unistellar 11.4\,cm telescopes and the multiband photometry of \cite{Li23}.
These prediscovery observations range from a few minutes to a few days before Itagaki's first report, and cover well before and after the explosion time of SN\,2023ixf; thus, the explosion time can be precisely estimated \citep{Hosseinzadeh2023,Hiramatsu2023}.
In this paper, we adopt the first-light time of SN\,2023ixf to be MJD = 60082.788$^{+0.02}_{-0.05}$ estimated by \cite{Li23}. 

Besides SN\,2023ixf, M101 also hosted a few other historical SNe, including SN\,1909A (type unknown), SN\,1951H (type unknown), SN\,1970G (Type II), and SN\,2011fe (Type Ia). Among these, the most important is SN\,Ia\,2011fe since it is one of the best-observed SNe and can be used for precise distance ($D$) estimation. \cite{Vinko2012} fit the multiband light curves of SN\,2011fe using both the MLCS2k2 and SALT2 methods, deriving  $D = 6.95 \pm 0.23$\,Mpc (distance modulus $\mu = 29.21 \pm 0.07$\,mag) from MLCS2k2 and $D = 6.46 \pm 0.21$\,Mpc ($\mu = 29.05 \pm 0.07$\,mag) from SALT2.

Given the proximity of M101, there are many other methods for measuring its distance. The most common and reliable one uses Cepheid variables. Over the past quarter century, several groups have independently measured the distance to M101 using Cepheids 
\citep{Freedman2001,Saha2006,Shappee2011}.
\cite{Riess2022}  reported the latest Cepheid distance of $6.85 \pm 0.15$\,Mpc ($\mu = 29.194 \pm 0.039$\,mag). The optical tip of the red giant branch (TRGB) method is another option for distance measurement; \cite{Beaton2019} found $D = 6.52 \pm 0.12_{\rm stat} \pm 0.15_{\rm sys}$ ($\mu = 29.07 \pm 0.04_{\rm stat} \pm 0.05_{\rm sys}$\,mag) using the latest TRGB method. Here for our absolute-magnitude analysis, we adopt $\mu = 29.194 \pm 0.039$\,mag from \cite{Riess2022}.

This paper is organized as follows. Section 2 contains a description of the observations and the data. In Section 3,
we discuss the evolution of the light curves and color curves, while in Section 4, we present the analysis of the spectroscopic data. Section 5  compares different methods to measure distances, and we conclude with a summary in Section 6.

\section{Observations and Data Reduction}~\label{sec:obs}
\subsection{Photometry}
We performed follow-up multiband observations of SN\,2023ixf with both the 0.76\,m Katzman Automatic Imaging Telescope (KAIT) and the 1\,m Nickel telescope as part of the Lick Observatory Supernova Search \citep[LOSS;][]{Filippenko2001}. $B$, $V$, $R$, and $I$ images of SN\,2023ixf were obtained with both telescopes; in addition, $Clear$-band (close to the $R$ band; see \citealt{Li2003}) images were obtained with KAIT. For the second and third nights, we also performed high-cadence observations with KAIT, rotating between the $B$, $V$, $R$, $I$, and $Clear$ bands with each exposure time being 60\,s (plus an overhead of $\sim 30$\,s), which gave a cadence of about 9\,min for each band.
(During the first night, unfortunately, winds at Lick Observatory were too strong for KAIT to be used; we had intended to conduct high-cadence photometry.)

All images were reduced using a custom pipeline\footnote{https://github.com/benstahl92/LOSSPhotPypeline} detailed by \citet[][]{Stahl2019}. Point-spread-function (PSF) photometry was obtained using {\tt DAOPHOT} \citep[][]{Stetson1987} from the {\tt IDL} Astronomy Users Library\footnote{http://idlastro.gsfc.nasa.gov/}. Owing to the small field of view of our images, only one reference star was available for calibration, namely star ``m'' from \citet[][ see their Fig. 1]{Henden2012}. The \citet{Landolt92} magnitudes of star ``m'' were transformed to the KAIT/Nickel natural system before calibration. Apparent magnitudes were all measured in the KAIT4/Nickel2 natural system, and the final results were transformed to the standard system using the local calibrator and color terms for KAIT4 and Nickel2 \citep[see][]{Stahl2019}.

\subsection{Spectroscopy} 

We acquired 42 spectra (see Table \ref{speclog} for a full log) of SN\,2023ixf, generally with the Kast double spectrograph \citep{miller_stone_1994} mounted on the Shane 3\,m telescope at Lick Observatory. Most observations utilized the $2''$-wide slit, 600/4310 grism, and 300/7500 grating (with a few exceptions). This instrument configuration has a combined wavelength range of $\sim 3500$-–10,500\,\AA\ and a spectral resolving power of $R \equiv \lambda/\Delta \lambda \approx 800$.  To minimize slit losses caused by atmospheric dispersion \citep{Filippenko1982}, the long slit was oriented at or near the parallactic angle. The data were reduced following standard techniques for CCD processing and spectrum extraction \citep{silverman_berkeley_2012} with  IRAF \citep{Tody1986} routines and custom Python and IDL codes\footnote{https://github.com/ishivvers/TheKastShiv}.  Low-order polynomial fits to comparison-lamp spectra were used to calibrate the wavelength scale, and small adjustments derived from night-sky emission lines in the target frames were applied. The spectra were flux calibrated using observations of appropriate spectrophotometric standard stars observed on the same night, at similar airmasses, and with an identical instrument configuration.
Among these spectra, some from the early phases have already been published by \cite{Zimmerman24} and \cite{Jacobson-Galan23} (see Table 1 for details), but they are reused here for our more detailed analysis.

\begin{deluxetable*}{cccccccc}
 \tabcolsep 2.0mm
 \tablewidth{0pt}
 \tablecaption{Log of Lick/Kast Spectroscopic Observations}
 \tablehead{
   \colhead{Time (UTC)} & \colhead{MJD} & \colhead{Phase$^a$ (d)} & \colhead{Grism} & \colhead{Grating} & \colhead{Exp. time (s)} & \colhead{Slit width ($''$)} & \colhead{Resolution (\AA)}
}
\startdata
\hline
20230520.183$^b$ & 60084.183  & 1.39   & 600/4310  & 300/7500   & 600/600    & 2.0  & 4.7/11.9   \\
20230520.196$^b$ & 60084.196  & 1.41   & 600/4310  & 600/5000   & 1200/1200  & 1.0  & 2.4/3.0    \\
20230520.367$^b$ & 60084.367  & 1.58   & 600/4310  & 1200/5000  & 3660/3600  & 1.0  & 2.4/1.5    \\
20230520.476$^b$ & 60084.476  & 1.69   & 600/4310  & 600/5000   & 1200/1200  & 1.0  & 2.4/3.0    \\
20230520.488$^b$ & 60084.488  & 1.7    & 600/4310  & 300/7500   & 600/600    & 2.0  & 4.7/11.9   \\
20230521.19$^c$  & 60085.19   & 2.4    & -         & -          & -          & -    & -/-        \\
20230522.202$^d$ & 60086.202  & 3.4    & 600/4310  & 600/7500   & 600/600    & 2.0  & 4.7/6.0    \\
20230523.231$^d$ & 60087.231  & 4.4    & 600/4310  & 600/7500   & 600/600    & 2.0  & 4.7/6.0    \\
20230524.313$^d$ & 60088.313  & 5.5    & 600/4310  & 600/7500   & 360/360    & 1.0  & 2.4/3.0    \\
20230525.197$^d$ & 60089.197  & 6.4    & 600/4310  & 600/7500   & 400/400    & 1.0  & 2.4/3.0    \\
20230527.210$^d$ & 60091.210  & 8.4    & 600/4310  & 600/7500   & 180/180    & 1.0  & 2.4/3.0    \\
20230528.189 & 60092.189  & 9.4    & 600/4310  & 300/7500   & 200/200    & 2.0  & 4.7/11.9   \\
20230528.196 & 60092.196  & 9.4    & 600/4310  & 600/5000   & 500/500    & 1.0  & 2.4/3.0    \\
20230528.435 & 60092.435  & 9.6    & 600/4310  & 300/7500   & 300/300    & 2.0  & 4.7/11.9   \\
20230528.445 & 60092.445  & 9.7    & 600/4310  & 600/5000   & 1200/1200  & 1.0  & 2.4/3.0    \\
20230530.247$^d$ & 60094.247  & 11.5   & 600/4310  & 600/7500   & 360/360    & 1.0  & 2.4/3.0    \\
20230531.193$^d$ & 60095.193  & 12.4   & 600/4310  & 600/7500   & 210/210    & 1.0  & 2.4/3.0    \\
20230601.194 & 60096.194  & 13.4   & 600/4310  & 600/7500   & 300/300    & 1.0  & 2.4/3.0    \\
20230602.258$^d$ & 60097.258  & 14.5   & 600/4310  & 300/7500   & 200/200    & 2.0  & 4.7/11.9   \\
20230610.262 & 60105.262  & 22.5   & 600/4310  & 300/7500   & 300/300    & 2.0  & 4.7/11.9   \\
20230619.246 & 60114.246  & 31.5   & 600/4310  & 300/7500   & 600/600    & 2.0  & 4.7/11.9   \\
20230622.266 & 60117.266  & 34.5   & 600/4310  & 300/7500   & 200/200    & 2.0  & 4.7/11.9   \\
20230626.242 & 60121.242  & 38.5   & 600/4310  & 300/7500   & 200/200    & 2.0  & 4.7/11.9   \\
20230710.233 & 60135.233  & 52.4   & 600/4310  & 300/7500   & 200/200    & 2.0  & 4.7/11.9   \\
20230719.319 & 60144.319  & 61.5   & 600/4310  & 300/7500   & 200/200    & 2.0  & 4.7/11.9   \\
20230811.246 & 60167.246  & 84.5   & 600/4310  & 300/7500   & 1200/1200  & 2.0  & 4.7/11.9   \\
20230824.232 & 60180.232  & 97.4   & 600/4310  & 300/7500   & 600/600    & 2.0  & 4.7/11.9   \\
20230908.158 & 60195.158  & 112.4  & 600/4310  & 300/7500   & 461/600    & 2.0  & 4.7/11.9   \\
20230916.144 & 60203.144  & 120.4  & -         & 300/7500   & -/360      & 2.0  & -  /11.9   \\
20231013.099 & 60230.099  & 147.3  & 600/4310  & 300/7500   & 300/300    & 0.5  & 1.2/3.0    \\
20231109.565 & 60257.565  & 174.8  & 600/4310  & 300/7500   & 1200/1200  & 2.0  & 4.7/11.9   \\
20231110.561 & 60258.561  & 175.8  & 600/4310  & 300/7500   & 1200/1200  & 2.0  & 4.7/11.9   \\
20231212.534 & 60290.534  & 207.7  & 600/4310  & 600/5000   & 2460/2400  & 1.0  & 2.4/3.0    \\
20240112.504 & 60321.504  & 238.7  & 600/4310  & 600/5000   & 3060/3000  & 1.0  & 2.4/3.0    \\
20240118.561 & 60327.561  & 244.8  & 600/4310  & 300/7500   & 2160/1400  & 2.0  & 4.7/11.9   \\
20240207.668$^e$ & 60347.668  & 264.9  & 600/4000  & 400/8500   & 250/250    & 1.0  & 4.7/8.6    \\
20240313.658$^e$ & 60382.658  & 299.9  & 600/4000  & 400/8500   & 180/180    & 1.0  & 4.7/8.6    \\
20240411.314 & 60411.314  & 328.5  & 600/4310  & 300/7500   & 2760/2700    & 2.0  & 4.7/11.9   \\
20240515.406 & 60445.406  & 362.6  & 600/4310  & 300/7500   & 2760/2700    & 2.0  & 4.7/11.9   \\
20240531.449 & 60461.449  & 378.7  & 600/4310  & 300/7500   & 3060/3000    & 2.0  & 4.7/11.9   \\
20240629.352 & 60490.352  & 407.6  & 600/4310  & 300/7500   & 3060/3000    & 2.0  & 4.7/11.9   \\
20240802.276 & 60524.276  & 441.5  & 600/4310  & 300/7500   & 3060/3000   & 2.0  & 4.7/11.9   \\
\enddata
\tablenotetext{a}{Relative to first-light time of MJD = 60082.788.}
\tablenotetext{b}{Already published by \cite{Zimmerman24}.}
\tablenotetext{c}{Adopted from \cite{Jacobson-Galan23} for analysis.}
\tablenotetext{d}{Already published by \cite{Jacobson-Galan23}, but with our independent reduction.}
\tablenotetext{e}{Taken with 10\,m Keck telescope + LRIS.}
\label{speclog}
\end{deluxetable*}

\section{Light-Curve Analysis}~\label{sec:lcanalysis}

\begin{figure}
    \centering
    \includegraphics[width=0.99\linewidth]{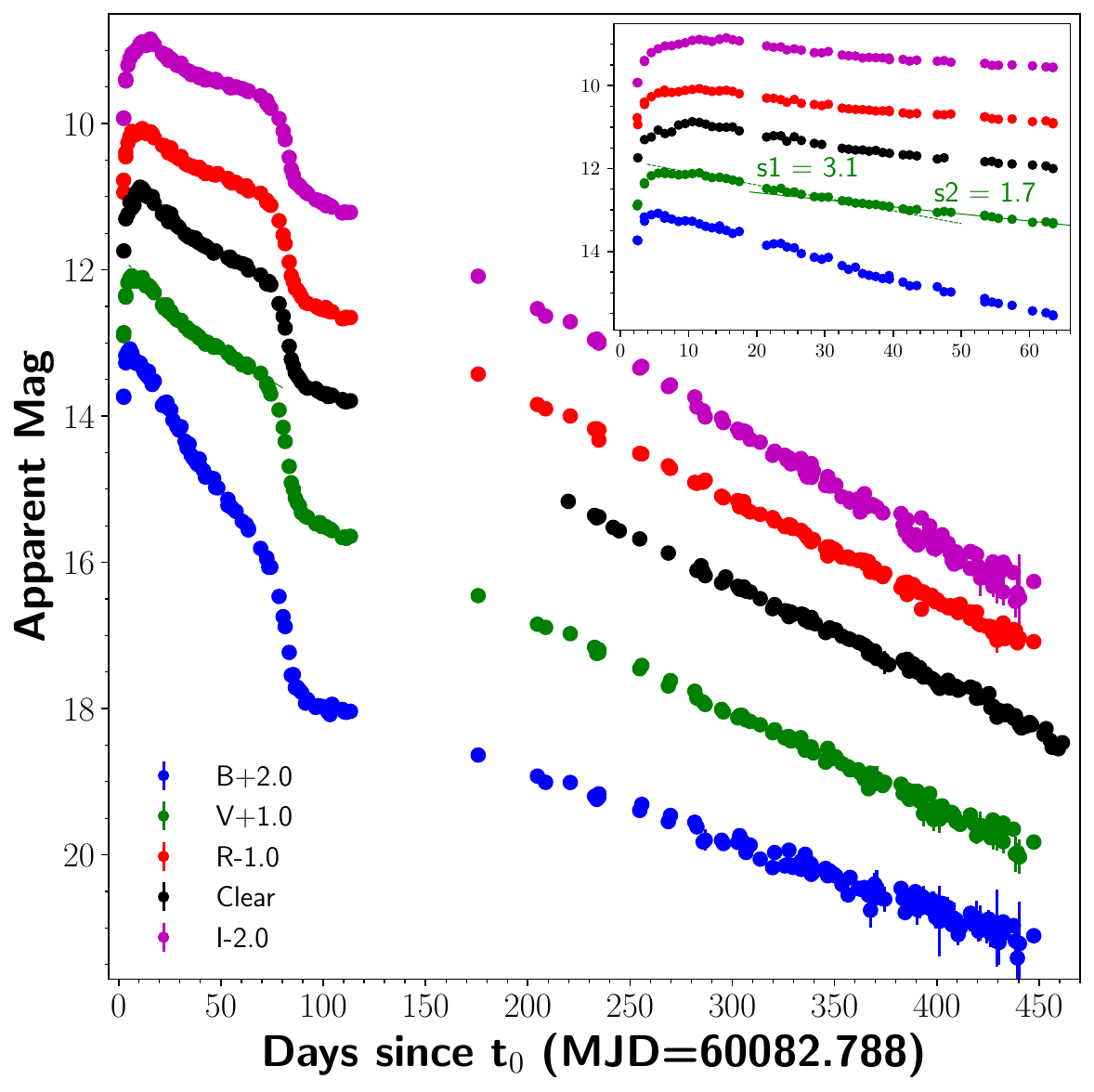}
    \caption{ 
    LOSS multiband light curves of SN\,2023ixf, color coded for different bands. Inset at top right shows the zoomed-in light curves at early phases. The $s1$ and $s2$ parameters are shown for the $V$-band fitting following \cite{anderson_characterizing_2014}. With an $s2$ value of 1.7\,mag, SN\,2023ixf is clearly in the fast-decliner (IIL) subclass instead of Type IIP.
    }~\label{fig:2023ixf_lc}
\end{figure}

Figure \ref{fig:2023ixf_lc} displays the full light curves of SN\,2023ixf from our observations. The brightness of SN\,2023ixf quickly rises to peak brightness in the first few days, especially in bluer bands, while the redder bands reach maximum brightness a bit later. There are noticeable fluctuations around the time of peak brightness in almost all bands, and the redder bands ($R$ and $I$) actually peaked on the bumps. These early-time fluctuations are likely caused by CSM interaction, which was also detected in the early-time spectra (see Sec. \ref{sec:specanalysis}).
We use a low-order polynomial to fit the multiband light curves and determine peak times of MJD = 60087.9, 60088.5, 60094.5, and 60097.3 (with 1$\sigma$ uncertainties of 0.5\,d) in the $B$, $V$, $R$, and $I$ bands (respectively), as well as respective apparent peak magnitudes of 11.08, 11.11, 11.05, and 10.88 (with 1$\sigma$ uncertainties of 0.05\,mag). By adopting a first-light time of MJD = 60082.788$^{+0.02}_{-0.05}$ following \cite{Li23}, this gives rise times of $t_{r,B} = 5.1$\,d, $t_{r,V} = 5.7$\,d, $t_{r,R} = 11.7$\,d, and $t_{r,I} = 14.5$\,d (with 1$\sigma$ uncertainties of $\sim 0.5$\,d). Our rise-time estimates are consistent with the results given by \cite{Jacobson-Galan23} in the $B$ and $V$ bands, but our $R$ and $I$  rise times are longer than their values, likely because our peak-time measurements in $R$ and $I$ landed on the bumps at later times.
\begin{figure}
    \centering
    \includegraphics[width=0.99\linewidth]{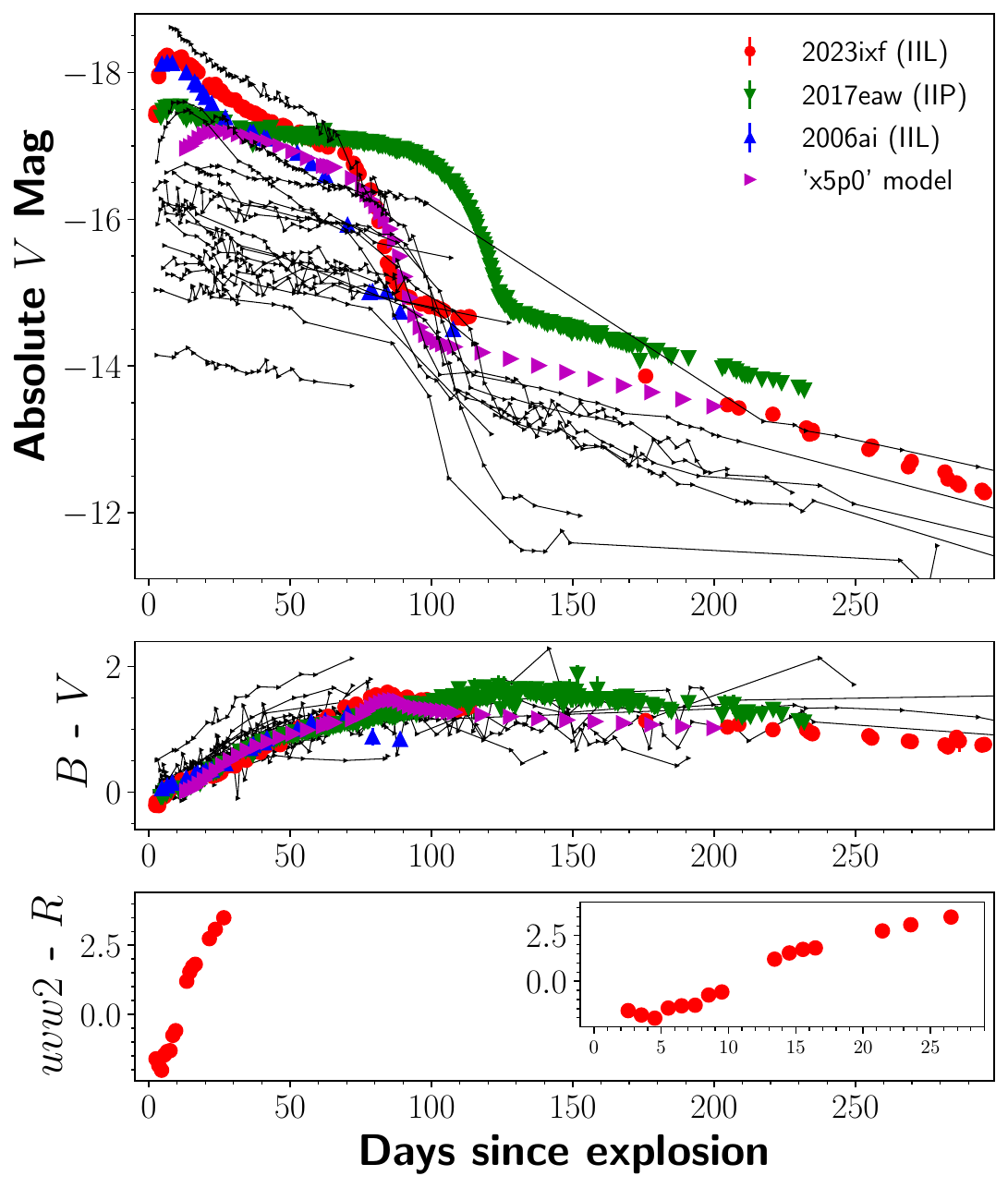}
    \caption{
    {\it Top panel:} Absolute $V$-band light curve of SN\,2023ixf (red) compared with the typical Type IIP SN\,2017eaw (green), the Type IIL SN\,2006ai, the model ``x5p0" (purple) from \cite{Hillier19} and a well-observed sample of SNe~II (black) from \cite{deJaeger2019}.
    SN\,2023ixf lies at the bright end of the Type II sample, similar to SN\,2006ai in absolute peak $V$-band magnitude, and exhibits a relatively short ``plateau'' phase (less than $\sim 70$ days). {\it Middle panel:} The $B - V$ color evolution of SN\,2023ixf and the comparison sample; all show very similar $B - V$ curves, with blue colors at the beginning evolving to redder colors. {\it Bottom panel:} The $uvw2 - R$ color evolution (with a zoomed-in panel) of SN\,2023ixf.
    }~\label{fig:2023ixf_lc_compare}
\end{figure}
We also fit the $s1$ and $s2$ parameters to the $V$ light curve (see Fig. \ref{fig:2023ixf_lc}) following \cite{anderson_characterizing_2014}, finding $s1 = 3.1 \pm 0.1$ and $s2 = 1.7 \pm 0.1$\,mag. This $s2$ value  clearly puts SN\,2023ixf in the fast-decliner (IIL) SN subclass instead of Type II plateau (IIP), at least according to their classification scheme.

To better compare with other SNe~II, we plot the absolute $V$-band light curve in Figure \ref{fig:2023ixf_lc_compare} along with that of other SNe. To derive the absolute magnitude, we adopt a distance modulus $\mu = 29.194 \pm 0.039$\,mag from \cite{Riess2022} and correct for both the Milky Way (MW) extinction of $E(B-V) = 0.008$\,mag \citep{schlafly_measuring_2011} and the host-galaxy extinction of $E(B-V) = 0.032$\,mag following \cite{VanDyk2023}. SN\,2023ixf reached a peak $V$ absolute magnitude of $-18.2 \pm 0.07$,
 at the bright end of SNe~II (see also \citealt{Teja2023}) as shown in Figure \ref{fig:2023ixf_lc_compare}, where a well-observed sample of SNe~II from \cite{deJaeger2019} is plotted for comparison (black lines). We also illustrate the typical Type IIP SN\,2017eaw \citep{dyk_type_2019} as well as the Type IIL SN\,2006ai \citep{Hiramatsu2021}. 
 It is interesting to note that SN\,2006ai not only has a similar peak absolute magnitude as SN\,2023ixf, but also a relatively short ``plateau'' phase (though both of them are SNe~IIL) of $\lesssim 70$\,days, compared to SN\,2017eaw which lasts $\sim 100$\,days before the light curve falls off the plateau. A shorter plateau phase usually indicates an H-rich envelope of lower mass in the progenitor before explosion \citep{Hillier19}. 
Model light curves for red supergiant (RSG) explosions with different H-rich envelope masses have been presented in a number of studies, including \cite{Morozova15} and \cite{Hillier19}. Model ``x5p0" from the latter matches closely the overall light curve of SN\,2023ixf apart from the underestimate in early-time brightness (see Fig. \ref{fig:2023ixf_lc_compare}). 
This arises because model ``x5p0" is a bare explosion (i.e., explosion within a vacuum), thus supporting the notion that the early-time luminosity boost stems from ejecta-CSM interaction.

The $B - V$ color evolution of SN\,2023ixf and the comparison sample is also shown in Figure \ref{fig:2023ixf_lc_compare} in the middle panel. SN\,2023ixf, SN\,2006ai, SN\,2017eaw, and the rest of the sample all have very similar $B - V$  evolution, with a blue color at the beginning evolving to a redder color. We plot the $uvw2 - R$ evolution in the bottom panel, where the $uvw2$-band data were adopted from \cite{Zimmerman24}; this shows that SN\,2023ixf reached its bluest UV/optical color $\sim 5$ days after explosion.

\begin{figure}
    \centering
    \includegraphics[width=0.99\linewidth]{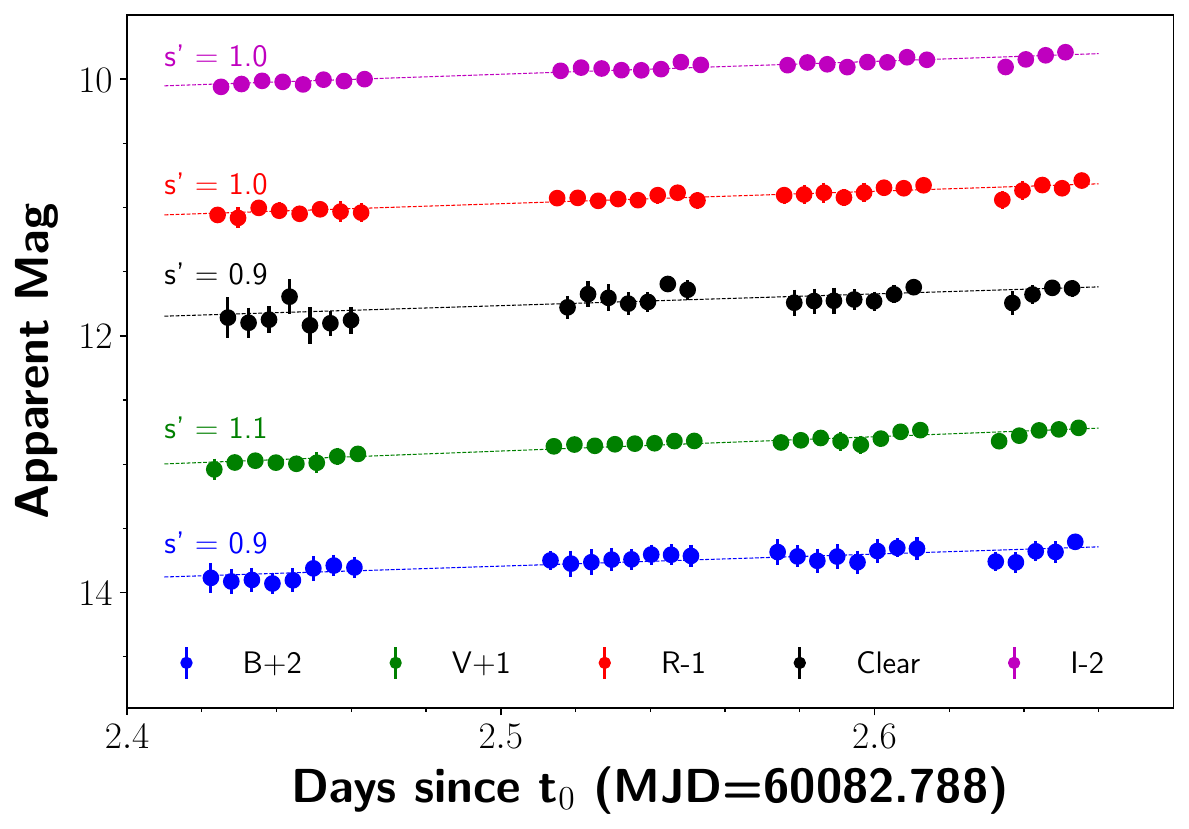}
    \includegraphics[width=0.99\linewidth]{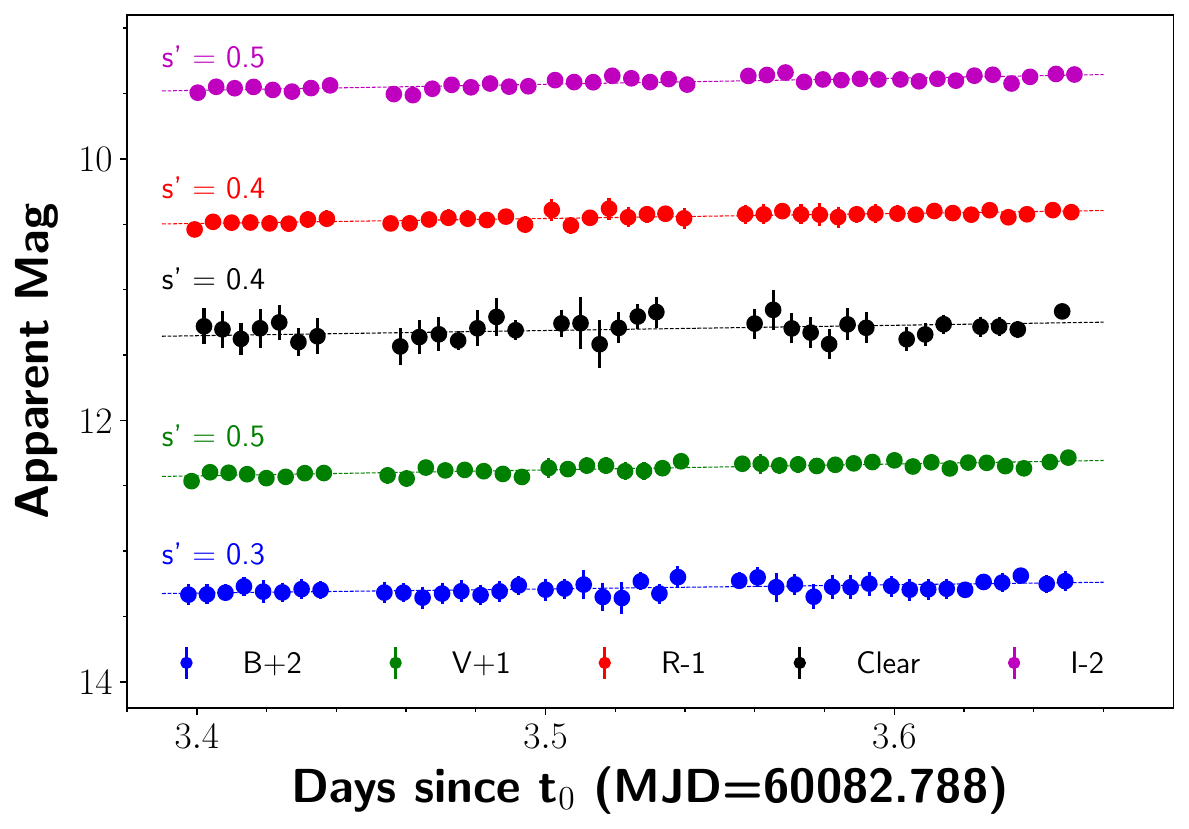}
    \caption{ 
    High-cadence KAIT light curves during Night 2 (top panel) and Night 3 (bottom panel). The smooth light curves show no evidence of rapid variability in both nights. The SN brightened steadily, with rates of $\sim 1.0$ and $\sim 0.4$\,mag per day  during Nights 2 and 3, respectively. }~\label{fig:2023ixf_lc_high_cadence}
\end{figure}

Figure \ref{fig:2023ixf_lc_high_cadence} shows our high-cadence light curves from Night 2 (top panel) and Night 3 (bottom panel). Such data can be used to search for rapid variability in SN light curves, as done for SN~2014J by \cite{Bonanos16}, who performed observations with a cadence of 2\,min in $B$ and $V$, finding evidence for rapid variability at the 0.02--0.05\,mag level on timescales of 15--60\,min. For SN\,2023ixf, we found no evidence of similar rapid variability in the second and third nights, since our high-cadence light curves (as shown in Fig. \ref{fig:2023ixf_lc_high_cadence}) are quite smooth, and also because our cadence is much lower. However, our light curves clearly show that the SN brightened steadily during both nights. Here we define $s'$ as the daily brightening rate; we find $s' = 1.0 \pm 0.1$\,mag\,day$^{-1}$ for all  bands in Night 2 and $s' = 0.4 \pm 0.1$\,mag\,day$^{-1}$ for all  bands in Night 3, so the SN brightened more quickly during the second night than the third night. A higher cadence with fewer filters may help reveal rapid variability (if it exists) in future observations of bright SNe.

\section{Spectral Analysis}~\label{sec:specanalysis}
\begin{figure}
    \centering
    \includegraphics[width=0.97\linewidth]{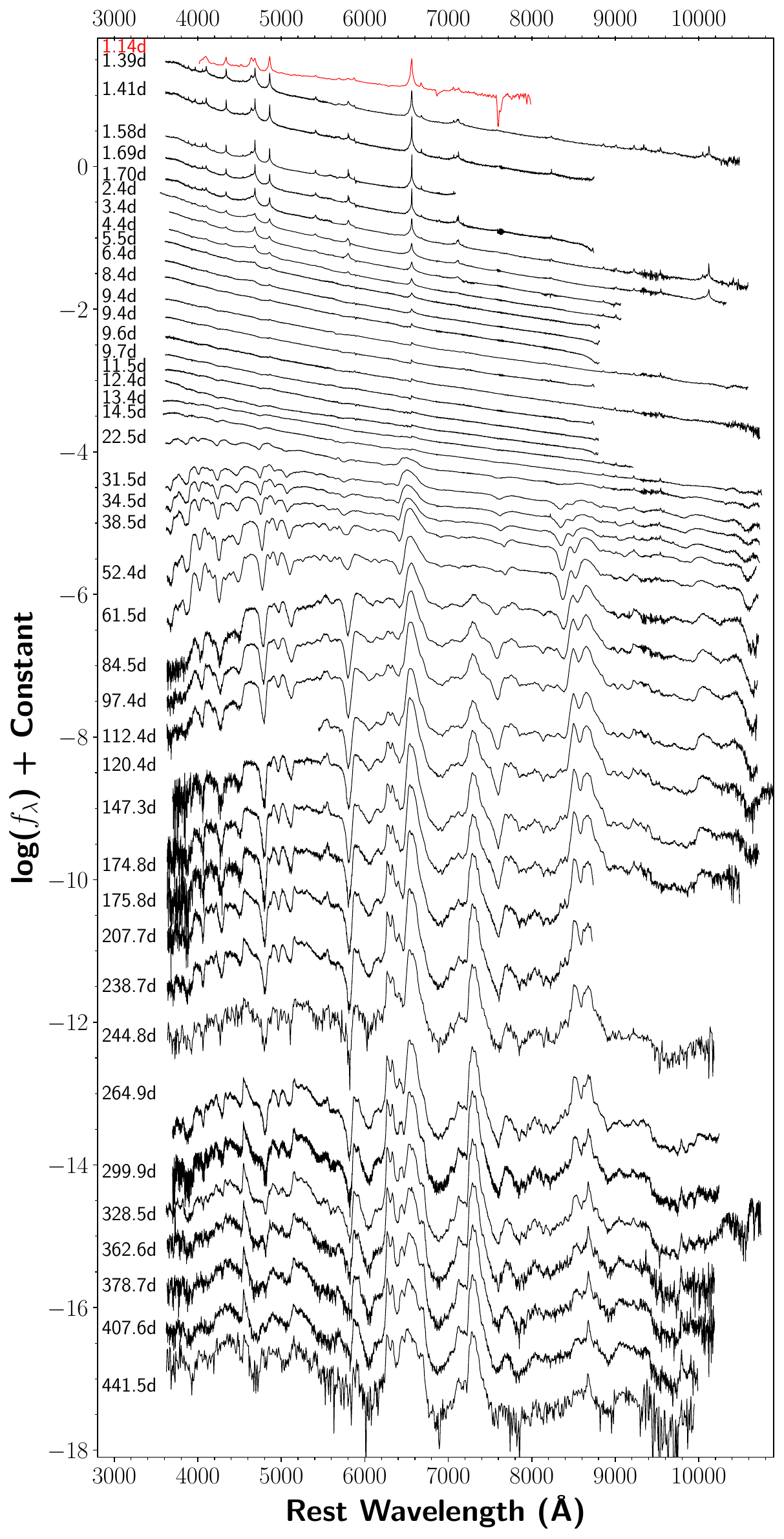}
    \caption{ 
    Lick/Kast observations of SN\,2023ixf, with a total of 42 spectra covering from day 1.4 to day 442 after explosion. The 1.14\,d spectrum (shown in red) was taken from \cite{perley23} for completeness.
    Distinct spectral evolution is seen. The first-week spectra present strong but narrow emission lines from the ionized CSM (CSM-dominated phase), which subsequently weakened;  the spectra appear to be featureless in the second week (CDS-dominated phase), though weak broad P~Cygni features  start  to emerge. After $\sim 3$ weeks, the spectra are similar to those of  other SNe~II with strong P~Cygni lines (ejecta-dominated phase). The spectra enter the nebular phase after $\sim 100$\,d and forbidden lines start to form at late times. 
    }~\label{fig:2023ixf_spec}
\end{figure}

Figure \ref{fig:2023ixf_spec} shows our full set of Kast spectra (a total of 42; see Table 1 for a full log) of SN\,2023ixf until 442 days after explosion. Our first spectrum was taken at MJD = 60084.183, merely 1.39 days after explosion.
We obtained nearly daily Kast spectra of SN\,2023ixf for the first two weeks after explosion. In addition, for the first-night observations, we obtained 5 spectra ranging from 1.4 days to 1.7 days with different grating resolution with the same telescope (note that \citealt{bostroem_early_2023} also obtained multiple spectra in the same night, but with different telescopes). These 5 spectra from the first-night observations were already  shown by \cite{Zimmerman24}; also, the 3.4\.d, 4.4\,d, 5.5\,d, 6.4\,d, and 8.4\,d spectra were presented by \cite{Jacobson-Galan23}, but we include them here for a more complete and detailed analysis.
Augmenting the spectral set, we plot the the classification spectrum of SN\,2023ixf at 1.14 days from the Liverpool telescope; this is the earliest reported spectrum of SN\,2023ixf \citep{perley23}.
These high-quality spectra allow us to perform thorough analysis of the spectral evolution of SN\,2023ixf.

Distinct spectral evolution was seen from day 1.4 to day 442 as shown in Figure \ref{fig:2023ixf_spec}. We can classify the evolution into four main phases, following the  suggestion given by \cite{Dessart16} for SN~1998S. During the first few days, the most distinct features are the strong but narrow emission lines from the ionized CSM; we call this the CSM-dominated phase. 
These emission features weakened after the first week, so the spectra appear to be blue and quasi-featureless in the second week; this is  the cold-dense-shell (CDS)-dominated phase, during which the only distinct features are from hydrogen, though weak and broad P~Cygni features started to emerge at the end of the second week. After $\sim 3$ weeks, during the ejecta-dominated phase, the spectra are similar to those of other SNe~II with strong P~Cygni lines. After $\sim 100$\,d, the spectra enter the nebular phase and forbidden lines start to form at late times. Detailed descriptions of each phase are given in the following sections.

\subsection{CSM-Dominated Phase}~\label{sec:earlyphaseionizationfeatures}
\begin{figure*}
    \centering
    \includegraphics[width=0.99\linewidth]{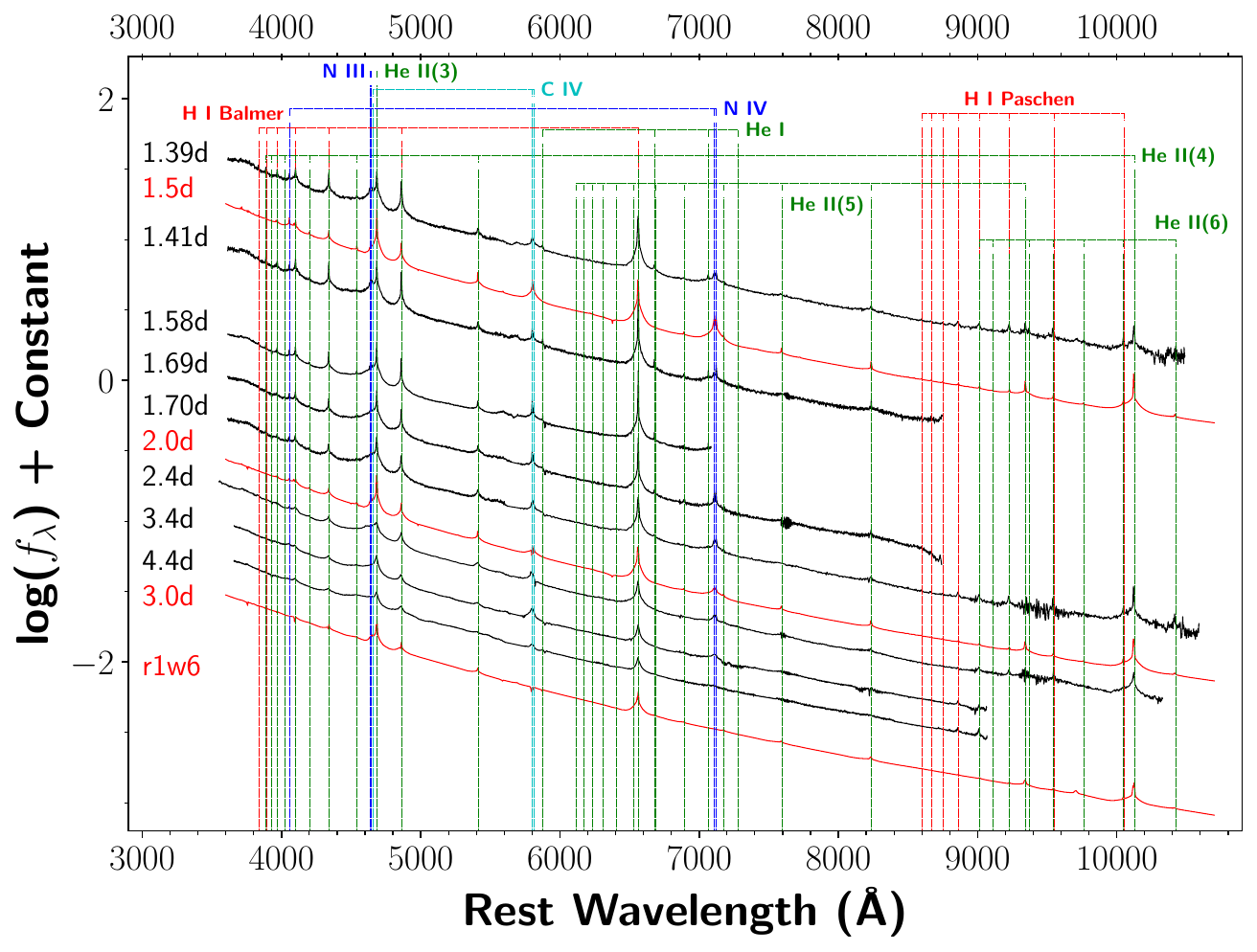}
    \caption{
    Lick/Kast spectra of SN\,2023ixf from the first 5 days after explosion during the CSM-dominated phase. Many species are detected and identified, including the \ion{H}{1} Balmer and Paschen series marked with red vertical lines; the \ion{He}{2} series from energy states of $n=3$, 4, 5, and 6, as well as a few \ion{He}{1} transitions marked with green vertical lines; and some \ion{C}{0}\ and \ion{N}{0}\ transitions marked with cyan and blue vertical lines.
    The red spectra were from model r1w6 at a nearby phase for comparison; see text for details.
    }~\label{fig:2023ixf_spec_first5days}
\end{figure*}

Figure \ref{fig:2023ixf_spec_first5days} presents spectra of SN\,2023ixf in the first 5 days during the CSM-dominated phase. During this phase, there are many both strong and weak narrow emission lines apparently caused by the ionized CSM, also reported by other groups \citep{Jacobson-Galan23,bostroem_early_2023,Hosseinzadeh2023,Hiramatsu2023,Teja2023,Yamanaka23}.
These lines are also accompanied by broad components which are caused by electron scattering.
We identified many species and found that most are from H and He.

The strongest emission in the CSM-dominated phase is the \ion{H}{1} Balmer series including H${\alpha}$, H${\beta}$, H${\gamma}$, H${\delta}$, H${\epsilon}$, H${\zeta}$, and H${\eta}$, owing to the large abundance of hydrogen. As expected, H${\alpha}$ is the strongest line, with other Balmer lines gradually weaker at higher energy states  (bluer wavelengths). In addition, the \ion{H}{1} Paschen series also appears in our spectra at 1.41\,d, 1.70\,d, and 2.4\,d. Although P${\alpha}$ $\lambda$12,157 is outside our wavelength coverage, P${\beta}$, P${\gamma}$, P${\delta}$, P${\epsilon}$, P${\zeta}$, and P${\eta}$ are clearly detected. However, the \ion{H}{1} Paschen series is much weaker compared with the \ion{H}{1} Balmer series. The \ion{H}{1} Balmer and Paschen series are marked with red vertical lines in Figure \ref{fig:2023ixf_spec_first5days}.

\begin{deluxetable}{crcrcrcr}
 \tabcolsep 0.2mm
 \tablewidth{0pt}
 \tablecaption{\ion{He}{2} air wavelengths for energy states $n=3$, 4, 5, and 6.}
 \tablehead{
   \colhead{$n=3$} & \colhead{$\lambda$ (\AA)} & \colhead{$n=4$} & \colhead{$\lambda$ (\AA)} & \colhead{$n=5$} & \colhead{$\lambda$ (\AA)} & \colhead{$n=6$} & \colhead{$\lambda$ (\AA)}                                                                                                                        
}
\startdata
$n'$  &         &  $n'$   &          & $n'$   &          & $n'$   &          \\
\hline
4     & 4685.70 &         &          &        &          &        &          \\      
5$^a$ & 3203.09 &  5      & 10123.59 &        &          &        &          \\      
      &         &  6      & 6560.09  & 6$^a$  & 18636.75 &        &          \\      
      &         &  7      & 5411.52  & 7$^a$  & 11626.41 & 7$^a$  & 30908.46 \\
      &         &  8      & 4859.32  & 8      & 9344.93  & 8$^a$  & 18743.33 \\
      &         &  9      & 4541.59  & 9      & 8236.79  & 9$^a$  & 14760.38 \\
      &         &  10     & 4338.68  & 10$^a$ & 7592.76  & 10$^a$ & 12812.83 \\
      &         &  11     & 4199.84  & 11     & 7177.53  & 11$^a$ & 11673.25 \\
      &         &  12     & 4100.05  & 12     & 6890.91  & 12$^a$ & 10933.62 \\
      &         &  13     & 4025.61  & 13     & 6683.21  & 13     & 10419.83 \\
      &         &  14$^a$ & 3968.44  & 14     & 6527.11  & 14     & 10045.27 \\
      &         &  15$^a$ & 3923.49  & 15     & 6406.39  & 15     & 9762.17  \\
      &         &         &          & 16     & 6310.86  & 16     & 9542.07  \\
      &         &         &          & 17$^a$ & 6233.83  & 17     & 9367.05  \\
      &         &         &          &        &          & 18     & 9225.25  \\
      &         &         &          &        &          & 19     & 9108.55  \\
      &         &         &          &        &          & 20$^a$ & 9011.23  \\
      &         &         &          &        &          & 21$^a$ & 8929.13  \\
      &         &         &          &        &          & 22$^a$ & 8859.17  \\
      &         &         &          &        &          & 23$^a$ & 8799.02  \\
\enddata
\tablenotetext{a}{Either not in our wavelength range or not obviously detected owing to low S/N.}
\label{HeIIlinewavelength}
\end{deluxetable}

Helium also exhibits some strong emission lines. In fact, \ion{He}{2} $\lambda$4686 (at an energy state of $n=3$) is the second-strongest feature during the CSM-dominated phase. Moreover, a few other \ion{He}{2} lines are detected from energy states $n=4$, 5, and 6, marked as green vertical lines in Figure \ref{fig:2023ixf_spec_first5days}. A list of  \ion{He}{2} wavelengths from different energy states is given in Table \ref{HeIIlinewavelength}, taken from the Atomic Data as part of CMFGEN\footnote{http://kookaburra.phyast.pitt.edu/hillier/web/CMFGEN.htm} \citep{Hillier12}. Such complete \ion{He}{2} series in SN spectra are presented here for the first time (to our knowledge), although some lines at higher energy states are too weak to be clearly detected. Higher signal-to-noise ratio (S/N) spectra could potentially reveal these weak lines in future SNe showing ionization features.
In addition to \ion{He}{2}, there are also a few \ion{He}{1} lines detected; though much weaker, they are still quite obvious in our first-night spectra, including \ion{He}{1} $\lambda$5876, $\lambda$6678, and $\lambda$7065.

Besides hydrogen and helium, a few \ion{C}{0}\ and \ion{N}{0}\ features (marked as cyan and blue vertical lines) were also identified in the early-time spectra. \ion{C}{4} $\lambda\lambda$5801, 5812 is not only detected, but even distinguished separately in our higher-resolution spectra on days 1.41, 1.58, and 1.69 (see also \citealt{Smith23} and \citealt{Dickinson24} with high-resolution spectroscopy).
The detection of \ion{C}{4} $\lambda$4658 is uncertain as it is blended with other lines.
No lower ionization state \ion{C}{3} lines
are obviously detected; \ion{C}{3} $\lambda\lambda$4647, 4650 might be present, but it is blended with \ion{C}{4} $\lambda$4658.
\ion{N}{4} lines are also clearly detected with \ion{N}{4} $\lambda$4058 and  $\lambda\lambda$7109, 7123.
Lower ionization state \ion{N}{3} $\lambda\lambda$4634, 4641 is visible, but \ion{N}{3} $\lambda$4687 is blended with \ion{He}{2} $\lambda$4686 and thus uncertain. 
Both \ion{N}{4} $\lambda\lambda$7109, 7123 and \ion{N}{3} $\lambda\lambda$4634, 4641 are also identified in our higher-resolution spectra on days 1.41, 1.58, and 1.69, similar to \ion{C}{4} $\lambda\lambda$5801, 5812 (\citealt{Smith23} also show high-resolution spectra).

\begin{figure*}
    \centering
    \includegraphics[width=0.32\linewidth]{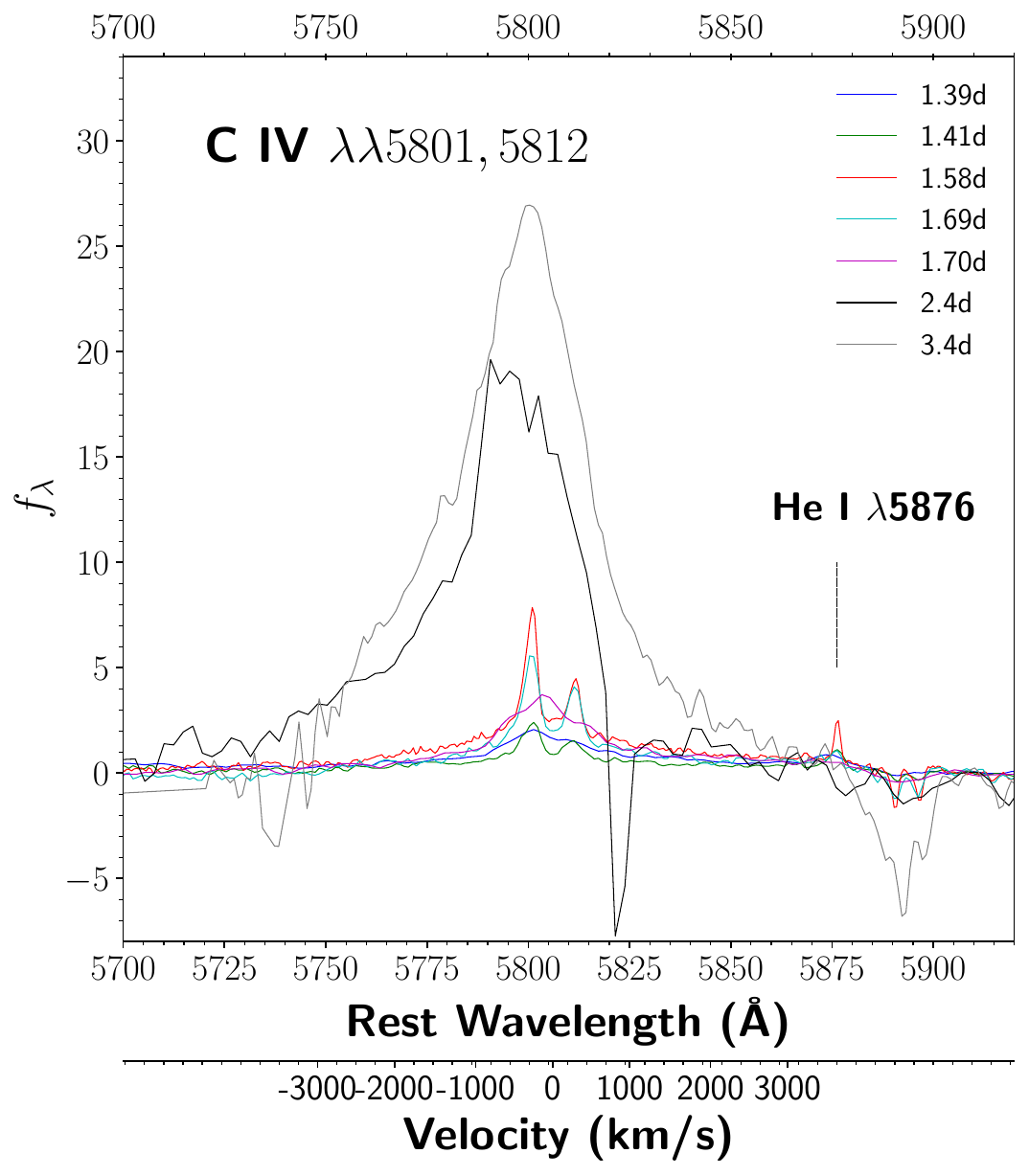}
    \includegraphics[width=0.316\linewidth]{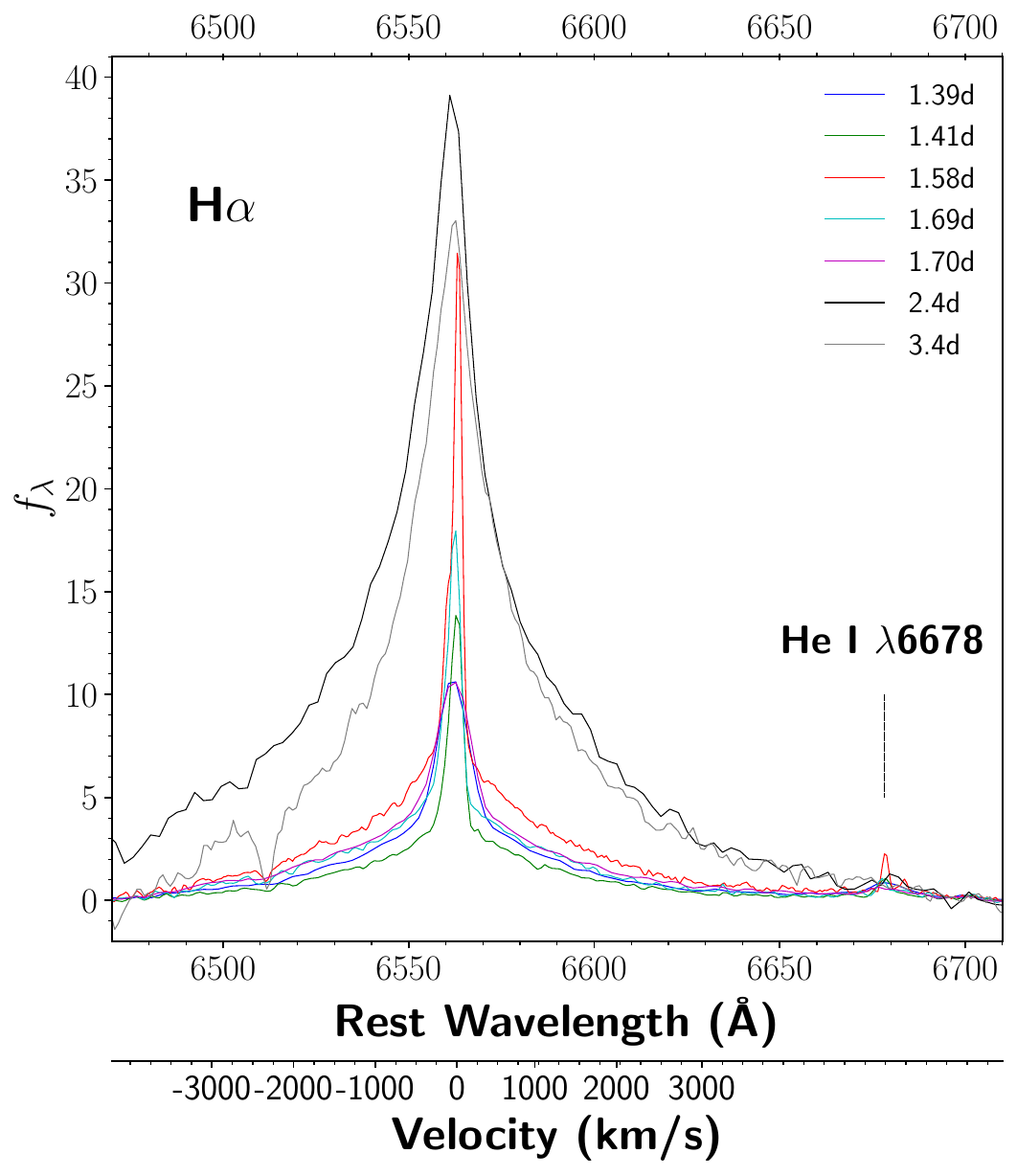}
    \includegraphics[width=0.33\linewidth]{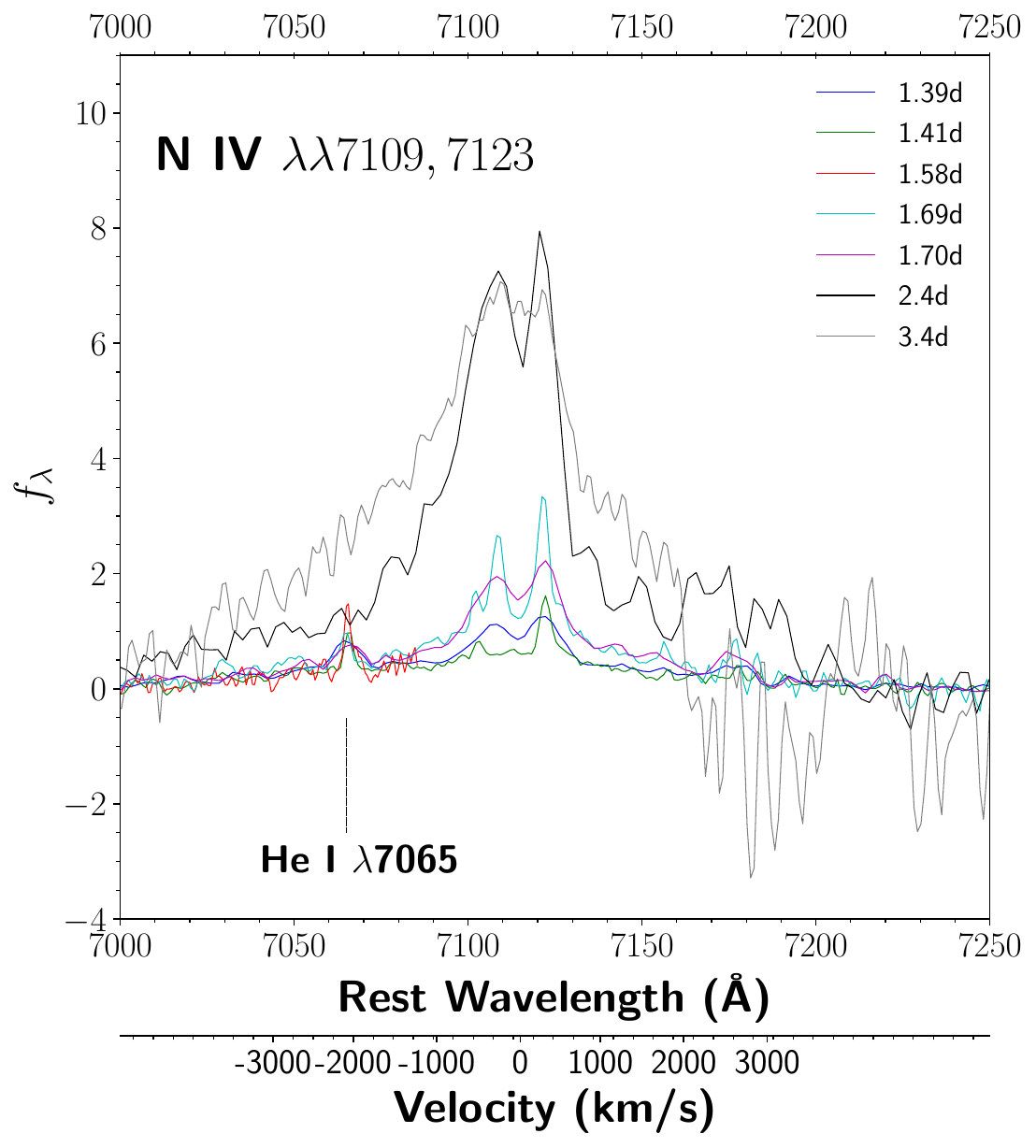}
    \caption{
    Zoomed-in regions of three detected \ion{He}{1} lines, $\lambda$5876 (left), $\lambda$6678 (middle), and $\lambda$7065 (right), respectively centered on \ion{C}{4} $\lambda\lambda$5801, 5812, H${\alpha}$, and \ion{N}{4} $\lambda$7115. All three \ion{He}{1} lines weakened and almost disappeared from day 1.41 to day 1.70, and by day 2.4 they vanished completely.
    }~\label{fig:2023ixf_spec_lines_HeI}
\end{figure*}

\begin{figure*}
    \centering
    \includegraphics[width=0.485\linewidth]{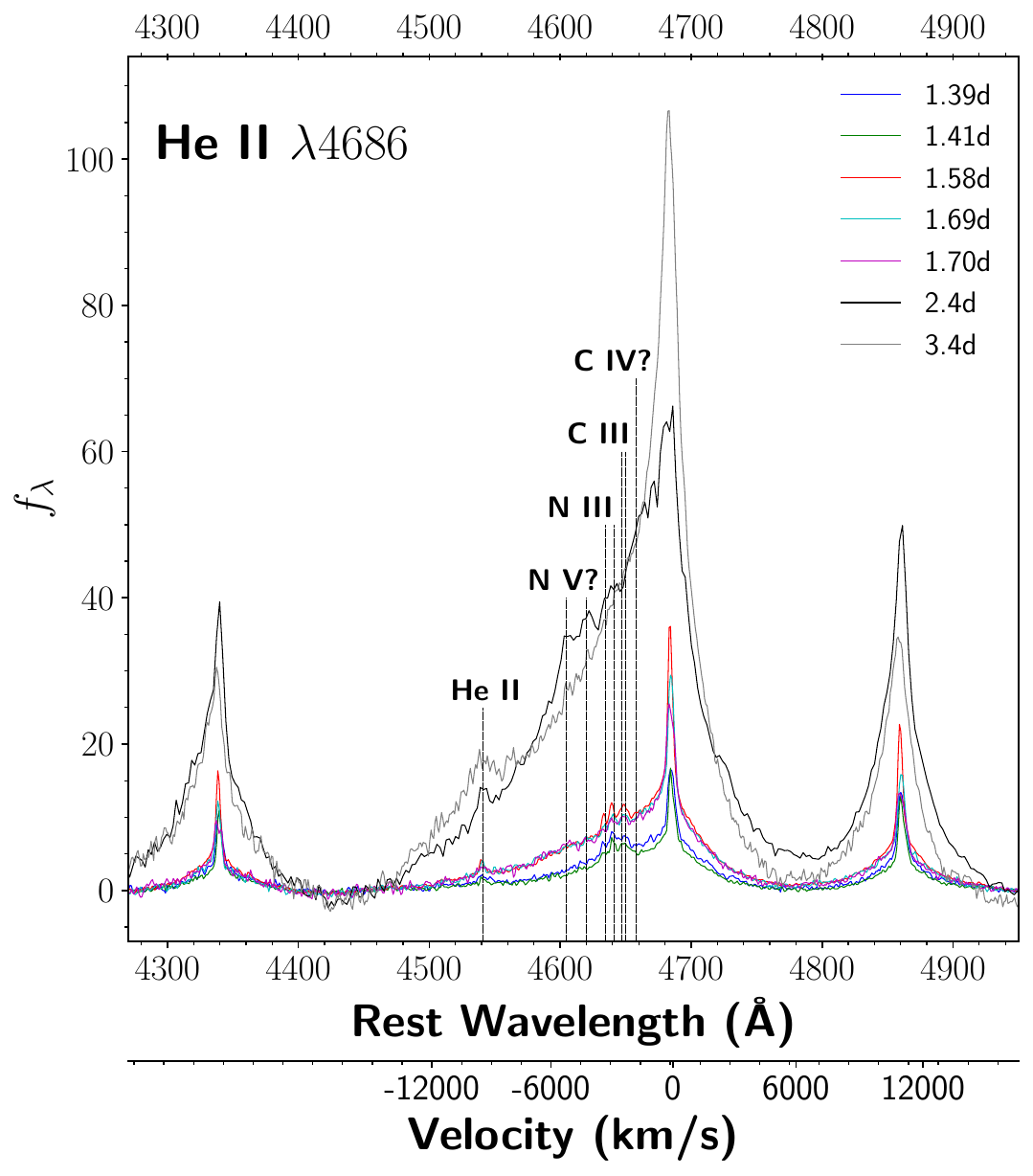}
    \includegraphics[width=0.49\linewidth]{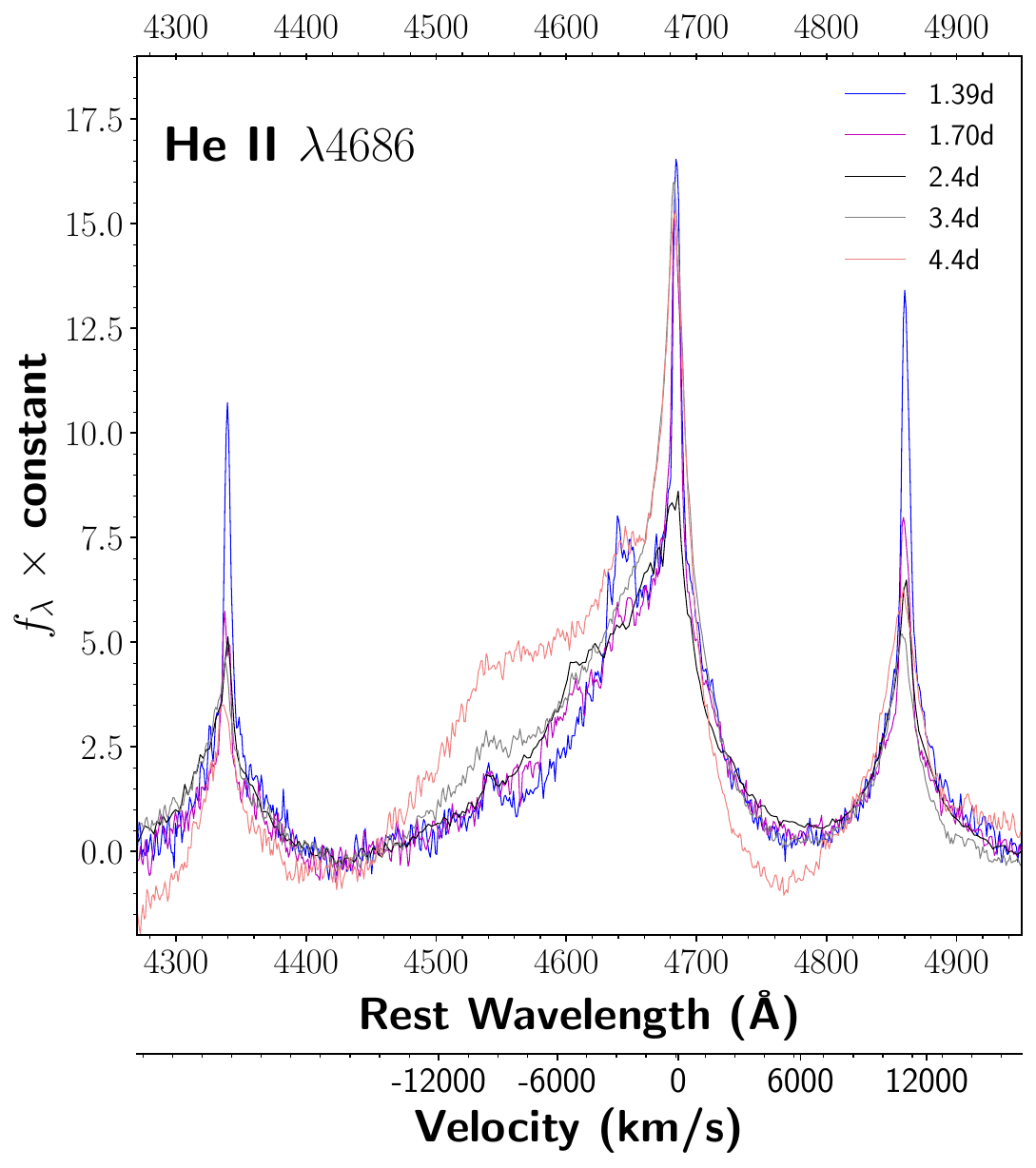}
    \caption{
    {\it Left:} Zoomed-in regions of \ion{He}{2} $\lambda$4686 along with H${\beta}$ and H${\gamma}$ centered on \ion{He}{2} $\lambda$4686. The blue-wing profile of the \ion{He}{2} $\lambda$4686 region is very complicated, with many lines blended together, including clear detections of \ion{C}{3} $\lambda\lambda$4647, 4650, \ion{N}{3} $\lambda\lambda$4634, 4641, and \ion{He}{2} $\lambda$4542, as well as possible \ion{C}{4} $\lambda$4658 and \ion{N}{5} $\lambda\lambda$4604, 4620. These lines boost the flux of the blue wing, makes the \ion{He}{2} profile very asymmetric. {\it Right:} The same region  but with spectra arbitrarily scaled to show that the emission blueward of \ion{He}{2} $\lambda$4686 strengthens over the first few days.
    }~\label{fig:2023ixf_spec_lines_NIII4634}
\end{figure*}

Interestingly, thanks to the total of 5 spectra taken during the first night of observations, we find that a few of the above-mentioned lines evolve quickly, showing intraday changes (see also \citealt{bostroem_early_2023}).
For example, all three detected \ion{He}{1} lines ($\lambda$5876, $\lambda$6678, and $\lambda$7065) weakened and almost disappeared from days 1.41 to 1.70, within 8\,hr. To better illustrate this, in Figure \ref{fig:2023ixf_spec_lines_HeI} we plot the three \ion{He}{1}  regions centered on \ion{C}{4} $\lambda\lambda$5801, 5812, H${\alpha}$, and \ion{N}{4} $\lambda$7115. By day 2.4 after explosion, all three \ion{He}{1} lines  disappeared completely.
In addition, the \ion{N}{3} $\lambda\lambda$4634, 4641 and \ion{C}{3} $\lambda\lambda$4647, 4650 lines (blended)  weakened quickly in the first two days, as shown in Figure \ref{fig:2023ixf_spec_lines_NIII4634} centered on \ion{He}{2} $\lambda$4686. The profile is very complicated, with many lines blended in the blue wing of \ion{He}{2} $\lambda$4686, including clear detections of \ion{C}{3} $\lambda\lambda$4647, 4650, \ion{N}{3} $\lambda\lambda$4634, 4641, and \ion{He}{2} $\lambda$4542, and possibly \ion{C}{4} $\lambda$4658 and \ion{N}{5} $\lambda\lambda$4604, 4620. These lines boost the flux of the blue wing (while the red wing has a normal shape), making the \ion{He}{2} profile very asymmetric. The emission blueward of \ion{He}{2} $\lambda$4686 even strengthens over the first few days, as shown in the right panel of Figure \ref{fig:2023ixf_spec_lines_NIII4634}.
Our highest-resolution spectrum at 1.58\,d is able to distinguish both components of the \ion{N}{3} $\lambda\lambda$4634, 4641 doublet, which was even stronger in the 1.14\,d spectrum from the Liverpool telescope, comparable to  \ion{He}{2} $\lambda$4686, but weakened significantly to nearly undetected by day 1.70.
Interestingly, there appears to be \ion{N}{5} $\lambda\lambda$4604, 4620 in the 2.4\,d spectrum, but not earlier nor later. If real, this is additional evidence for intraday changes of the quickly evolving lines.

\subsection{CDS-Dominated Phase}~\label{sec:blackbodyphase}

\begin{figure}
    \centering
    \includegraphics[width=0.99\linewidth]{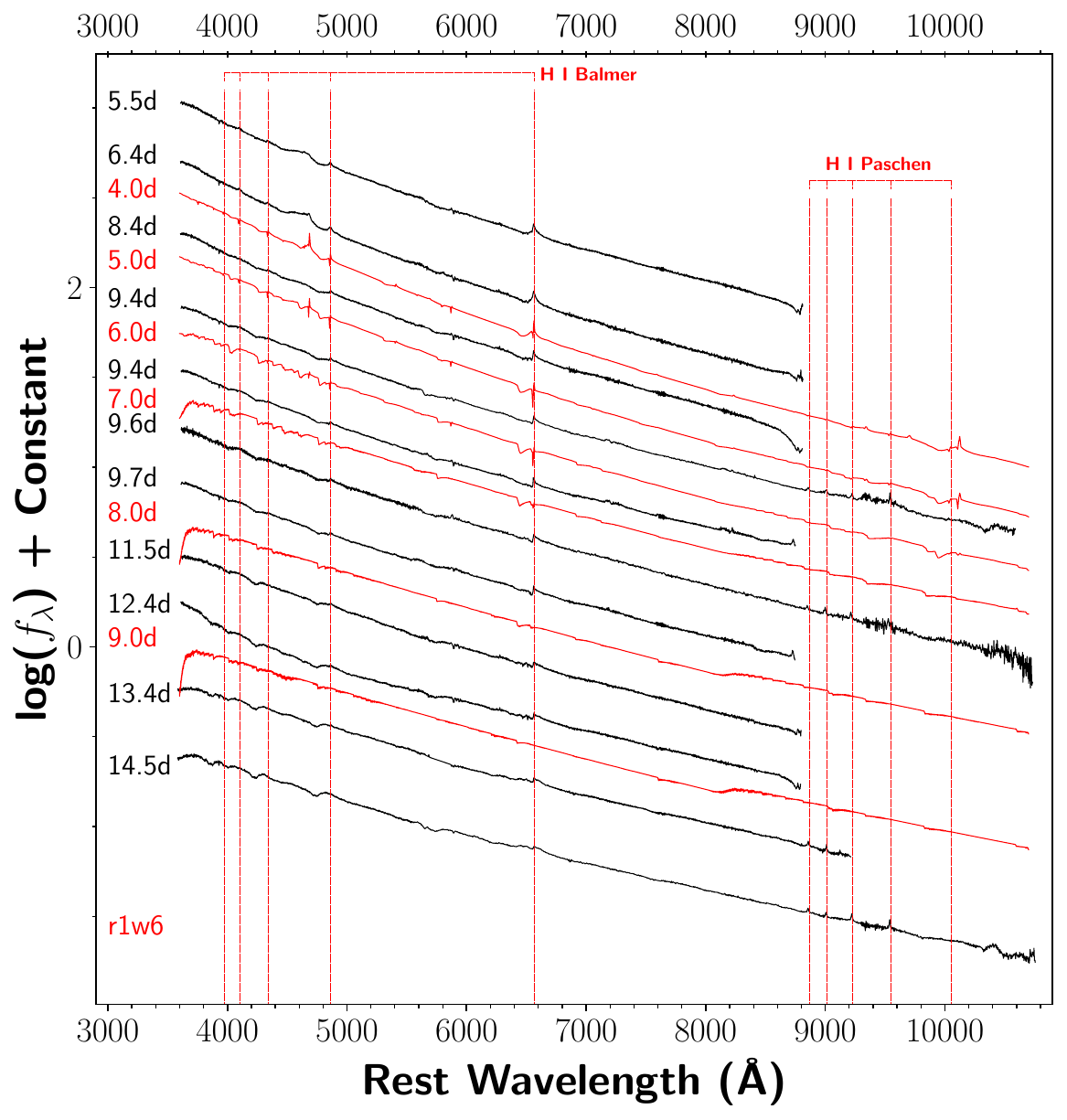}
    \caption{
    Lick/Kast spectra of SN\,2023ixf from the second week after explosion in the CDS-dominated phase, during which they become featureless; almost all emission lines faded away after about a week, except H${\alpha}$ (other \ion{H}{1} Balmer and Paschen series lines are quite weak and are marked as red vertical lines). The H${\alpha}$ line starts showing a low-velocity (less than 1000\,km\,s$^{-1}$) P~Cygni profile on day 8, and a broad P~Cygni profile starts to emerge in the bluer region (H${\beta}$, H${\gamma}$, H${\delta}$, etc.) on day 9.4. The spectra shown in red spectra are from the r1w6 model at a nearby phase for comparison; see text for details. Note that the drop in the model at the blue edge after 7 d corresponds to the Balmer edge, at which the model predicts erroneously a sharp jump.
    }~\label{fig:2023ixf_spec_middlefewdays}
\end{figure}

\begin{figure}
    \centering
    \includegraphics[width=0.99\linewidth]{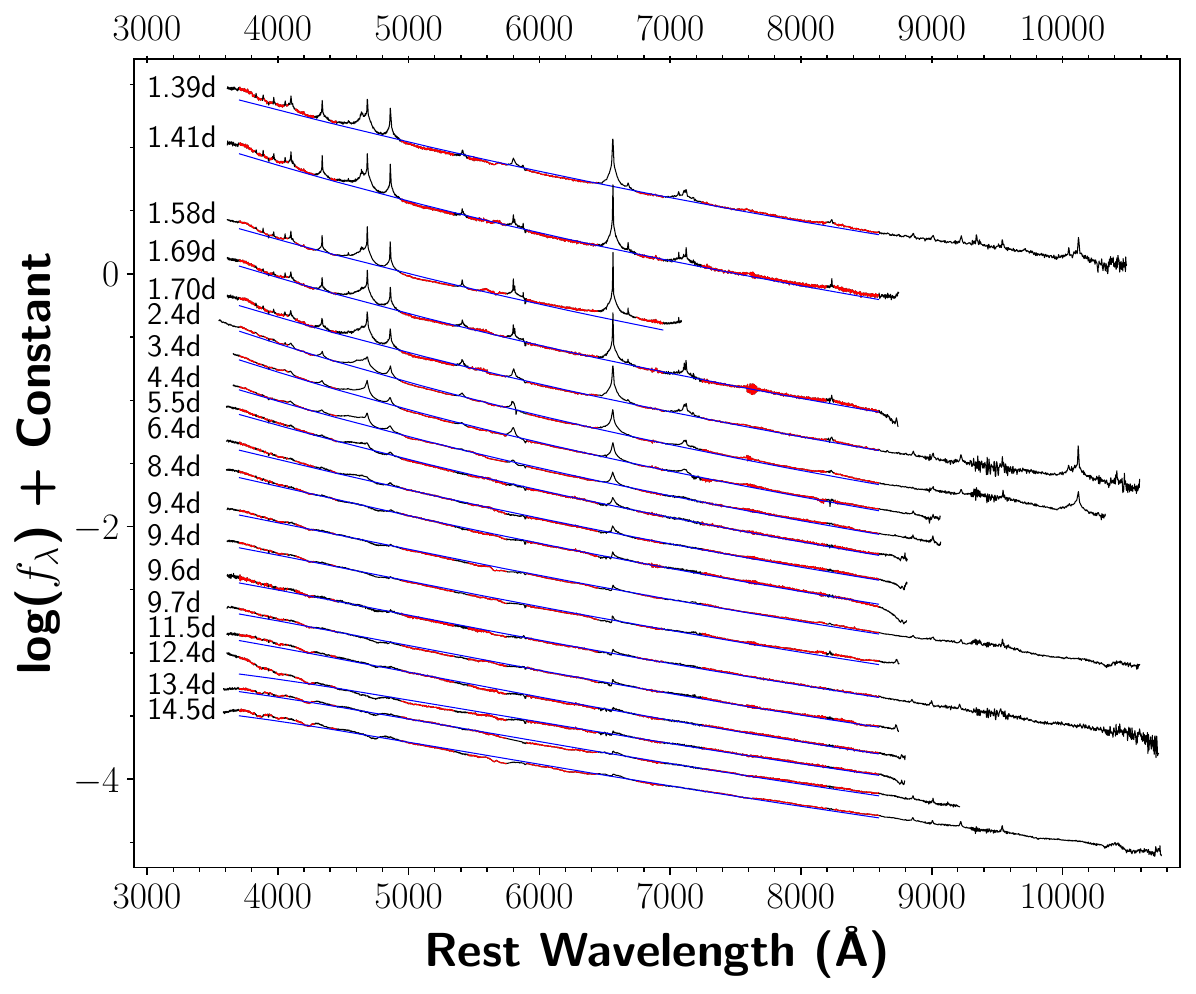}
    \includegraphics[width=0.99\linewidth]{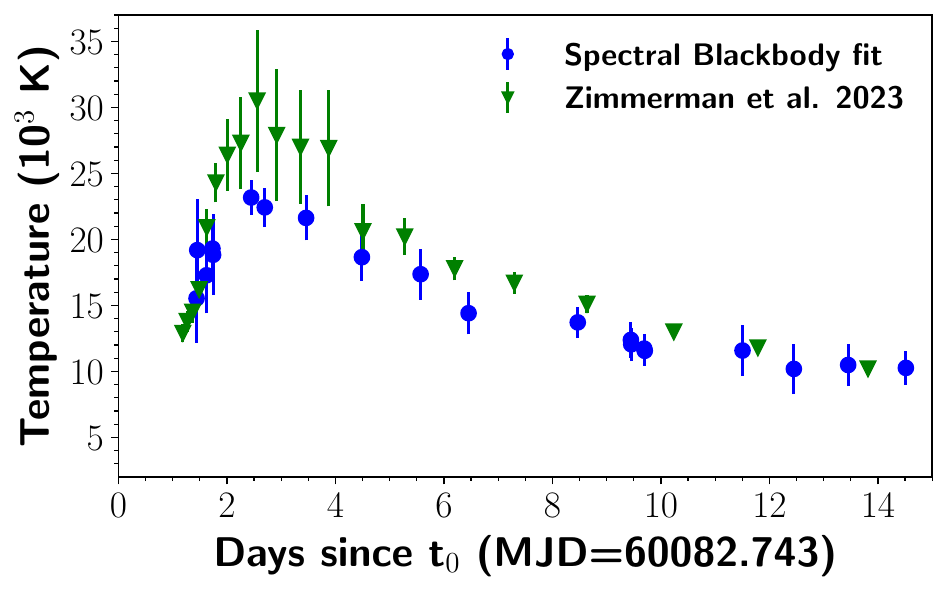}
    \caption{
    {\it Top:} Blackbody fitting to the observed spectra, where blue lines show the results of the blackbody function and red indicates the selected fitting region after excluding emission lines and beyond 8600\,\AA\ (black).
    {\it Bottom:} Fitted temperature (blue circle) evolution with time. The ejecta temperature rises quickly in the first few days, reached peak around 24,000\,K at $\sim 2.5$\,days after explosion. It then gradually drops in the next two weeks. Similar fitting results from \cite{Zimmerman24} are also plotted for comparison (green triangle), showing consistent results.
    }~\label{fig:2023ixf_spec_BBfit}
\end{figure}

Overall, almost all the emission features faded away after about a week (except for the \ion{H}{1} Balmer series, mainly H${\alpha}$), and the spectra became quasi-featureless with a blue continuum during the second week, indicating that the ejecta are CDS dominated at this time, as shown in Figure \ref{fig:2023ixf_spec_middlefewdays}. Because of this, one can try to fit the quasi-featureless spectra with a blackbody function to estimate the temperature of the ejecta, given the reliable relative flux calibration. To do so, we first correct the spectral extinction by adopting a total $E(B-V) = 0.040$\,mag as mentioned above. We then exclude regions with obvious emission lines (we applied the same blackbody fitting to the spectra from the CSM-dominated phase, too). Finally, in order to keep consistent for spectra with different wavelength coverage at the red end, we restrict the fitting region to be below 8600\,\AA\ (the blue-end coverage is almost the same in all spectra). A blackbody function was applied to fit the spectra and the results are shown in Figure \ref{fig:2023ixf_spec_BBfit}; the top panel indicates how well the blackbody function (blue) matches the selected data (red) by excluding the emission lines and the region beyond 8600\,\AA\ (black), and the bottom panel shows the evolution of the fitted temperature (blue circle) with time.

Our results demonstrate that the blackbody temperature rises quickly during the first few days, with $\sim 16,000$\,K at 1.4\,d and quickly reaching peak around 24,000\,K about 2.5\,days after explosion. The temperature then gradually drops in the next two weeks, and by day 14 it is $\sim 13,000$\,K. For comparison, we also plot the temperature fitting results from \cite{Zimmerman24} (green triangles), where they use the photometric fitting method (including UV  photometry). Our results are generally consistent with theirs, though it appears that our fitted temperatures are slightly lower; however, our results are higher than the fitted temperatures given by \cite{zhangj2023} (see their Fig. S3, middle panel).

\begin{figure}
    \centering
    \includegraphics[width=0.99\linewidth]{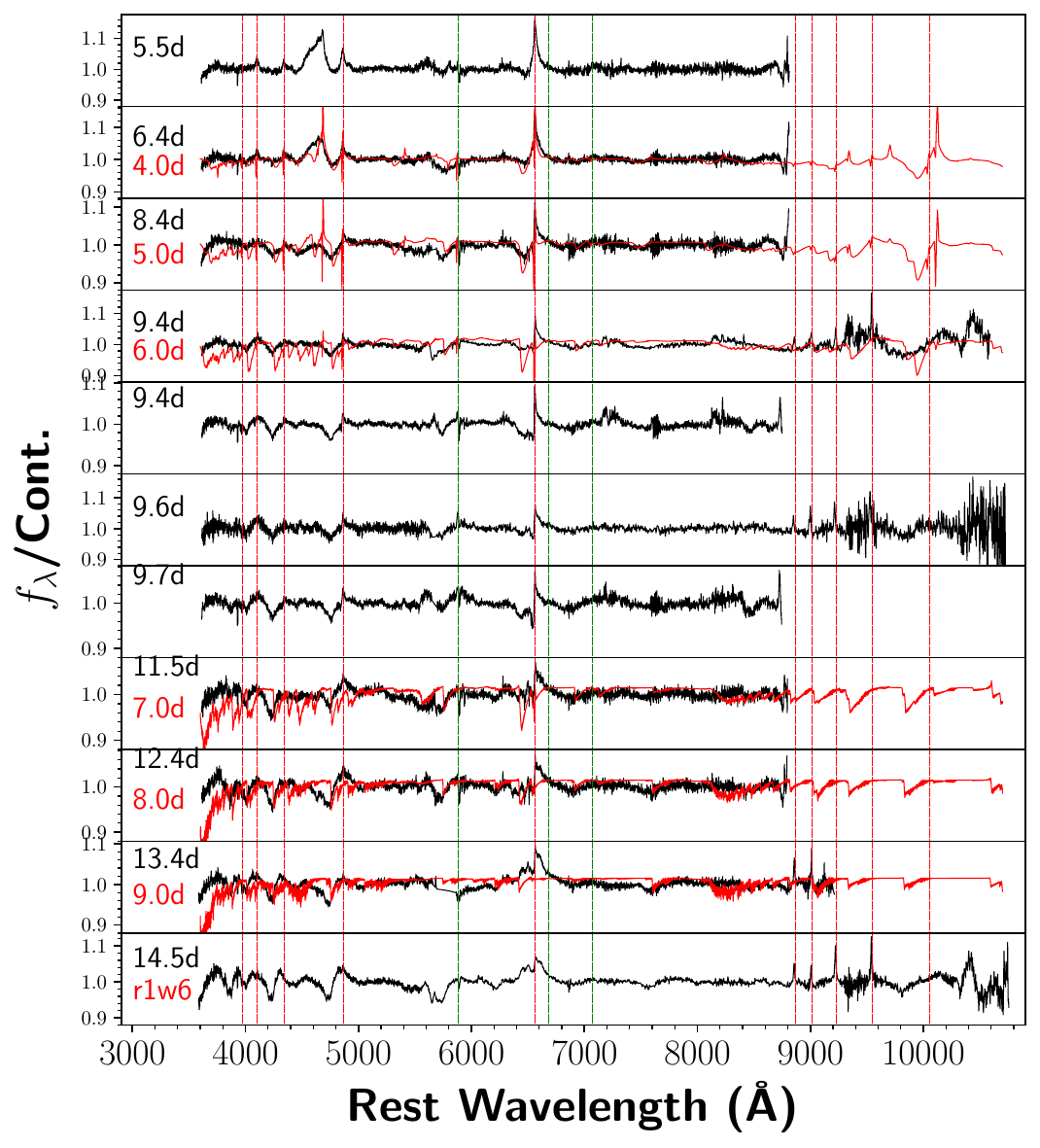}
    \caption{
    Spectra at the same epochs as in Figure \ref{fig:2023ixf_spec_BBfit}, but  normalized by dividing by the continuum. Red vertical lines mark the \ion{H}{1} Balmer and Paschen series, and green vertical lines mark \ion{He}{1} $\lambda$5876, $\lambda$6678, and $\lambda$7065.
    Red spectra are the same as given in Fig. \ref{fig:2023ixf_spec_middlefewdays}. }~\label{fig:2023ixf_spec_middlefewdays_flatspectral}
\end{figure}

In order to better reveal the morphology of the weak lines in the quasi-featureless spectra, we adopt a normalization procedure following that of \cite{Leonard00} (see their Fig. 5): the spectra are divided by the continuum, which was fitted with several-order splines by excluding obvious emission and absorption regions. The normalized spectra are shown in Figure \ref{fig:2023ixf_spec_middlefewdays_flatspectral}.

\begin{figure}
    \centering
    \includegraphics[width=0.99\linewidth]{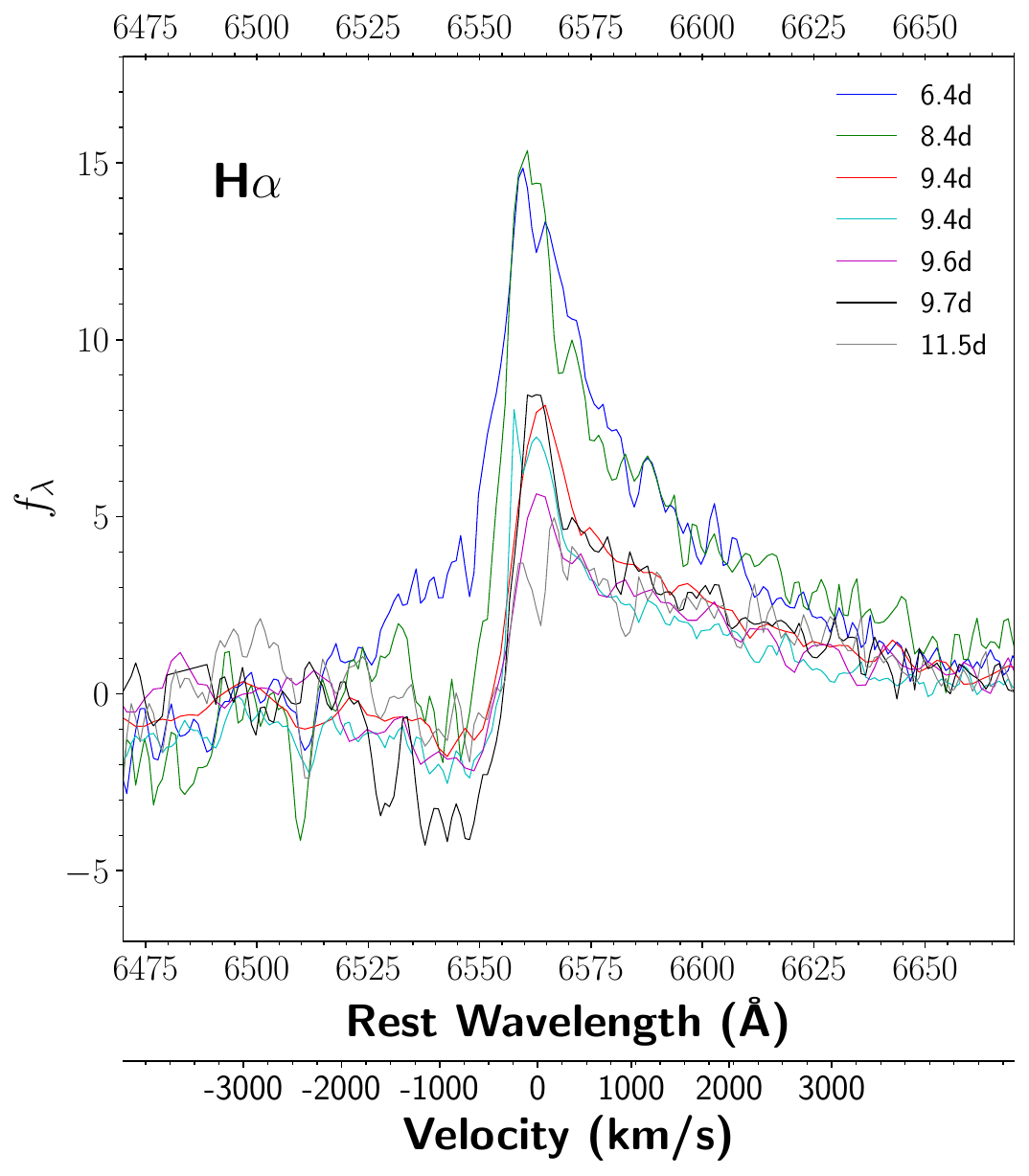}
    \caption{
    Zoomed-in regions of H${\alpha}$ profile from days 6.4 to 11.5. The H${\alpha}$ line starts to show a low-velocity P~Cygni profile on day 8.4 with velocity $< 1000$\,km\,s$^{-1}$, indicated by the blue absorption minimum.
    }~\label{fig:2023ixf_spec_lines_Halpha2}
\end{figure}

\begin{figure*}
    \centering
    \includegraphics[width=0.326\linewidth]{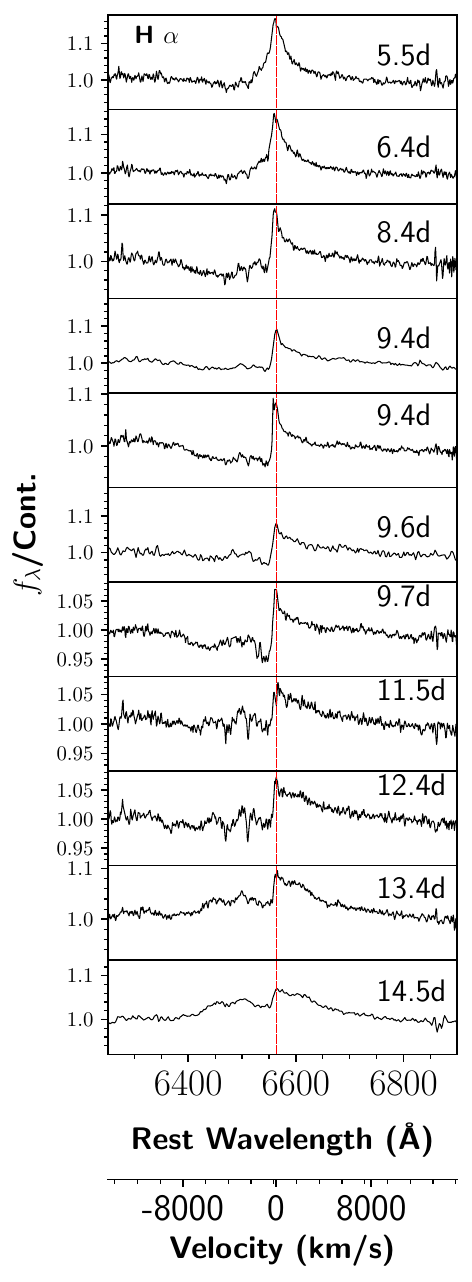}
    \includegraphics[width=0.326\linewidth]{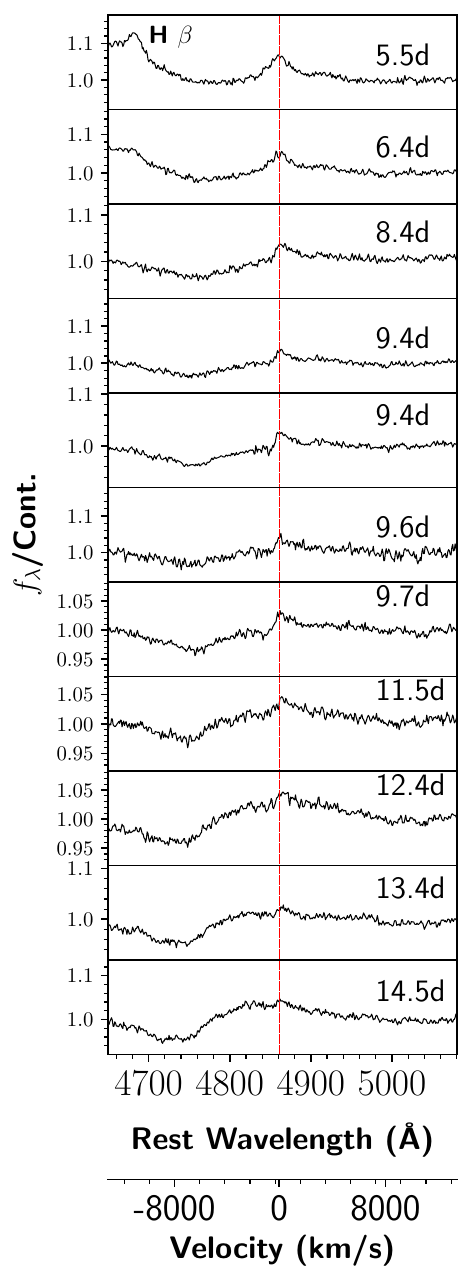}
    \includegraphics[width=0.33\linewidth]{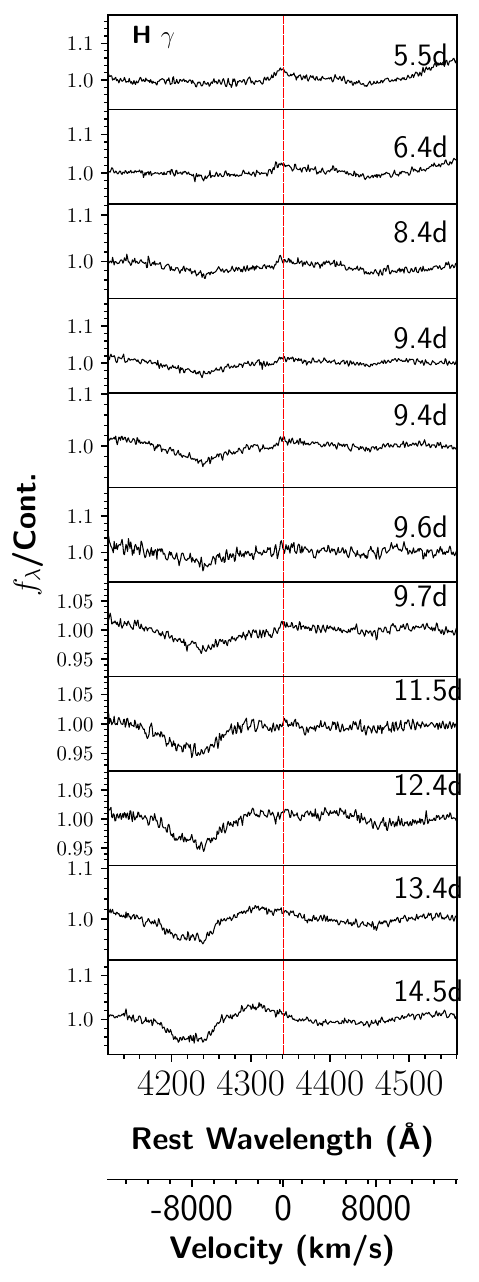}
    \caption{
    Zoomed-in regions of H${\alpha}$ (left panel), H${\beta}$ (middle panel) and H${\gamma}$ (right panel) profile from 5.5 to 14.5 days. Broad P-Cygni profile ($v >$ 5000 km\,s$^{-1}$) starts to emerge in H${\beta}$ and H${\gamma}$ (but not H${\alpha}$) since day 8.4.
    }~\label{fig:2023ixf_spec_flat_lines_HalphabetaProfile}
\end{figure*}

During this phase, one interesting feature is that the H${\alpha}$ line starts to show a low-velocity P~Cygni profile on day 8.4, as shown in Figure \ref{fig:2023ixf_spec_middlefewdays_flatspectral} (also see Figs. \ref{fig:2023ixf_spec_lines_Halpha2} and  \ref{fig:2023ixf_spec_flat_lines_HalphabetaProfile}). The velocity from the blue absorption minimum is $< 1000$\,km\,s$^{-1}$ and lasts for about a week until day 14.5 (before a gap of our spectral coverage in the third week). Likely caused by radiative acceleration \citep{Dessart24}, it is still unclear whether this mechanism could last for such a long time (see Sec. \ref{sec:modelandphysics}).
However, there are no such low-velocity P~Cygni features in the other \ion{H}{1} Balmer lines beyond H${\alpha}$ (perhaps a weak signature in H${\beta}$ on day 9; see the middle panel in Fig. \ref{fig:2023ixf_spec_flat_lines_HalphabetaProfile}). Instead, a broad P~Cygni profile ($v > 5000$\,km\,s$^{-1}$) starts to emerge in H${\beta}$, H${\gamma}$, and H${\delta}$ on day 8.4 (see next section for velocity measurements).

The H${\alpha}$ line at this phase does not exhibit a broad P~Cygni profile, probably because the CDS at this phase is still optically thick.
The \ion{H}{1} Paschen series, however, remains  as narrow emission until day 14.5. Unlike the \ion{H}{1} Balmer lines, no low-velocity P~Cygni nor broad P~Cygni profiles are clearly seen in The \ion{H}{1} Paschen series.

\begin{figure}
    \centering
    \includegraphics[width=0.485\linewidth]{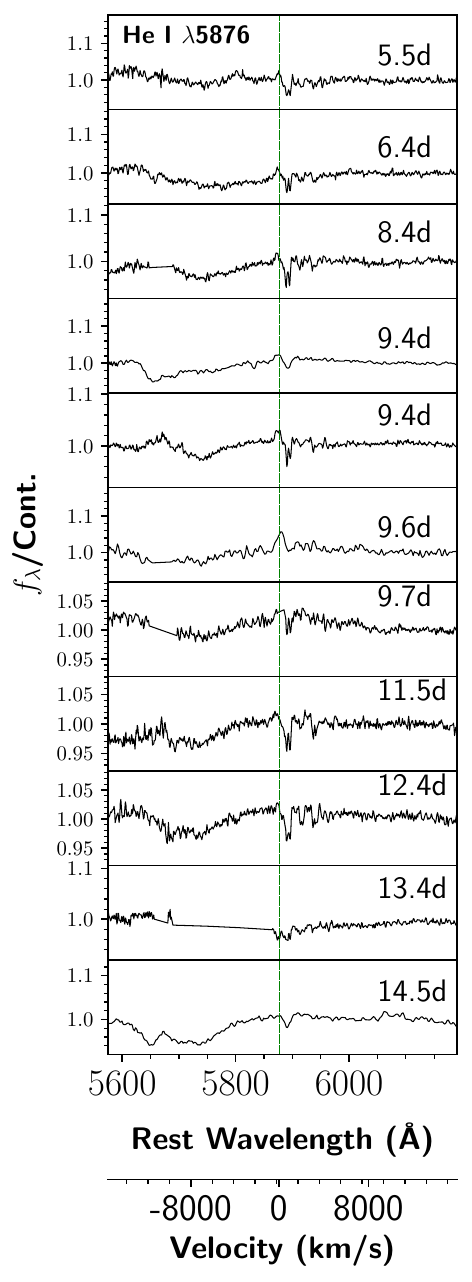}
    \includegraphics[width=0.495\linewidth]{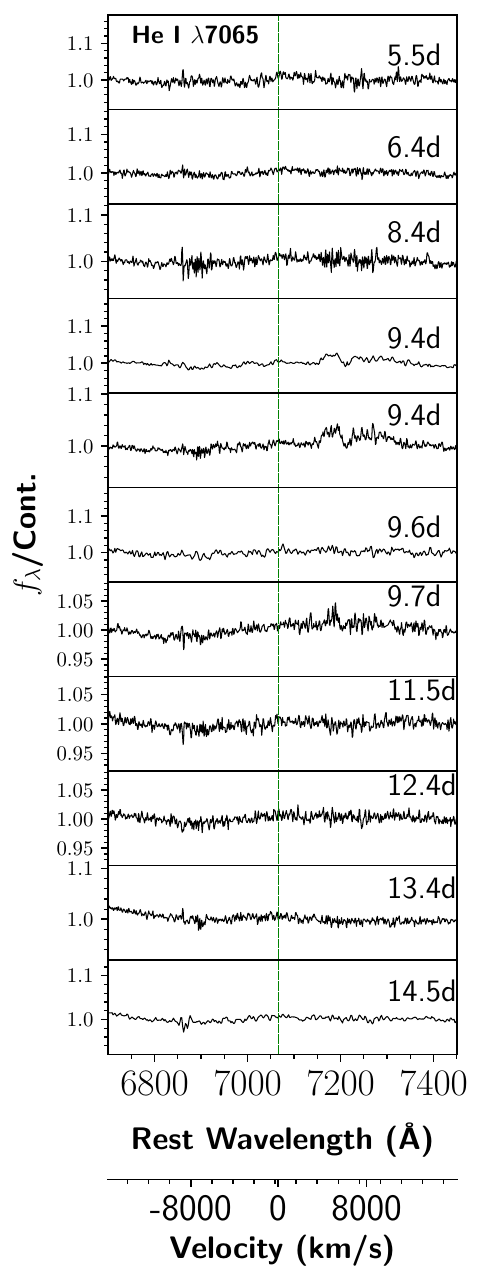}
    \caption{
    Zoomed-in regions of \ion{He}{1} $\lambda$5876 {\it (left)} and \ion{He}{1} $\lambda$7065 {\it (right)} from 5.5 to 14.5 days. The \ion{He}{1} $\lambda$5876 profile show obvious absorption at a velocity of $\sim 7000$\,km\,s$^{-1}$ around day 8.
    }~\label{fig:2023ixf_spec_lines_HeI5876HeI7065Profile}
\end{figure}

Interestingly, the \ion{He}{1} lines which earlier disappeared (narrow emission lines) also start to be visible again, with broad P~Cygni profiles around day 8, as shown in Figure \ref{fig:2023ixf_spec_lines_HeI5876HeI7065Profile}. The \ion{He}{1} $\lambda$5876 profile in the left panel shows obvious absorption at a  velocity of $\sim 7000$\,km\,s$^{-1}$, though the absorption in the \ion{He}{1} $\lambda$7065 profile (right panel) is not  significant. 
Also, there appears to be another bluer absorption in \ion{He}{1} $\lambda$5876 at 14.5\,d with a velocity around 12,000\,km\,s$^{-1}$, which seems a bit high but not  impossible.

\subsection{Ejecta-Dominated Phase}~\label{sec:photospherephase}

\begin{figure*}
    \centering
    \includegraphics[width=0.99\linewidth]{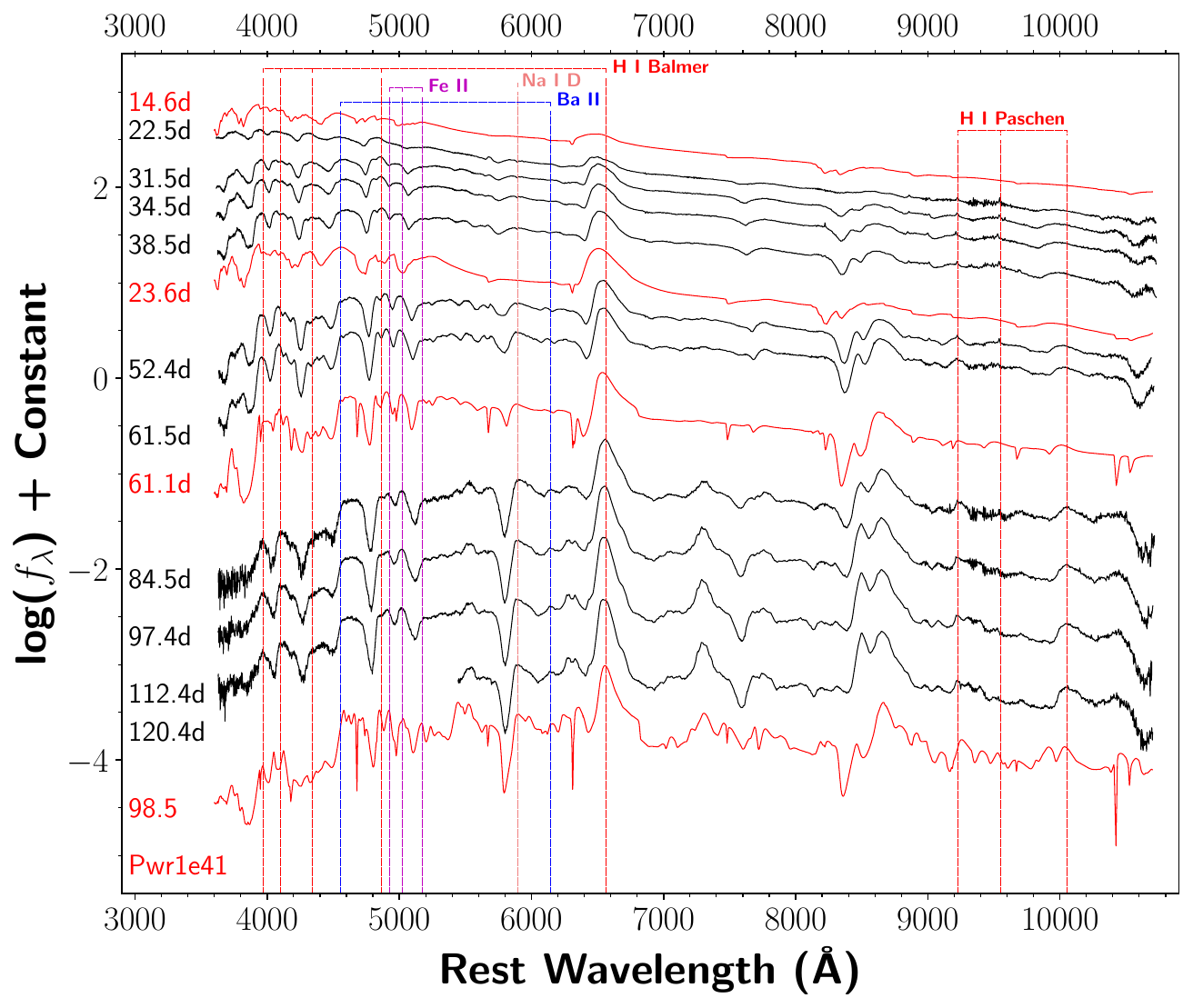}
    \caption{
    Kast spectra of SN\,2023ixf after 3 weeks (from days 22.5 to 120) in the ejecta-dominated phase, during which the spectra show many broad P~Cygni features that are seen in typical normal SNe~II. The red spectra were from the Pwr1e41 model at a nearby phase for comparison; see text for details.
    }~\label{fig:2023ixf_spec_lastfewdays}
\end{figure*}

Owing to a lack of spectral coverage during the third week after explosion, the spectrum taken after 3 weeks (day 22.5 and thereafter) shows that the SN is ejecta dominated; there are many broad P~Cygni features that are seen in normal SNe~II. Figure \ref{fig:2023ixf_spec_lastfewdays} shows the spectra from days 22.5 to 120.4. Broad P~Cygni profile of \ion{H}{1} Balmer lines are very obvious, as well as a few other lines such as \ion{Fe}{2} $\lambda\lambda\lambda$4924, 5018, 5169.

\begin{figure}
    \centering
    \includegraphics[width=0.99\linewidth]{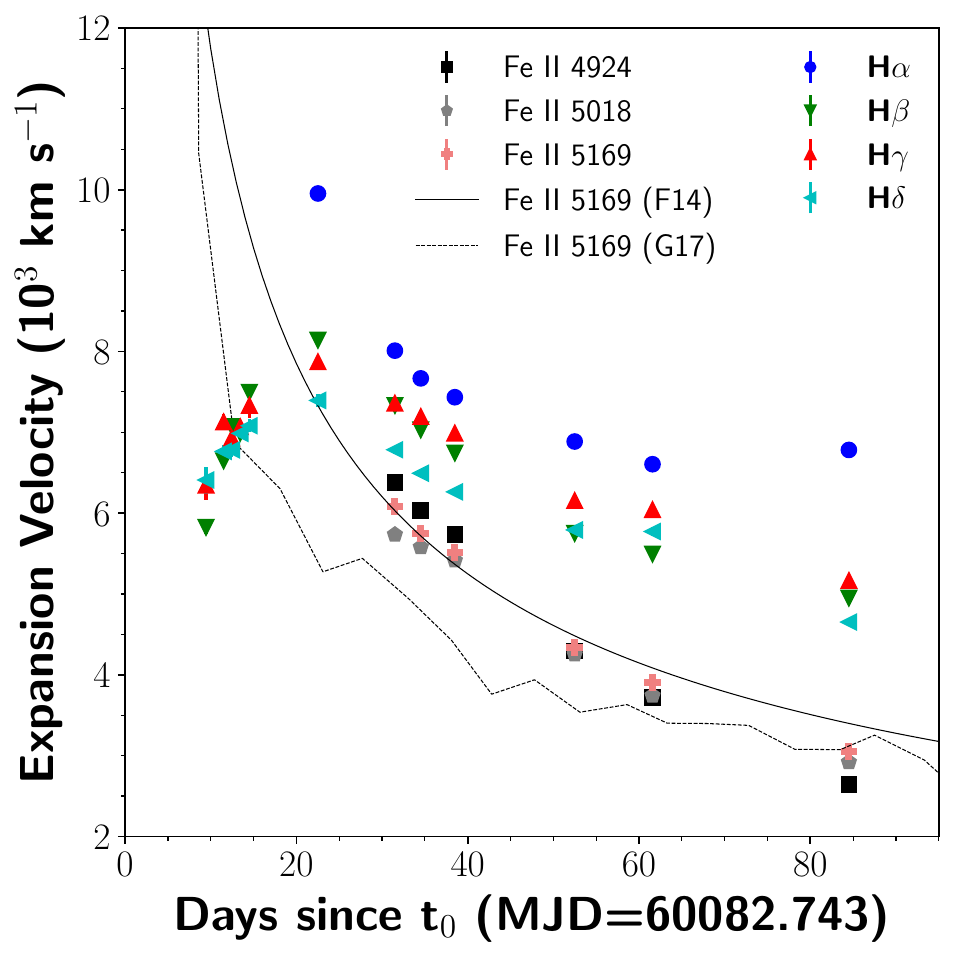}
    \caption{
    Velocity evolution measured from the absorption minimum of the broad P~Cygni profile. Dashed black and solid black lines shows the \ion{Fe}{2} $\lambda$5169 velocity evolution from \cite[G17]{gutierrez_type_2017} and \cite[F14]{faran_photometric_2014}, respectively. The evolution curve after 3 weeks (power-law shape) is typical of normal SNe~II. However, the velocities of H${\beta}$, H${\gamma}$, and H${\delta}$ (emerging earlier than H${\alpha}$) at early times actually rise, which is not commonly seen in normal SNe~II.
    }~\label{fig:2023ixf_spec_velocity}
\end{figure}

We measure the velocity from the blue absorption minimum of the broad P~Cygni profile, and the results are shown in Figure \ref{fig:2023ixf_spec_velocity}. The velocity evolution after 3 weeks is typical of all lines compared with other normal SNe~II; they all decline gradually with a power-law shape, with the H${\alpha}$ velocity higher than that of other lines. There is a dip blueward of H${\alpha}$ absorption from days 22.5 to 38.5, likely due to \ion{Si}{2} $\lambda$6355 at $\sim 5500$\,km\,s$^{-1}$ (unlikely to be high-velocity H${\alpha}$, which would be at $\sim 15,000$\,km\,s$^{-1}$). \ion{Si}{2} $\lambda$6355 is also predicted by the model (see model spectrum at 23.6\,d in Fig. \ref{fig:2023ixf_spec_lastfewdays}).
However, the velocity of \ion{Si}{2} $\lambda$6355 ($\sim 5500$\,km\,s$^{-1}$) appears to be a bit low compared with H${\alpha}$, so it is possible this is caused by some other line.
The \ion{Fe}{2} $\lambda\lambda\lambda$4924, 5018, 5169 lines also evolve in a manner  similar to that of \cite{faran_photometric_2014} as shown in Figure \ref{fig:2023ixf_spec_velocity}, though slightly higher than the results of \cite{gutierrez_type_2017}.

However, it is interesting to see that the velocities of H${\beta}$, H${\gamma}$, and H${\delta}$ (emerging earlier than H${\alpha}$) at early times actually rise in the second week and reach their peak at $\sim 20$ days, $\sim 8000$\,km\,s$^{-1}$. Such a velocity rise at early times is very unusual in SNe~II, though it is predicted by some models (see, e.g., Fig. 9 of \citealt{Dessart10}).
This is likely not related to the unshocked CSM because the velocities are too high. Instead, it is a projection effect caused by sphericity \citep{Dessart05b,Dessart10}.

\subsection{Nebular Phase}~\label{sec:nebulaphase}

\begin{figure*}
    \centering
    \includegraphics[width=0.90\linewidth]{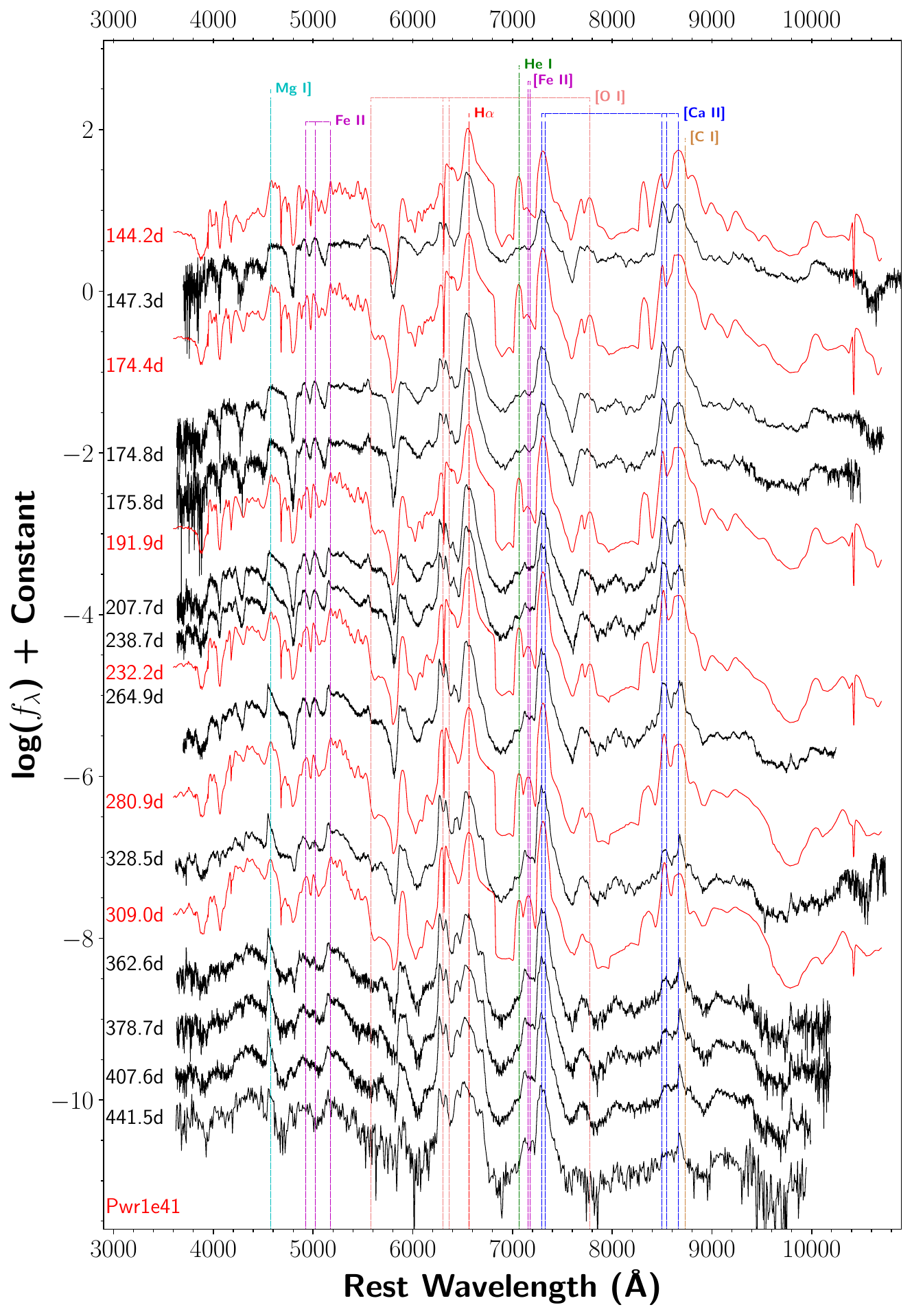}
    \caption{
    Kast spectra of SN\,2023ixf in the nebular phase when the ejecta start to become optically thin; strong forbidden emission  lines start to emerge, including [\ion{O}{1}], \ion{Mg}{1}], and \ion{Ca}{2}. Spectra shown in red ae from the Pwr1e41 model at a nearby phase for comparison.
    }~\label{fig:2023ixf_spec_nebula}
\end{figure*}

After $\sim 100$\,d, the spectra enter the nebular phase and forbidden lines start to form at late times.
Although \ion{H}{1} Balmer lines remain the strongest features (mostly still absorption, as well as absorption from \ion{Fe}{2} $\lambda\lambda\lambda$4924, 5018, 5169), strong emission from forbidden lines also starts to emerge, such as [\ion{O}{1}] $\lambda\lambda$6300, 6364 and $\lambda$7774, \ion{Mg}{1}] $\lambda$4571, [\ion{Ca}{2}] $\lambda\lambda$7291, 7323, and \ion{Ca}{2} $\lambda\lambda\lambda$8498, 8542, 8662, as shown in Figure \ref{fig:2023ixf_spec_nebula}.

\subsection{Models and Physics}~\label{sec:modelandphysics}
\begin{figure}
    \centering
    \includegraphics[width=0.99\linewidth]{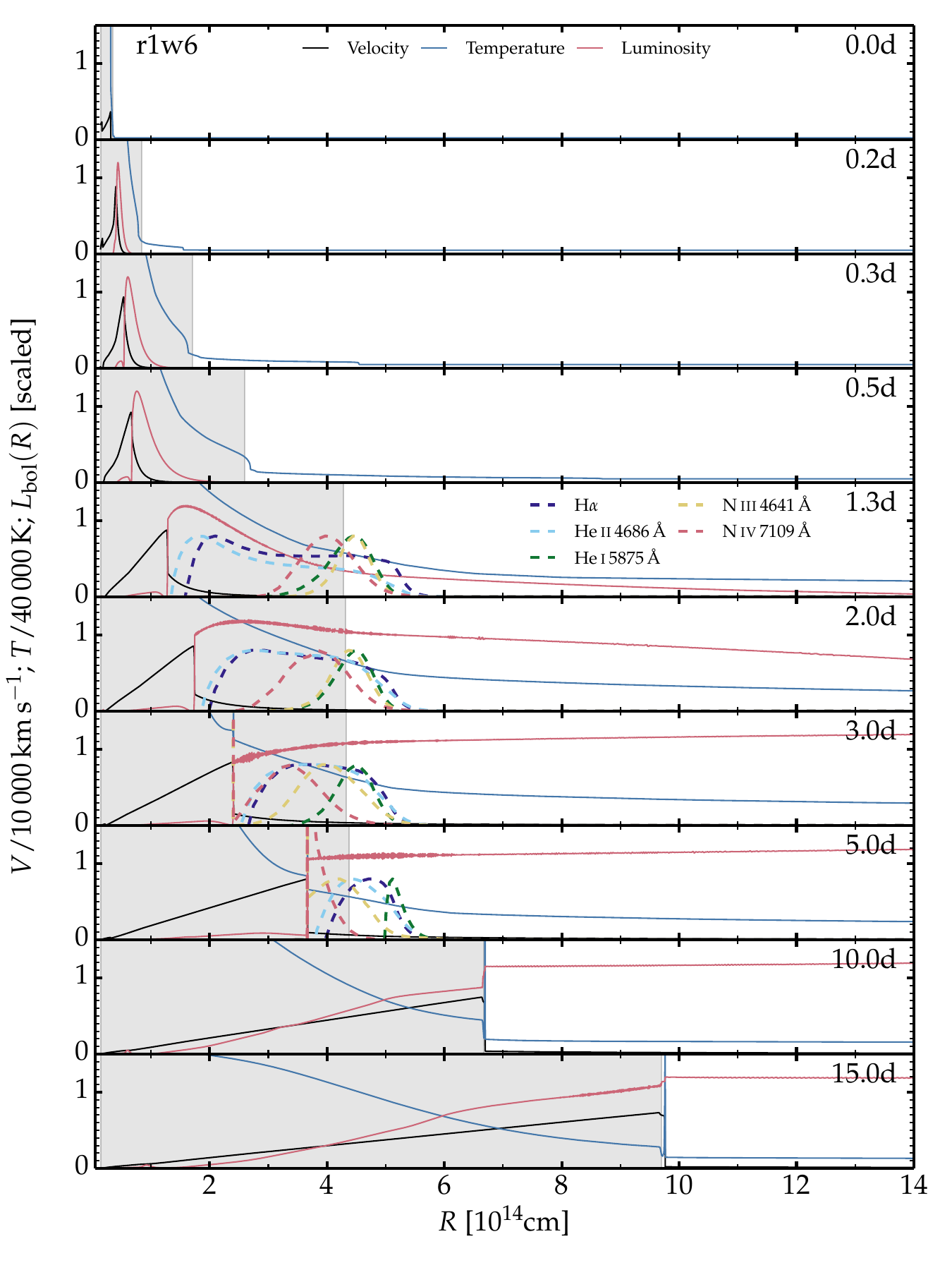}
    \caption{
    Montage showing the evolution of the ejecta/CSM properties from the radiation hydrodynamical simulation of model r1w6 developed by D17, together with the regions of formation of various lines as computed with CMFGEN. The shaded area represents regions with $\tau_{\rm es} > 2/3$.
    }~\label{fig:2023ixf_spec_r1w6_model_montage}
\end{figure}

In order to produce the strongly ionized features \citep{Gal-Yam14} seen in the early-time spectra of SN\,2023ixf, a certain amount of dense CSM is required surrounding the progenitor star, which could extend the shock-breakout time to days (compared to a timescale of hours if without dense CSM, thus difficult to detect).
As the radiation leaks from the shock, UV photons are produced from the cooling of the shock-heated CSM, which then ionizes the CSM, forming these features. Such models were studied/simulated by \citet[][hereafter D17]{Dessart17}, where a set with different parameter spaces (radius, mass-loss rate, ejecta mass, etc.) were explored after explosion from an RSG progenitor.
In fact, the progenitor of SN\,2023ixf has been directly identified and confirmed to be an RSG \citep[e.g.,][]{Kilpatrick23,Soraisam23,VanDyk2023,Jencson23,Pledger23,Niu23,Xiang23,Qin23}, with an effective temperature, luminosity, and initial mass of $\sim 3450$\,K, $\sim 9.3 \times 10^4\, L_{\odot}$, and 12--15\,$M_{\odot}$, respectively \citep{VanDyk2023}. In the D17 models, r1w4 and r1w6 are the two showing strongly ionized features at early times, but  r1w6 matches the evolution of SN\,2023ixf better since the r1w4 model evolves too fast, so here we focus more on r1w6. For the r1w6 model, the following parameters were set: progenitor star radius $R_\star = 501\,R_\odot$ (i.e., $\sim 3.5 \times 10^{13}$\,cm), mass-loss rate $\dot M = 10^{-2}\,M_\odot$\,yr$^{-1}$, extended CSM mass $M_{\rm ext} = 3.04 \times 10^{-2}\,M_\odot$ up to a distance of $R_{\rm CSM} \approx 5 \times 10^{14}$\,cm, and ejecta mass $M_{\rm ejecta} = 12.52\,M_\odot$ (see Table 1 and Sec. 3 in \citealt{Dessart17} for more details). A montage showing the evolution of the ejecta/CSM properties from the radiation hydrodynamical simulation of model r1w6 \citep{Dessart17}
is plotted in Figure \ref{fig:2023ixf_spec_r1w6_model_montage}, where black, blue, and red lines respectively show the velocity, temperature, and luminosity along the radial distance at various epochs. The shaded area represents regions with $\tau_{\rm es} > 2/3$.

The r1w6 model spectra\footnote{Those used in this paper are slightly updated compared with the original r1w6 model given by D17 in which only \ion{H}{0}, \ion{He}{0}, \ion{CNO}{0}, and \ion{Fe}{0} were included. The updated models presented here include all important metals up to Nickel together with a more extended model atom for up to five ions per species (\ion{Fe}{1} to \ion{Fe}{5}) as well as number of levels etc.}
at 1.5\,d, 2.0\,d, and 3.0\,d are plotted in Figure \ref{fig:2023ixf_spec_first5days} in the CSM-dominated phase compared with the observed spectra. Overall, the model spectra match the observed spectra quite satisfactorily, including most of the major features such as the \ion{H}{1} Balmer and Paschen series, all the \ion{He}{2} series, \ion{C}{4}, \ion{N}{3}, and \ion{N}{4}. However, there do exist some mismatches between the model spectra and the observed spectra. (1) As mentioned in Section \ref{sec:earlyphaseionizationfeatures}, there are three \ion{He}{1} lines ($\lambda$5876, $\lambda$6678, and $\lambda$7065) detected in the first-night spectra (days 1.41 to 1.70), yet they are not predicted (at least not obvious) in the model spectra\footnote{The \ion{He}{1} lines are predicted in a slight variant of the r1w6 model presented by \cite{Jacobson-Galan23}; see their Figure 2.},
and they evolve quickly, showing intraday changes and almost disappearing within 8\,hr. The disappearance of the \ion{He}{1} lines is likely because of the quick increase of the temperature during the second day after explosion.
As shown in Figure \ref{fig:2023ixf_spec_BBfit}, the temperature rises quickly in the first 3 days, with $\sim 16,000$\,K at 1.4\,d and quickly reaches peak around 24,000\,K at $\sim 2.5$ days after explosion.
The \ion{He}{1} lines require lower ionization energy (temperature); thus, at $t<1.7$\,days, with lower temperatures, some \ion{He}{1} lines formed and are detected in our spectra. As the temperature increase quickly and reaches its peak at $\sim 2.5$\,d, most of the helium was ionized to \ion{He}{2}, which requires a higher ionization energy (temperature) than \ion{He}{1}, so \ion{He}{1} almost disappears after two days, as predicted by the models too.
(2) There appears to be \ion{N}{5} $\lambda\lambda$4604, 4620 in the 2.4\,d spectrum, but not before nor after. If real, this is likely also because of the temperature changes.
Typically, owing to temperature/ionization stratification in the CSM, the \ion{N}{5} forms in the deepest layers of the CSM (i.e., closer to the shock where temperature is higher) as \ion{N}{5} requires a very high ionization energy/temperature (there could also be a greater diversity in ionization if the CSM is asymmetric). At day 2.4, the temperature reached its peak, meeting the requirements to form \ion{N}{5} lines (but not before the peak), and the temperature also decreases after the peak, so the \ion{N}{5} disappeared thereafter. 
(3) The blue wing of \ion{He}{2} $\lambda$4686 is stronger and more asymmetric compared with the red wing, differing from the model spectra and were seen in some SNe \citep{Jacobson-Galan24}. Although the blue wing of \ion{He}{2} $\lambda$4686 is very complicated with many lines blended together (see Fig. \ref{fig:2023ixf_spec_lines_NIII4634}), the flux in this region still seems to be too strong compared with the model spectra.
This flux is therefore likely from the blueshifted emission of the fast-moving dense shell or the ejecta. The emission blueward of \ion{He}{2} $\lambda$4686 even strengthens in the first few days (see the right panel of Fig. \ref{fig:2023ixf_spec_lines_NIII4634}), supporting the fact that we are seeing progressively more \ion{He}{2} $\lambda$4686 emission from the dense shell, which appears blueshifted at infinity due to optical-depth effects.
The blueshifted emission eventually disappears after a week  because \ion{He}{2} recombines as the temperature drops.

\begin{figure}
    \centering
    \includegraphics[width=0.99\linewidth]{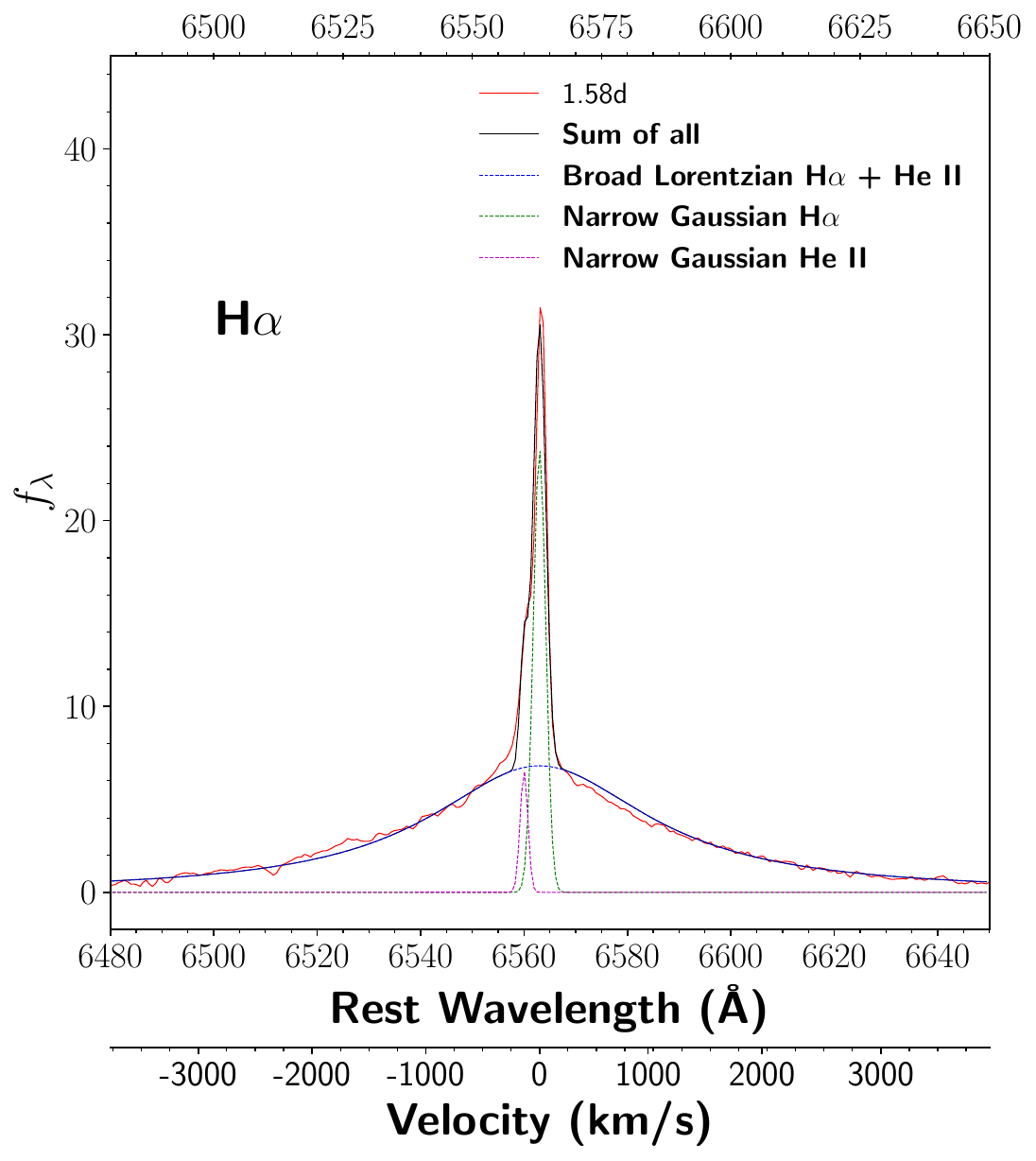}
    \caption{
    H${\alpha}$ profile from our higher-resolution spectra at day 1.58, where \ion{He}{2} $\lambda$6560 is distinguished from H${\alpha}$.  Three components are used to fit the profile: a narrow Gaussian component for \ion{He}{2} $\lambda$6560 (dashed cyan line), a narrow Gaussian component for H${\alpha}$ (dashed green line), and a wide Lorentzian component for both H${\alpha}$ and \ion{He}{2} $\lambda$6560 (dashed blue line).
    }~\label{fig:2023ixf_spec_lines_HalphaD1.62}
\end{figure}

During the CSM-dominated phase, in the higher-resolution spectrum at 1.58\,d we are able to  distinguish \ion{He}{2} $\lambda$6560 from the blue wing of H${\alpha}$, as shown in Figure \ref{fig:2023ixf_spec_lines_HalphaD1.62}.
The ionized H${\alpha}$ luminosity can be used to estimate the mass-loss rate before explosion \citep{Ofek13}. To do so, we follow the steps presented by \cite{zhangj2023}.
We fit the H${\alpha}$ profile with three components: a narrow Gaussian component for \ion{He}{2} $\lambda$6560 (dashed cyan line), a narrow Gaussian component for H${\alpha}$ (dashed green line), and a wide Lorentzian component for both H${\alpha}$ and \ion{He}{2} $\lambda$6560 (dashed blue line) as shown in Figure \ref{fig:2023ixf_spec_lines_HalphaD1.62}.
We only adopt one wide component for both H${\alpha}$ and \ion{He}{2} $\lambda$6560 because it is impossible to distinguish the two wide components given the small difference in central wavelength but very wide wings.
We estimate the intrinsic FWHM of H${\alpha}$ to be $80 \pm 10$\,km\,s$^{-1}$, after removing by quadrature the instrumental FWHM of $\sim 102$\,km\,s$^{-1}$ (estimated from the night-sky emission lines) from the original measurement of $\sim 129$\,km\,s$^{-1}$.
But note that the intrinsic FWHM of H${\alpha}$ measured at this time may have already been changed because of radiative acceleration, so it is more appropriate to consider this estimate as an upper limit of the intrinsic FWHM.
The flux ratio between narrow H${\alpha}$ and \ion{He}{2} $\lambda$6560 is $\sim 6:4$.
The wide Lorentzian component gives a FWHM of $\sim 2400$\,km\,s$^{-1}$, much broader than the narrow component.
According to the relation  $L_{\rm{H}\alpha}\,\approx\,2\,\times\,10^{39}\,\dot{\rm M}^2_{0.01}\,v^{-2}_{\rm{w},500}\,\beta\,\rm{r^{-1}_{15}\,erg\,s^{-1}}$ \citep{Ofek13}, we estimate the narrow H${\alpha}$ luminosity to be $2.04 \times 10^{38}$\,erg\,s$^{-1}$. 
At 1.58\,d we estimate the inner CSM distance (before swept up by the ejecta) to be $\sim 1.4 \times 10^{14}$\,cm (see method in next paragraph), so the mass-loss rate is found to be $\dot M \approx 4 \times 10^{-4}\, M_{\odot}$\,yr$^{-1}$.
But note that the broad H${\alpha}$ component carries much more flux\footnote{Measured to be about 10 times more, but only 60\% is contributed by the broad H${\alpha}$ component, and 40\% is contributed by the broad \ion{He}{2} $\lambda$6560 component, assuming the flux ratio is the same as for the narrow component.} compared to the narrow H${\alpha}$ component; therefore, if counting the entire H${\alpha}$ luminosity, the estimated mass-loss rate would also be 6 times higher, $\dot M \approx 2.4 \times 10^{-3}\,M_{\odot}$\,yr$^{-1}$. This value is lower than the mass-loss rate adopted by the r1w6 model of $\dot{M} = 0.01\, \rm M_{\odot}\,yr^{-1}$, likely because the calculation ignored the optical-depth effects in H$\alpha$.

The H${\alpha}$ profile broadens quickly after the first 2 days, as shown in the middle panel of Figure \ref{fig:2023ixf_spec_lines_HeI}. The FWHM measured at 2.4\,d and 3.4\,d was $\sim 500$\,km\,s$^{-1}$. Interestingly, the FWHM measured from the wide Lorentzian component remains the same at $\sim 2400$\,km\,s$^{-1}$, but still much broader than the narrow component.

\begin{figure}
    \centering
    \includegraphics[width=0.402\linewidth]{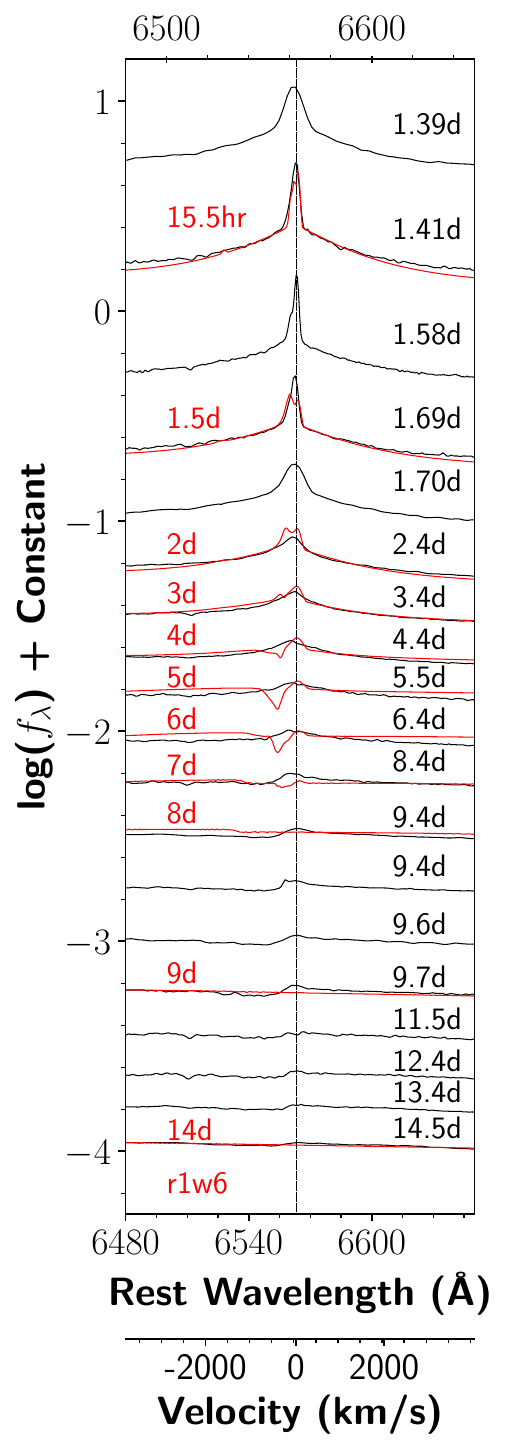}
    \includegraphics[width=0.58\linewidth]{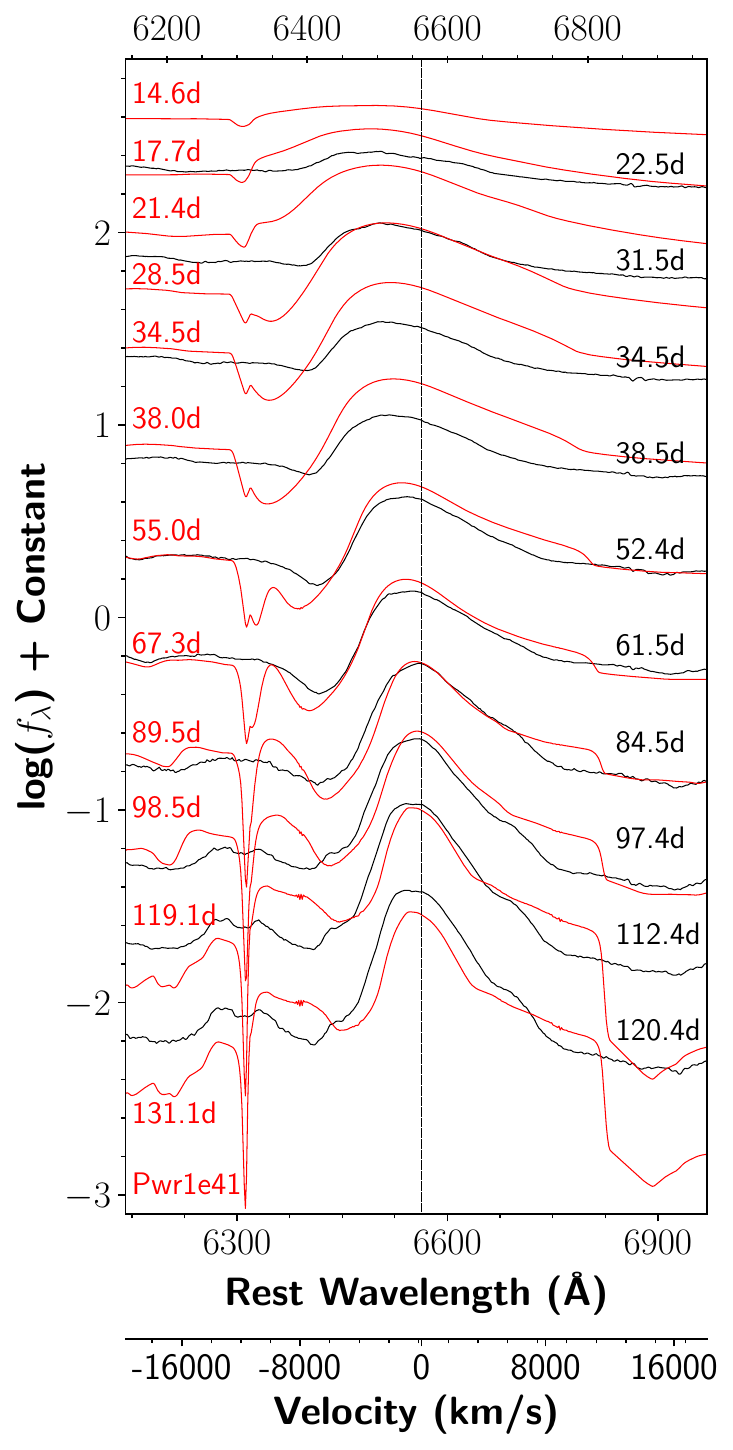}
    \caption{
    {\it Left:} Early-time evolution (first 2 weeks) of the H${\alpha}$ profile compared with the r1w6 model \citep{Dessart17} at similar phases. The FWHM of H${\alpha}$ increases quickly after first 2 days. After a week, H${\alpha}$ starts to show a P~Cygni profile with a minimum absorption velocity of $\sim 1000$\,km\,s$^{-1}$, lasting for more than a week. {\it Right:} Later-time evolution (after 3 weeks) of H${\alpha}$ compared with the Pw1e41 model \citep{Dessart22} at similar phases. The H${\alpha}$ profile starts to show a boxy shape at day 84.5, clearly on the red side, but with an edge velocity of only $\sim 7000$\,km\,s$^{-1}$), much smaller than the $\sim 10,000$\,km\,s$^{-1}$ ejecta velocity. This is direct evidence for an asymmetric dense shell.
    }~\label{fig:2023ixf_spec_lines_HalphaProfile}
\end{figure}

After the first week of the SN explosion, the H${\alpha}$ starts to show a P~Cygni profile with a minimum absorption velocity of $\sim 1000$\,km\,s$^{-1}$, as shown in Figure \ref{fig:2023ixf_spec_lines_Halpha2} (see also left panel of Fig. \ref{fig:2023ixf_spec_lines_HalphaProfile}).
The P~Cygni profile persists for more than a week (during this period, both the observed and r1w6 model spectra are mostly featureless as shown in Fig. \ref{fig:2023ixf_spec_middlefewdays}). The broadening of H${\alpha}$ from $\sim 100$\,km\,s$^{-1}$ to $\sim 1000$\,km\,s$^{-1}$ is unlikely to be caused by
any outbursts or enhanced winds immediately before the explosion, but rather by radiative acceleration\footnote{An alternative explanation could be that we are seeing broad, boxy emission from the interaction with lower-density CSM, and that emission partially fills the H${\alpha}$ absorption. If the CSM is asymmetric and clumpy, this broad emission could be bumpy.}
as suggested by \citet{Zimmerman24}.
However, the strong radiative acceleration mechanism requires the wind to have relatively low density,
which means there is likely a big drop in the density profile following the shock breakout from the dense CSM. Since the P~Cygni feature of H${\alpha}$ was detected
as late as 14.5\,d (after which the CSM would be swept up by the ejecta), by assuming a ejecta velocity of 10,000\,km\,s$^{-1}$ (as inferred from the highest H${\alpha}$ velocity in Fig. \ref{fig:2023ixf_spec_velocity}), one can calculate a distance of $\sim 1.3 \times 10^{15}$\,cm; the low-density CSM surrounding the progenitor was presented at least up to this distance.
Similarly, one can also try to estimate the distance of the dense CSM from the shock breakout time, which is roughly equal to the temperature peak time, because once the shock breaks out, the CSM starts to cool. From the temperature evolution curve shown in Figure \ref{fig:2023ixf_spec_BBfit}, we estimate the peak time to be day 2.5, thus giving a distance of $\sim 2.2 \times 10^{14}$\,cm. This is about 6 times the progenitor radius from the model. 

The drop in the CSM radial density profile  indicates an enhanced episode of mass loss at a certain time prior to the final core collapse. Assuming a wind velocity of 80\,km\,s$^{-1}$ (see above and Fig. \ref{fig:2023ixf_spec_lines_HalphaD1.62}),
the promptly ejected matter would require $\sim 320$ days to reach a distance of $\sim 2.2 \times 10^{14}$\,cm. Since the typical RSG wind velocity is much smaller ($\sim 20$\,km\,s$^{-1}$) than 80\,km\,s$^{-1}$, the actual delay time between the precursor event and the SN explosion could thus be longer, up to a few years.
This could be consistent with the eruptive mass-loss scenario given by \cite{Hiramatsu2023}, who favor eruptions releasing 0.3-–1\,$M_{\odot}$ of the envelope at $\sim 1$\,yr before explosion, although other possibilities are also discussed by \cite{Hiramatsu2023}. 
However, such a high-mass eruption before the explosion
could be at odds with the appearance of SN~IIn-like spectral signatures
because the photons cannot escape (see \citealt{Dessart23}); thus, a gradually enhanced episode of mass loss is more plausible.
We also remark that the enhanced mass loss before the explosion inferred from the early-time spectral evolution of SN\,2023ixf is  corroborated by the semi-analytic fits to its early chromatic luminosity evolution \citep{Li23}. The latter demand the presence of a dense CSM shell confined within a distance of a few $10^{14}$ to $10^{15}$\,cm from the progenitor star, which is most likely to be produced through elevated mass loss at a rate of $\sim 10^{-3}$ to $10^{-2}\,M_{\odot}$\,yr$^{-1}$ a few years before the SN explosion.

At the same (CDS-dominated) phase, the \ion{He}{1} lines (which disappeared earlier) also start showing broad P~Cygni profiles around day 8, as shown in Figures \ref{fig:2023ixf_spec_middlefewdays_flatspectral} and  \ref{fig:2023ixf_spec_lines_HeI5876HeI7065Profile}.
\ion{He}{1} $\lambda$5876 has a velocity of 7000\,km\,s$^{-1}$ (\ion{He}{1} $\lambda$7065 is not very significant), similar to the velocity of H${\beta}$, H${\gamma}$, and H${\delta}$ at the same phase. 
The emergence of broad \ion{He}{1} is also predicted by the r1w6 model (Fig. \ref{fig:2023ixf_spec_middlefewdays_flatspectral}, red spectra) as the ejecta cool down and proceed to the recombination phase, though the model predicts the line to be present earlier (emerging on day 5 and disappearing by day 9).

On day 22, the broad P~Cygni profile is well-developed and dominates the spectra. By this time, since the r1w6 model has no corresponding spectra at similar phases, we plot the shock-powered model Pwr1e41 \citep{Dessart22} instead, as shown in Figure \ref{fig:2023ixf_spec_lastfewdays}. In the Pwr1e41 model, a dense shell at 11,000\,km\,s$^{-1}$ was placed (which is very close to the ejecta velocity of $\sim 10,000$\,km\,s$^{-1}$ measured from H${\alpha}$) to mimic the interaction of CSM at the earliest times and then a constant power of 10$^{41}$\,erg\,s$^{-1}$. Since the model was developed in 1D, both the dense shell and the shock power are distributed in a  spherical shell at 11,000 km\,s$^{-1}$. We adopt the Pwr1e41 model here because it matches best with the observed spectra, including most of the major P~Cygni features. 

We also note that starting from day 84.5,
a box-shaped component underlying the H${\alpha}$ profile emerges as indicated by the notch on its right side.
The boxy profile is an indication of the continued interaction between the ejecta and the CSM confined within a radially extended shell.
However, the boxy shape edge in the observed spectra shows a velocity of only $\sim 7000$\,km\,s$^{-1}$ (see more details in Fig. \ref{fig:2023ixf_spec_lines_HalphaProfile}, right panel), much smaller than the $\sim 10,000$\,km\,s$^{-1}$ ejecta velocity measured from H${\alpha}$, which suggests an asymmetric dense shell or shock power (asymmetric wind). 
The Pwr1e41 model also clearly predicts the boxy profile at a higher velocity of 11,000\,km\,s$^{-1}$ (model input parameter), on both the red and blue sides. On the blue side, the box-shaped signature in the model spectra is
a strong absorption feature at 11,000\,km\,s$^{-1}$, which persists in the model spectra since day 15 (see Fig. \ref{fig:2023ixf_spec_lines_HalphaProfile}, right panel). However, such a strong feature is not clearly seen in the observed spectra, at least not before day 97 (there is a weak absorption feature after day 97 at similar velocity, but unclear whether it is related to the boxy profile). Instead, there is a boxy shape at $\sim 6500$\,km\,s$^{-1}$, similar to the red side.
If these two boxy profiles on both the blue and red side are connected (since they are at similar velocity), it would suggest a possible axial symmetry, as in a bipolar explosion.\footnote{Note that it could also be possible that the CDS of SN\,2023ixf is actually only at $\sim 7000$\,km\,s$^{-1}$, but this conflicts with the velocity measured from H${\alpha}$ of $\sim 10,000$\,km\,s$^{-1}$.} In any case, since their velocities are much smaller than the ejecta velocity, it is an indication of an asymmetric dense shell, which may be related to the polarization. In fact, polarization of SN\,2023ixf has been detected and reported at early phases \citep{Vasylyev23}, and it further  indicates that the CSM is indeed asymmetric.
The asymmetry may help explain the observed lower velocity, as we might be seeing only the projected velocity along our line of sight, which is much lower than the model CDS velocity at 11,000 km\,s$^{-1}$.

\begin{figure*}
    \centering
    \includegraphics[width=0.49\linewidth]{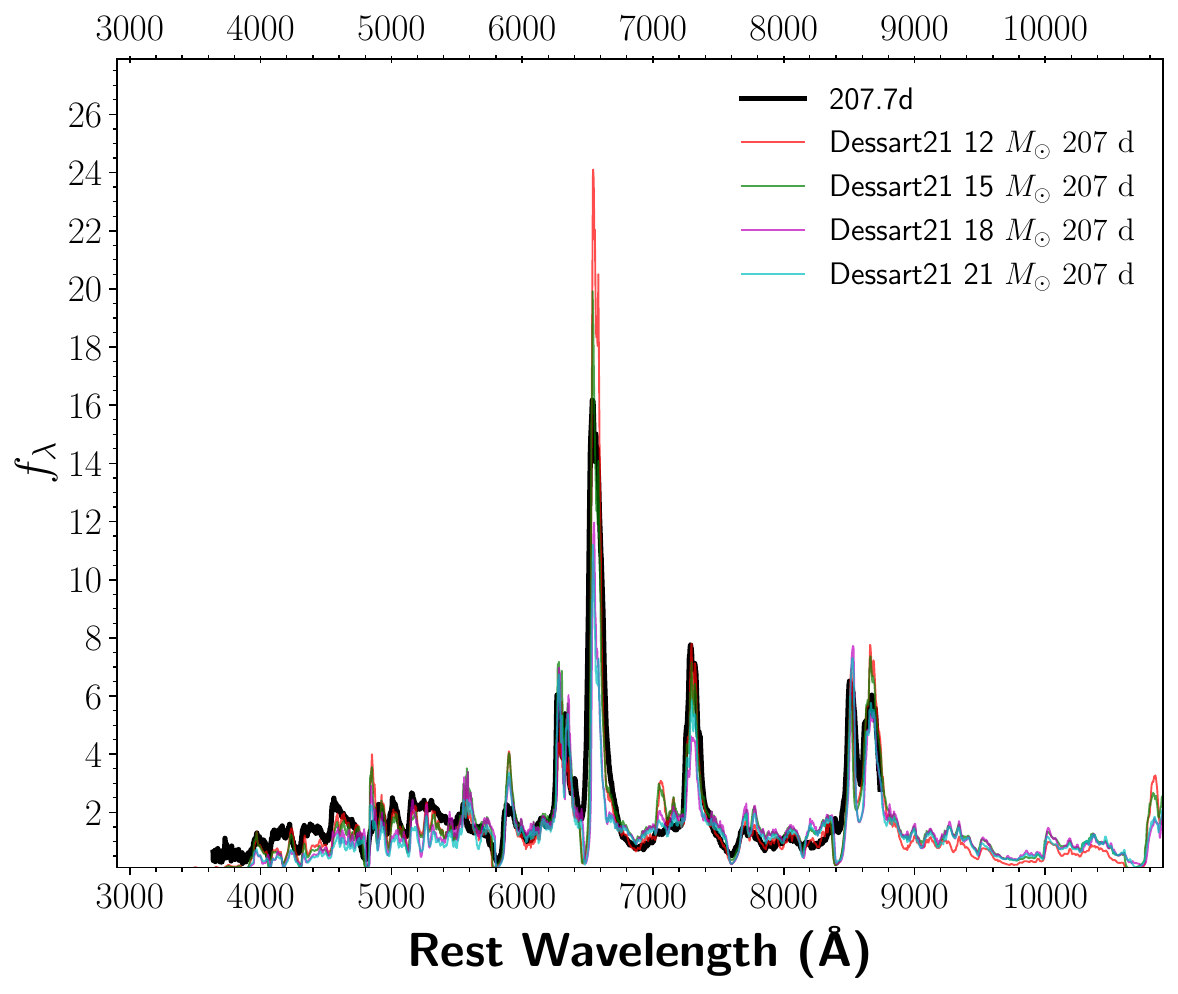}
    \includegraphics[width=0.49\linewidth]{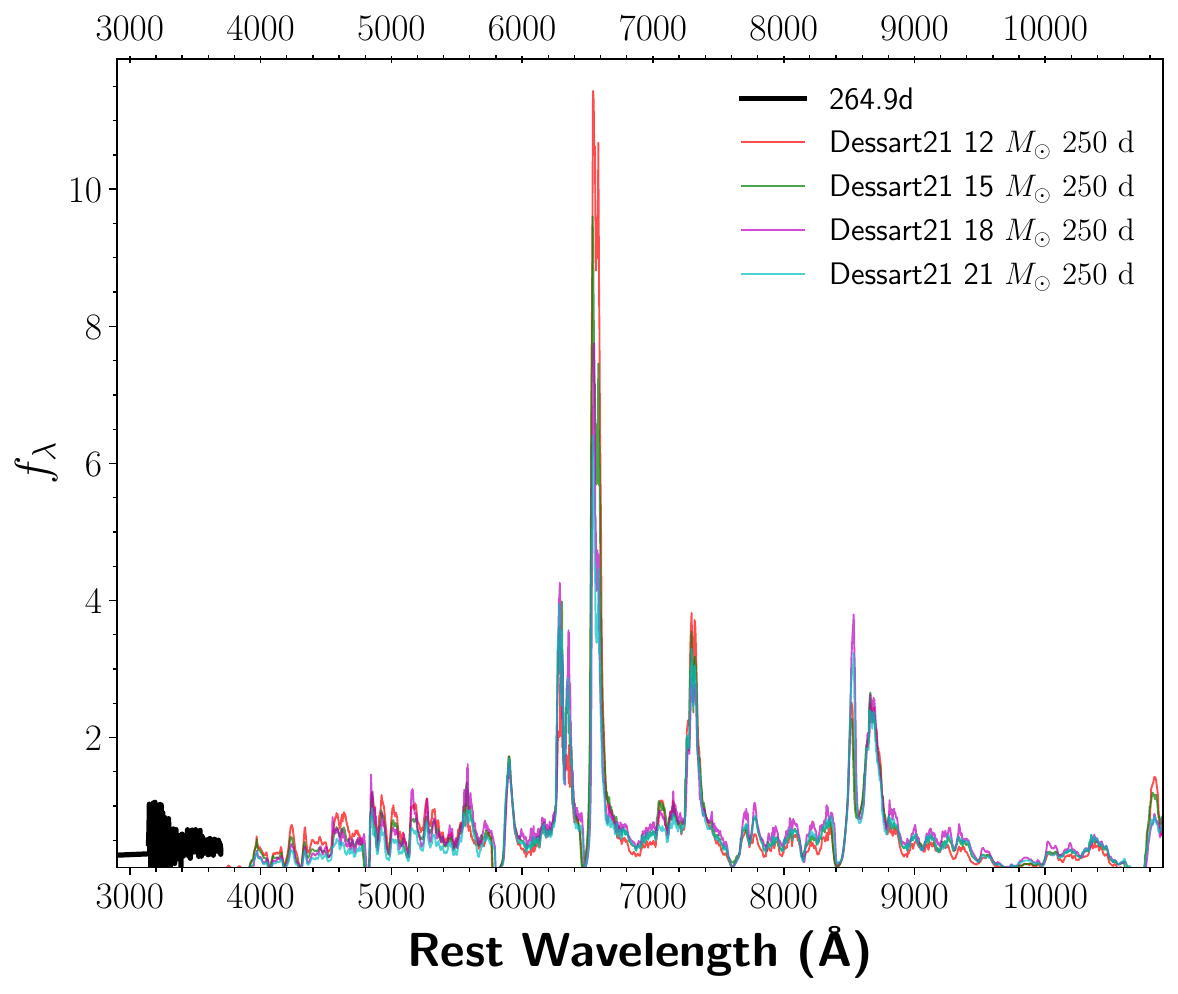}
    \includegraphics[width=0.49\linewidth]{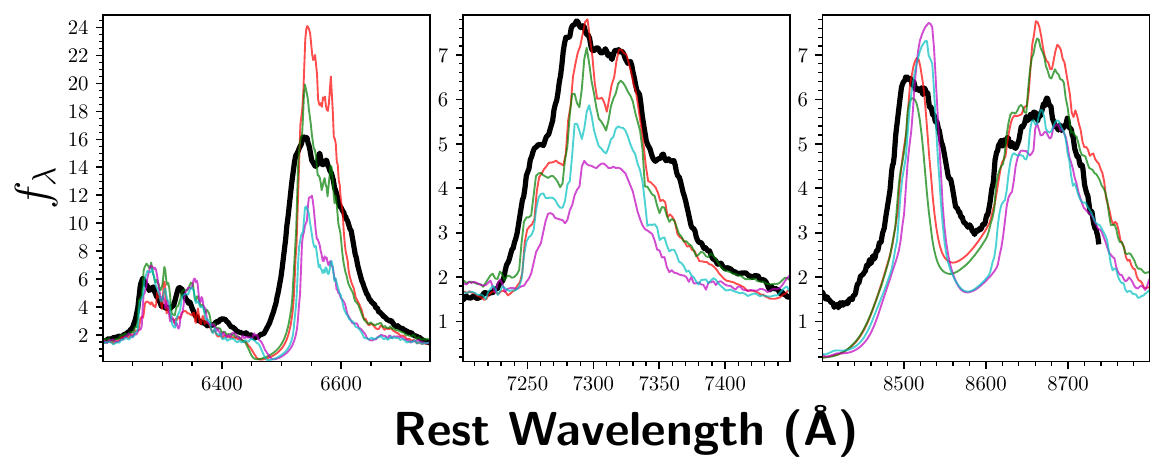}
    \includegraphics[width=0.49\linewidth]{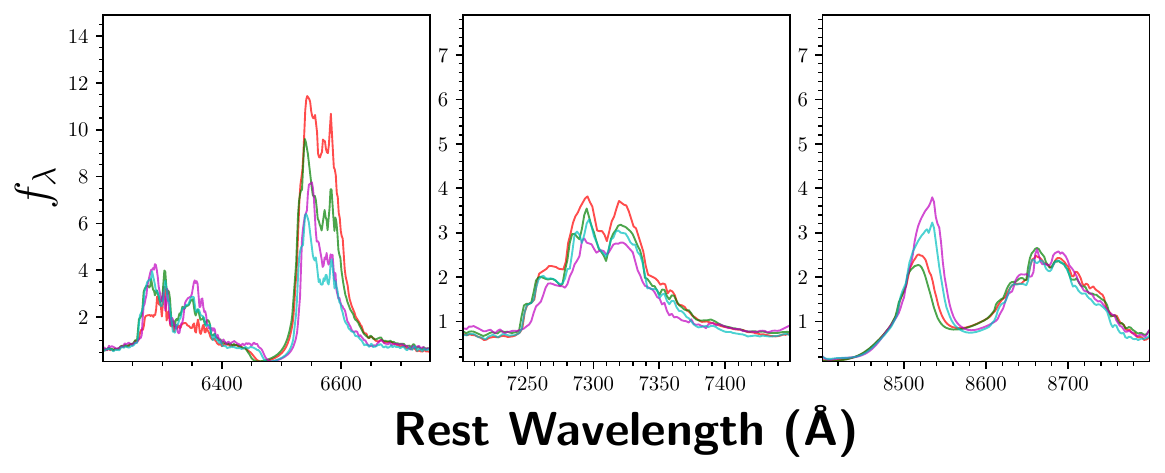}
    \includegraphics[width=0.49\linewidth]{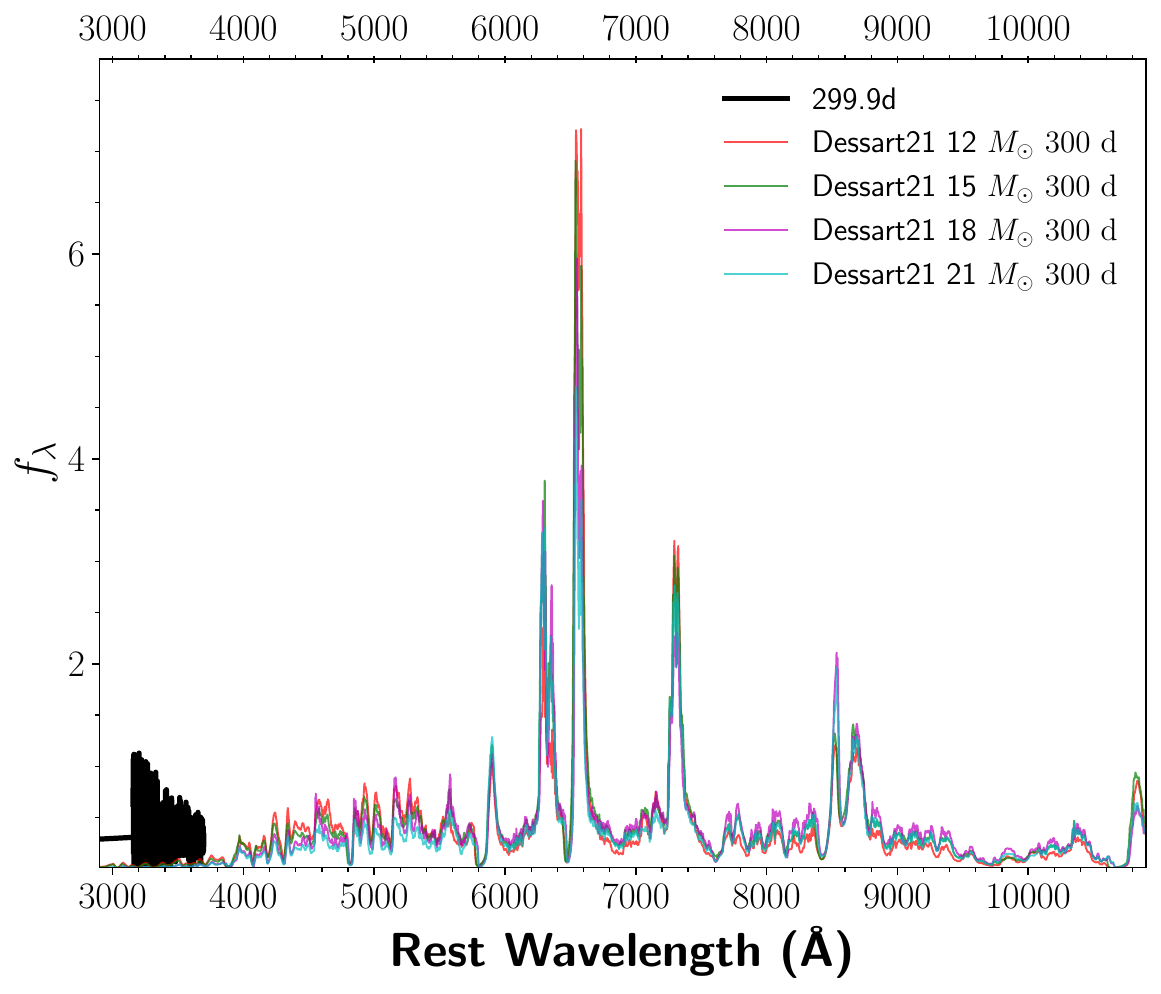}
    \includegraphics[width=0.49\linewidth]{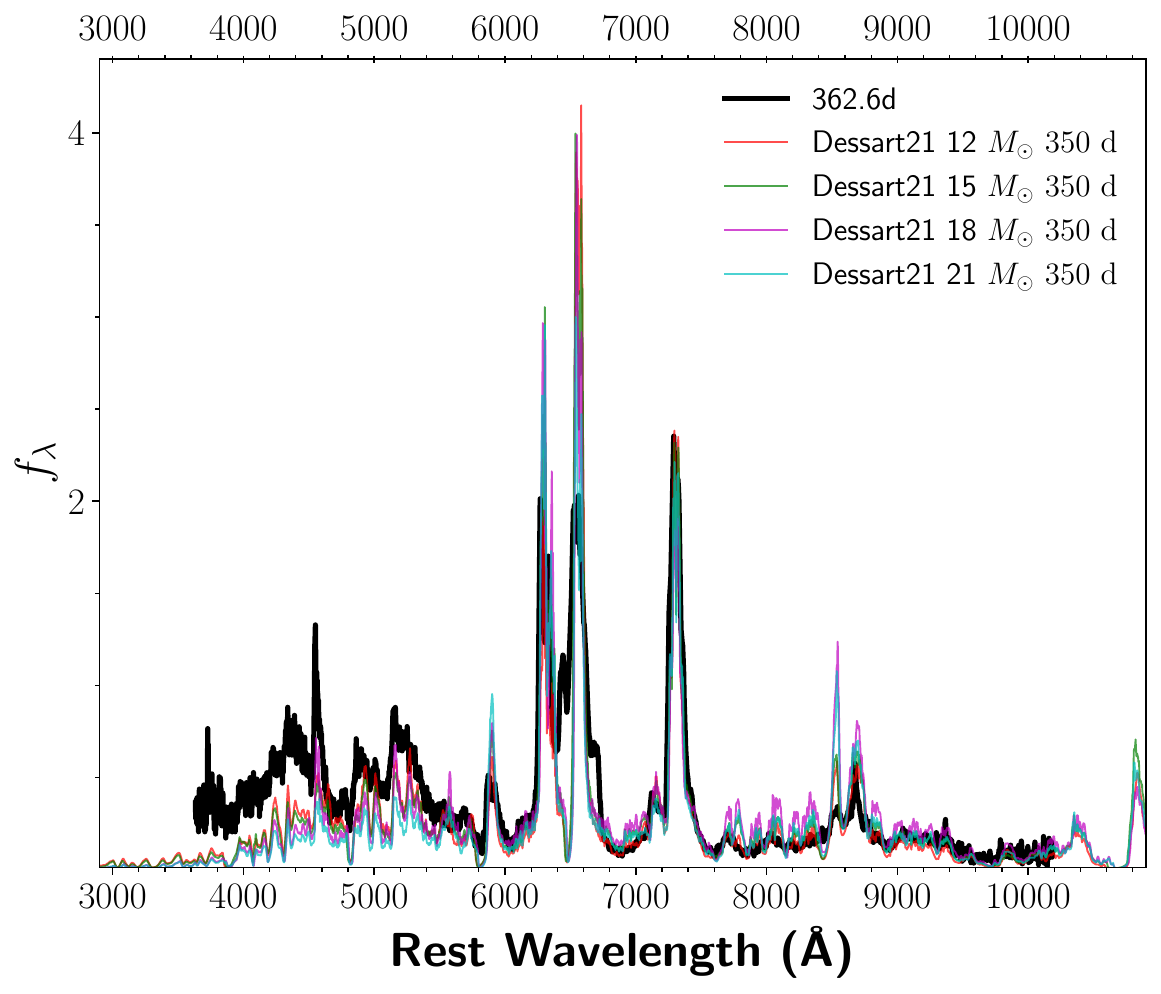}
    \includegraphics[width=0.49\linewidth]{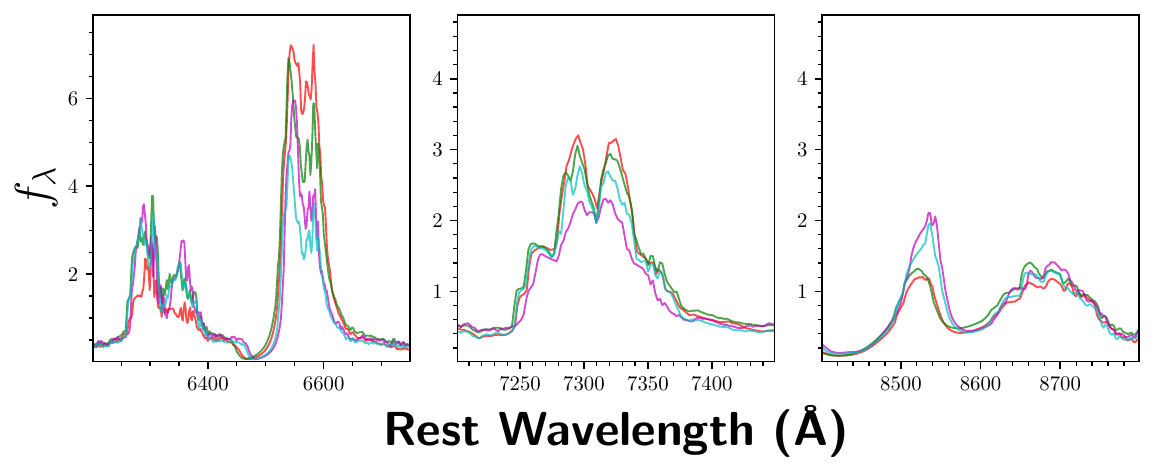}
    \includegraphics[width=0.49\linewidth]{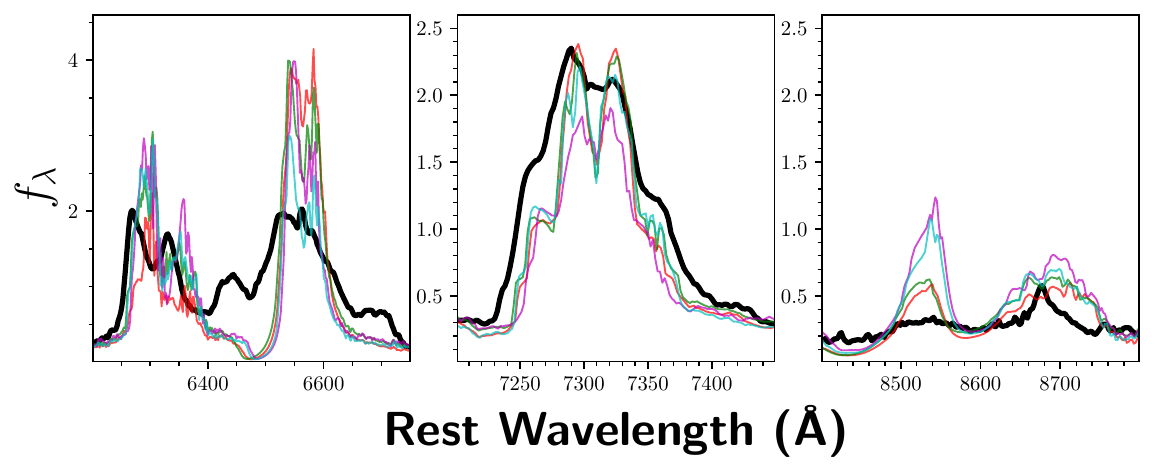}
    \caption{
    Spectra of SN\,2023ixf in the nebular phase at days 207.7, 264.9, 299.9, and 362.6, compared with models \citep{Dessart21} having different masses. Each bottom panel shows zoomed-in regions of the three strongest lines to reveal details of the comparisons.
    }~\label{fig:2023ixf_spec_nebula_modelcompare_207d}
\end{figure*}

The nebular spectra can help us estimate the progenitor mass by comparing the observed spectra with the model spectra. We obtained several nebular spectra of SN\,2023ixf at different phases.
The model spectra were constructed based on the ejecta/progenitor models given by \cite{Dessart21} and \cite{Sukhbold16}. Figure \ref{fig:2023ixf_spec_nebula_modelcompare_207d} shows the comparison at four epochs\footnote{Note that \cite{Dessart21}  only showed model spectra at 350\,d, other epochs model spectra were computed as part of this work.}: 207.7\,d, 264.9\,d, 299.9\,d, and 362.6\,d. It appears that the $15\,M_{\odot}$ model matches best with the observed nebular spectrum (followed by the $12\,M_{\odot}$ model) for almost all the epochs. This is also consistent with the direct progenitor analysis by \cite{VanDyk2023}, who constrained the initial mass of the progenitor candidate from $12\,M_{\odot}$ to $15\,M_{\odot}$.
Comparing the four phases, we notice that the spectral flux decreased for all the lines
except [\ion{Ca}{2}] $\lambda\lambda$7291, 7323 (middle of bottom zoomed-in panels for each phase in Fig. \ref{fig:2023ixf_spec_nebula_modelcompare_207d}), which actually became stronger relative to \ion{Ca}{2} $\lambda\lambda\lambda$8498, 8542, 8662 (right side of bottom zoomed-in panels for each phase in Fig. \ref{fig:2023ixf_spec_nebula_modelcompare_207d}). This is reasonable; as the ejecta grow more optically thin, the forbidden lines of \ion{Ca}{2} strengthen relative to the permitted \ion{Ca}{2} triplet.
Also, note that the \ion{Fe}{2} emission below 5500\,\AA\ is much stronger than in the model spectra; this might come from supersolar metallicity, or greater $^{56}$Ni mass, or (more likely) extra emission from interaction.
Similarly, the nebular [\ion{Ca}{2}] $\lambda\lambda$7291, 7323 doublet on days 264.9  and 299.9 is also stronger than in the model spectra, which generally implies greater $^{56}$Ni since the line forms primarily from the $^{56}$Ni-rich material.
Most lines also show broad emission, implying a sizable CDS contributing H${\alpha}$ and perhaps some \ion{Ca}{2}.
In addition, the ``bridge'' between H${\alpha}$ and [\ion{O}{1}] $\lambda\lambda$6300, 6364 is much stronger in the nebular spectra (left side of bottom zoomed-in panels for each phase in Fig. \ref{fig:2023ixf_spec_nebula_modelcompare_207d}), a clear signature of a broad H${\alpha}$ emission from interaction. 
This means that some H${\alpha}$ emission is present at high blueshifted velocities, while it seems less present on the opposite, redshifted/receding part of the ejecta, 
linking to evidence of asymmetry at earlier times (i.e., asymmetry still prevails far from the star, although it is unclear whether this is the same geometry).
In any case, the nebular spectrum until 362.6\,d and even 441.5\,d suggests that SN\,2023ixf continues to interact (ever since shock breakout) with CSM produced by wind mass loss prior to the explosion.

\section{Distance Measurement}~\label{sec:disfitanalysis}

M101 is a unique galaxy for testing our current methods to measure extragalatic distances. We have precise distance measurements from Cepheids \citep{Freedman2001,Saha2006,Shappee2011,Riess2022}, TRGB\citep{rizzi07,Shappee2011,Beaton2019,anand22}, and SNe~Ia \citep{Vinko2012,Riess2022}. The most recent measurements from those three techniques are all in good agreement. Using Cepheids, \citet{Riess2022} obtained a distance modulus of $29.194 \pm 0.03$\,mag; with TRGB, \citet{anand22} obtained $29.08 \pm 0.05$\, mag; and with SN 2011fe, \cite{Riess2022} measured $29.04 \pm 0.12$\,mag.



However, the distance to M101 can also be derived using SN\,2023ixf. Even if SNe~II display a large range of peak luminosities, it has been demonstrated that they are excellent distance indicators. Using theoretical and empirical methods to calibrate them  \citep{review}, we can measure distances with a precision of $\sim 15$\% \citep{dejaeger20}.

We focus our effort on the standard candle method \citep[SCM;][]{hamuy_distance_2001}, which is the most common and most accurate empirical technique used to derive SN~II distances. Developed by \citet{hamuy_distance_2001}, it is based on the correlation between the SN absolute magnitude and two observables: the expansion velocity and the color. The expansion velocity is generally determined from the blueshift of Fe~II $\lambda 5169$ or H$\beta$  P~Cygni features. Using this technique, we obtain a distance of $28.67\pm0.14$~mag. This value is smaller than those derived from other techniques (Cepheids, TRGB, SNe~Ia). The difference can be explained by the presence of strong CSM interaction \citep{Hiramatsu2023}, which boosts the luminosity above that of a  normal SN~II. The SCM is not able to correct for this effect, so the derived M101 distance from SN\,2023ixf is likely too small. However, even if the distance is somewhat inaccurate, we can try to include SN\,2023ixf as a calibrator to derive the Hubble-Lema{\^i}tre constant through the distance-ladder technique. Adding SN\,2023ixf to the 13 calibrators from \citet{dejaeger22b}, and using its Cepheid-based distance, we obtain H$_{0}=73.1^{+3.68}_{-3.50}$\,km\,s$^{-1}$\,Mpc$^{-1}$ (statistical only) instead of  H$_0= 75.4^{+3.8}_{-3.7}$\,km\,s$^{-1}$\,Mpc$^{-1}$ \citep{dejaeger22b}. 
As expected, the uncertainties decrease because the number of calibrators increases (13 vs. 14), and the new value is smaller because SN\,2023ixf is more luminous than the average SN~II. 

More recently, a theoretical method called the tailored expanding photosphere method has been developed by \citet{voglphd}. It is based on the expanding photosphere method \citep{kirshner_distances_1974} but avoids the systematic distance uncertainties from the dilution factors. This method has been applied to several objects, yielding distance uncertainties smaller than 5\% \citep{2023A&A...672A.129C}. However, it cannot be used for SN\,2023ixf because the radiative-transfer codes of \citet{voglphd} do not include the strong CSM interaction seen in this object.

\section{Conclusion and Discussion}~\label{sec:conclusion}

We present both photometric and spectroscopic observations of SN\,2023ixf covering from early times to the nebular phase.
The light curves show that SN\,2023ixf is in the fast decliner (IIL) subclass with a relatively short ``plateau'' phase (less than $\sim 70$ days), indicating an H-rich envelope of lower mass in the progenitor before explosion. It reached a peak $V$-band absolute magnitude of $-18.2 \pm 0.07$\,mag, thus putting it at the bright end of SNe~II.

Optical spectra show that SN\,2023ixf transitioned from Type IIn to a typical Type II SN after three weeks. Early-time spectra of SN\,2023ixf exhibit strong, narrow emission lines from the ionized CSM. We identified many species that produce these lines and found that most of them lines are from H and He, including \ion{He}{2} series from energy states of $n=4$, 5, and 6. The emission features weakened after the first week and the spectra appear to be blue and quasi-featureless in the second week, and can be fitted with a blackbody. After $\sim 3$ weeks, the spectra are similar to those of other SNe~II, with strong P~Cygni features, and the SN entered the nebular phase after about 6 months.

We compare observed spectra of SN\,2023ixf with various model spectra in order to understand the physics behind SN\,2023ixf. There is likely to be dense CSM surrounding the progenitor up to a distance of $\sim 2.2 \times 10^{14}$\,cm, followed by lower density CSM to distances of  at least $\sim 1.3 \times 10^{15}$\,cm. Shortly before exploding, the progenitor may have had a  mass-loss rate of up to $\dot M \approx 2.4 \times 10^{-3}\,M_{\odot}$\,yr$^{-1}$. Our nebular spectra match best with a $15\,M_{\odot}$ model spectrum.

SN\,2023ixf is used as a distance indicator by fitting the light curves; we obtain a distance modulus of $28.67 \pm 0.14$\,mag, slightly smaller than that derived from other techniques (Cepheids, TRGB, SN~Ia), possibly because of strong CSM interaction in SN\,2023ixf. By including SN\,2023ixf as a calibrator and using its Cepheid distance, we obtain H$_{0}=73.1^{+3.68}_{-3.50}$\,km\,s$^{-1}$\,Mpc$^{-1}$ (statistical only), slightly smaller (because SN\,2023ixf is more luminous than the average SN~II) but still in good agreement with H$_0= 75.4^{+3.8}_{-3.7}$\,km\,s$^{-1}$\,Mpc$^{-1}$ given by \cite{dejaeger22b}.

\section{acknowledgments} 
A major upgrade of the Kast spectrograph on the Shane 3\,m telescope at Lick Observatory, led by Brad Holden, was made possible through gifts from the Heising-Simons Foundation, William and Marina Kast, and the University of California Observatories.
KAIT and its ongoing operation were made possible by donations from Sun Microsystems, Inc., the Hewlett-Packard Company, AutoScope Corporation, Lick Observatory, the U.S. National Science Foundation, the University of California, the Sylvia \& Jim Katzman Foundation, and the TABASGO Foundation. 
Research at Lick Observatory is partially supported by a gift from Google. 
Some of the data presented herein were obtained at the W. M. Keck Observatory, which is operated as a scientific partnership among the California Institute of Technology, the University of California, and NASA; the observatory was made possible by the generous financial support of the W. M. Keck Foundation.
We appreciate the expert assistance of the staffs at Lick and Keck Observatories.  

We thank the following Lick Observatory Graduate Workshop participants for help obtaining Nickel photometry on two nights and  Shane spectra on one night:
Fangyi Cao, Casey Carlile, Madalyn Johnson,  Zhexing Li, Ye Lin, Patrick Maloney, Michael McDonald, Pradyumna Sadhu, Loraine Sandoval Ascencio, Sogul Sanjaripour, Niloofar Sharei, Hurum Maksora Tohfa, and Michael Wozniak.

Generous financial support was provided to A.V.F.'s supernova group at U.C. Berkeley by the Christopher R. Redlich Fund, Steven Nelson, Sunil Nagaraj, Landon Noll, Sandy Otellini, Gary and Cynthia Bengier, Clark and Sharon Winslow, Alan Eustace, William Draper, Timothy and Melissa Draper, Briggs and Kathleen Wood, and Sanford Robertson (S.S.V. is a Steven Nelson Graduate Fellow in Astronomy, K.C.P. was a Nagaraj-Noll-Otellini Graduate Fellow in Astronomy, W.Z. is a Bengier-Winslow-Eustace Specialist in Astronomy, T.G.B. is a Draper-Wood-Robertson Specialist in Astronomy, Y.Y. was a Bengier-Winslow-Robertson Fellow in Astronomy).
Numerous other donors to his group and/or 
research at Lick Observatory include Michael Antin, Duncan Beardsley, Jim Connelly and Anne Mackenzie, Curt and Shelley Covey, Byron and Allison Deeter, Arthur Folker, Tom and Dana Grogan, 
Heidi Gerster, Harvey Glasser, Charles and Gretchen Gooding, Tim and Judi Hachman, Alan and Gladys Hoefer, George Hume, Stephen Imbler, Michael Kast and Rebecca Lyon, Lata Krishnan and Ajay Shah, Walter Loewenstern, Rand Morimoto and Anna Henderson, Edward Oates, Doug Ogden, Jon and Susan Reiter, Cat Rondeau, Laura Sawczuk and Luke Ellis, Stan Schiffman, Richard Sesler, Justin and Seana Stephens, Ilya Strebulaev and Anna Dvornikova, Diane Tokugawa and Alan Gould, David and Joanne Turner, David and Malin Walrod, Gerry and Virginia Weiss, Janet Westin and Mike McCaw, David and Angie Yancey.
A.S. acknowledges support from NASA/HST grant GO-17216.
L.D. acknowledges access to the HPC resources of TGCC under the allocation 2023 -- A0150410554 on Irene-Rome made by GENCI, France.
\bigskip

\software{{Astropy \citep{astropy:2013, astropy:2018}, 
IDL Astronomy user's library \citep{landsman_idl_1993}}} 
\bigskip
\newpage

\restartappendixnumbering
    

\bibliographystyle{aasjournal}
\bibliography{2023ixf}

\begin{thebibliography}{}
\expandafter\ifx\csname natexlab\endcsname\relax\def\natexlab#1{#1}\fi
\providecommand{\url}[1]{\href{#1}{#1}}
\providecommand{\dodoi}[1]{doi:~\href{http://doi.org/#1}{\nolinkurl{#1}}}
\providecommand{\doeprint}[1]{\href{http://ascl.net/#1}{\nolinkurl{http://ascl.net/#1}}}
\providecommand{\doarXiv}[1]{\href{https://arxiv.org/abs/#1}{\nolinkurl{https://arxiv.org/abs/#1}}}

\bibitem[{{Anand} {et~al.}(2022){Anand}, {Tully}, {Rizzi}, {Riess}, \&
  {Yuan}}]{anand22}
{Anand}, G.~S., {Tully}, R.~B., {Rizzi}, L., {Riess}, A.~G., \& {Yuan}, W.
  2022, \apj, 932, 15, \dodoi{10.3847/1538-4357/ac68df}

\bibitem[{Anderson {et~al.}(2014)Anderson, González-Gaitán, Hamuy,
  Gutiérrez, Stritzinger, E, Phillips, Schulze, Antezana, Bolt, Campillay,
  Castellón, Contreras, Jaeger, Folatelli, Förster, Freedman, González,
  Hsiao, Krzemiński, Krisciunas, Maza, McCarthy, Morrell, Persson, Roth,
  Salgado, Suntzeff, \& Thomas-Osip}]{anderson_characterizing_2014}
Anderson, J.~P., González-Gaitán, S., Hamuy, M., {et~al.} 2014, ApJ, 786, 67,
  \dodoi{10.1088/0004-637X/786/1/67}

\bibitem[{{Astropy Collaboration} {et~al.}(2013){Astropy Collaboration},
  {Robitaille}, {Tollerud}, {Greenfield}, {Droettboom}, {Bray}, {Aldcroft},
  {Davis}, {Ginsburg}, {Price-Whelan}, {Kerzendorf}, {Conley}, {Crighton},
  {Barbary}, {Muna}, {Ferguson}, {Grollier}, {Parikh}, {Nair}, {Unther},
  {Deil}, {Woillez}, {Conseil}, {Kramer}, {Turner}, {Singer}, {Fox}, {Weaver},
  {Zabalza}, {Edwards}, {Azalee Bostroem}, {Burke}, {Casey}, {Crawford},
  {Dencheva}, {Ely}, {Jenness}, {Labrie}, {Lim}, {Pierfederici}, {Pontzen},
  {Ptak}, {Refsdal}, {Servillat}, \& {Streicher}}]{astropy:2013}
{Astropy Collaboration}, {Robitaille}, T.~P., {Tollerud}, E.~J., {et~al.} 2013,
  \aap, 558, A33, \dodoi{10.1051/0004-6361/201322068}

\bibitem[{{Astropy Collaboration} {et~al.}(2018){Astropy Collaboration},
  {Price-Whelan}, {Sip{\H{o}}cz}, {G{\"u}nther}, {Lim}, {Crawford}, {Conseil},
  {Shupe}, {Craig}, {Dencheva}, {Ginsburg}, {Vand erPlas}, {Bradley},
  {P{\'e}rez-Su{\'a}rez}, {de Val-Borro}, {Aldcroft}, {Cruz}, {Robitaille},
  {Tollerud}, {Ardelean}, {Babej}, {Bach}, {Bachetti}, {Bakanov}, {Bamford},
  {Barentsen}, {Barmby}, {Baumbach}, {Berry}, {Biscani}, {Boquien}, {Bostroem},
  {Bouma}, {Brammer}, {Bray}, {Breytenbach}, {Buddelmeijer}, {Burke},
  {Calderone}, {Cano Rodr{\'\i}guez}, {Cara}, {Cardoso}, {Cheedella}, {Copin},
  {Corrales}, {Crichton}, {D'Avella}, {Deil}, {Depagne}, {Dietrich}, {Donath},
  {Droettboom}, {Earl}, {Erben}, {Fabbro}, {Ferreira}, {Finethy}, {Fox},
  {Garrison}, {Gibbons}, {Goldstein}, {Gommers}, {Greco}, {Greenfield},
  {Groener}, {Grollier}, {Hagen}, {Hirst}, {Homeier}, {Horton}, {Hosseinzadeh},
  {Hu}, {Hunkeler}, {Ivezi{\'c}}, {Jain}, {Jenness}, {Kanarek}, {Kendrew},
  {Kern}, {Kerzendorf}, {Khvalko}, {King}, {Kirkby}, {Kulkarni}, {Kumar},
  {Lee}, {Lenz}, {Littlefair}, {Ma}, {Macleod}, {Mastropietro}, {McCully},
  {Montagnac}, {Morris}, {Mueller}, {Mumford}, {Muna}, {Murphy}, {Nelson},
  {Nguyen}, {Ninan}, {N{\"o}the}, {Ogaz}, {Oh}, {Parejko}, {Parley}, {Pascual},
  {Patil}, {Patil}, {Plunkett}, {Prochaska}, {Rastogi}, {Reddy Janga},
  {Sabater}, {Sakurikar}, {Seifert}, {Sherbert}, {Sherwood-Taylor}, {Shih},
  {Sick}, {Silbiger}, {Singanamalla}, {Singer}, {Sladen}, {Sooley},
  {Sornarajah}, {Streicher}, {Teuben}, {Thomas}, {Tremblay}, {Turner},
  {Terr{\'o}n}, {van Kerkwijk}, {de la Vega}, {Watkins}, {Weaver}, {Whitmore},
  {Woillez}, {Zabalza}, \& {Astropy Contributors}}]{astropy:2018}
{Astropy Collaboration}, {Price-Whelan}, A.~M., {Sip{\H{o}}cz}, B.~M., {et~al.}
  2018, \aj, 156, 123, \dodoi{10.3847/1538-3881/aabc4f}

\bibitem[{{Beaton} {et~al.}(2019){Beaton}, {Seibert}, {Hatt}, {Freedman},
  {Hoyt}, {Jang}, {Lee}, {Madore}, {Monson}, {Neeley}, {Rich}, \&
  {Scowcroft}}]{Beaton2019}
{Beaton}, R.~L., {Seibert}, M., {Hatt}, D., {et~al.} 2019, \apj, 885, 141,
  \dodoi{10.3847/1538-4357/ab4263}

\bibitem[{{Bonanos} \& {Boumis}(2016)}]{Bonanos16}
{Bonanos}, A.~Z., \& {Boumis}, P. 2016, \aap, 585, A19,
  \dodoi{10.1051/0004-6361/201425412}

\bibitem[{{Bostroem} {et~al.}(2023){Bostroem}, {Pearson}, {Shrestha}, {Sand},
  {Valenti}, {Jha}, {Andrews}, {Smith}, {Terreran}, {Green}, {Dong},
  {Lundquist}, {Haislip}, {Hoang}, {Hosseinzadeh}, {Janzen}, {Jencson},
  {Kouprianov}, {Paraskeva}, {Meza Retamal}, {Reichart}, {Arcavi}, {Bonanos},
  {Coughlin}, {Dobson}, {Farah}, {Galbany}, {Guti{\'e}rrez}, {Hawley}, {Hebb},
  {Hiramatsu}, {Howell}, {Iijima}, {Ilyin}, {Jhass}, {McCully}, {Moran},
  {Morris}, {Mura}, {M{\"u}ller-Bravo}, {Munday}, {Newsome}, {Pabst}, {Ochner},
  {Gonzalez}, {Pastorello}, {Pellegrino}, {Piscarreta}, {Ravi}, {Reguitti},
  {Salo}, {Vink{\'o}}, {de Vos}, {Wheeler}, {Williams}, \&
  {Wyatt}}]{bostroem_early_2023}
{Bostroem}, K.~A., {Pearson}, J., {Shrestha}, M., {et~al.} 2023, \apjl, 956,
  L5, \dodoi{10.3847/2041-8213/acf9a4}

\bibitem[{{Bruch} {et~al.}(2023){Bruch}, {Gal-Yam}, {Yaron}, {Chen},
  {Strotjohann}, {Irani}, {Zimmerman}, {Schulze}, {Yang}, {Kim}, {Bulla},
  {Sollerman}, {Rigault}, {Ofek}, {Soumagnac}, {Masci}, {Fremling}, {Perley},
  {Nordin}, {Cenko}, {Ho}, {Adams}, {Adreoni}, {Bellm}, {Blagorodnova},
  {Burdge}, {De}, {Dekany}, {Dhawan}, {Drake}, {Duev}, {Graham}, {Graham},
  {Jencson}, {Karamehmetoglu}, {Kasliwal}, {Kulkarni}, {Miller}, {Neill},
  {Prince}, {Riddle}, {Rusholme}, {Sharma}, {Smith}, {Sravan}, {Taggart},
  {Walters}, \& {Yan}}]{Bruch23}
{Bruch}, R.~J., {Gal-Yam}, A., {Yaron}, O., {et~al.} 2023, \apj, 952, 119,
  \dodoi{10.3847/1538-4357/acd8be}

\bibitem[{{Chufarin} {et~al.}(2023){Chufarin}, {Potapov}, {Ionov}, {Korotkiy},
  {Nazarov}, \& {Sokolovsky}}]{Chufarin2023}
{Chufarin}, V., {Potapov}, N., {Ionov}, I., {et~al.} 2023, Transient Name
  Server AstroNote, 150, 1

\bibitem[{{Cs{\"o}rnyei} {et~al.}(2023){Cs{\"o}rnyei}, {Vogl}, {Taubenberger},
  {Fl{\"o}rs}, {Blondin}, {Cudmani}, {Holas}, {Kressierer}, {Leibundgut}, \&
  {Hillebrandt}}]{2023A&A...672A.129C}
{Cs{\"o}rnyei}, G., {Vogl}, C., {Taubenberger}, S., {et~al.} 2023, \aap, 672,
  A129, \dodoi{10.1051/0004-6361/202245379}

\bibitem[{{de Jaeger} \& {Galbany}(2023)}]{review}
{de Jaeger}, T., \& {Galbany}, L. 2023, arXiv e-prints, arXiv:2305.17243,
  \dodoi{10.48550/arXiv.2305.17243}

\bibitem[{{de Jaeger} {et~al.}(2022){de Jaeger}, {Galbany}, {Riess}, {Stahl},
  {Shappee}, {Filippenko}, \& {Zheng}}]{dejaeger22b}
{de Jaeger}, T., {Galbany}, L., {Riess}, A.~G., {et~al.} 2022, \mnras, 514,
  4620, \dodoi{10.1093/mnras/stac1661}

\bibitem[{de~Jaeger {et~al.}(2019)de~Jaeger, Zheng, Stahl, Filippenko, Brink,
  Bigley, Blanchard, Blanchard, Bradley, Cargill, Casper, Cenko, Channa, Choi,
  Clubb, Cobb, Cohen, de~Kouchkovsky, Ellison, Falcon, Fox, Fuller,
  Ganeshalingam, Gould, Graham, Halevi, Hayakawa, Hestenes, Hyland, Jeffers,
  Joubert, Kandrashoff, Kelly, Kim, Kim, Kumar, Leonard, Li, Lowe, Lu, Mason,
  McAllister, Mauerhan, Modjaz, Molloy, Perley, Pina, Poznanski, Ross,
  Shivvers, Silverman, Soler, Stegman, Taylor, Tang, Wilkins, Wang, Wang, Yuk,
  Yunus, \& Zhang}]{dejaeger_berkeley_2019}
de~Jaeger, T., Zheng, W., Stahl, B.~E., {et~al.} 2019, MNRAS, 490, 2799,
  \dodoi{10.1093/mnras/stz2714}

\bibitem[{{de Jaeger} {et~al.}(2019){de Jaeger}, {Zheng}, {Stahl},
  {Filippenko}, {Brink}, {Bigley}, {Blanchard}, {Blanchard}, {Bradley},
  {Cargill}, {Casper}, {Cenko}, {Channa}, {Choi}, {Clubb}, {Cobb}, {Cohen}, {de
  Kouchkovsky}, {Ellison}, {Falcon}, {Fox}, {Fuller}, {Ganeshalingam}, {Gould},
  {Graham}, {Halevi}, {Hayakawa}, {Hestenes}, {Hyland}, {Jeffers}, {Joubert},
  {Kandrashoff}, {Kelly}, {Kim}, {Kim}, {Kumar}, {Leonard}, {Li}, {Lowe}, {Lu},
  {Mason}, {McAllister}, {Mauerhan}, {Modjaz}, {Molloy}, {Perley}, {Pina},
  {Poznanski}, {Ross}, {Shivvers}, {Silverman}, {Soler}, {Stegman}, {Taylor},
  {Tang}, {Wilkins}, {Wang}, {Wang}, {Yuk}, {Yunus}, \& {Zhang}}]{deJaeger2019}
{de Jaeger}, T., {Zheng}, W., {Stahl}, B.~E., {et~al.} 2019, \mnras, 490, 2799,
  \dodoi{10.1093/mnras/stz2714}

\bibitem[{{de Jaeger} {et~al.}(2020){de Jaeger}, {Galbany},
  {Gonz{\'a}lez-Gait{\'a}n}, {Kessler}, {Filippenko}, {F{\"o}rster}, {Hamuy},
  {Brown}, {Davis}, {Guti{\'e}rrez}, {Inserra}, {Lewis}, {M{\"o}ller},
  {Scolnic}, {Smith}, {Brout}, {Carollo}, {Foley}, {Glazebrook}, {Hinton},
  {Macaulay}, {Nichol}, {Sako}, {Sommer}, {Tucker}, {Abbott}, {Aguena},
  {Allam}, {Annis}, {Avila}, {Bertin}, {Bhargava}, {Brooks}, {Burke}, {Carnero
  Rosell}, {Carrasco Kind}, {Carretero}, {Costanzi}, {Crocce}, {da Costa}, {De
  Vicente}, {Desai}, {Diehl}, {Doel}, {Drlica-Wagner}, {Eifler}, {Estrada},
  {Everett}, {Flaugher}, {Fosalba}, {Frieman}, {Garc{\'\i}a-Bellido},
  {Gaztanaga}, {Gruen}, {Gruendl}, {Gschwend}, {Gutierrez}, {Hartley},
  {Hollowood}, {Honscheid}, {James}, {Kuehn}, {Kuropatkin}, {Li}, {Lima},
  {Maia}, {Menanteau}, {Miquel}, {Palmese}, {Paz-Chinch{\'o}n}, {Plazas},
  {Romer}, {Roodman}, {Sanchez}, {Scarpine}, {Schubnell}, {Serrano},
  {Sevilla-Noarbe}, {Soares-Santos}, {Suchyta}, {Swanson}, {Tarle}, {Thomas},
  {Tucker}, {Varga}, {Walker}, {Weller}, \& {Wilkinson}}]{dejaeger20}
{de Jaeger}, T., {Galbany}, L., {Gonz{\'a}lez-Gait{\'a}n}, S., {et~al.} 2020,
  \mnras, \dodoi{10.1093/mnras/staa1402}

\bibitem[{{Dessart}(2024)}]{Dessart24}
{Dessart}, L. 2024, arXiv e-prints, arXiv:2405.04259,
  \dodoi{10.48550/arXiv.2405.04259}

\bibitem[{{Dessart} \& {Hillier}(2005)}]{Dessart05b}
{Dessart}, L., \& {Hillier}, D.~J. 2005, \aap, 439, 671,
  \dodoi{10.1051/0004-6361:20053217}

\bibitem[{{Dessart} \& {Hillier}(2010)}]{Dessart10}
---. 2010, \mnras, 405, 2141, \dodoi{10.1111/j.1365-2966.2010.16611.x}

\bibitem[{{Dessart} \& {Hillier}(2022)}]{Dessart22}
---. 2022, \aap, 660, L9, \dodoi{10.1051/0004-6361/202243372}

\bibitem[{{Dessart} {et~al.}(2017){Dessart}, {Hillier}, \& {Audit}}]{Dessart17}
{Dessart}, L., {Hillier}, D.~J., \& {Audit}, E. 2017, \aap, 605, A83,
  \dodoi{10.1051/0004-6361/201730942}

\bibitem[{{Dessart} {et~al.}(2016){Dessart}, {Hillier}, {Audit}, {Livne}, \&
  {Waldman}}]{Dessart16}
{Dessart}, L., {Hillier}, D.~J., {Audit}, E., {Livne}, E., \& {Waldman}, R.
  2016, \mnras, 458, 2094, \dodoi{10.1093/mnras/stw336}

\bibitem[{{Dessart} {et~al.}(2021){Dessart}, {Hillier}, {Sukhbold}, {Woosley},
  \& {Janka}}]{Dessart21}
{Dessart}, L., {Hillier}, D.~J., {Sukhbold}, T., {Woosley}, S.~E., \& {Janka},
  H.~T. 2021, \aap, 652, A64, \dodoi{10.1051/0004-6361/202140839}

\bibitem[{{Dessart} \& {Jacobson-Gal{\'a}n}(2023)}]{Dessart23}
{Dessart}, L., \& {Jacobson-Gal{\'a}n}, W.~V. 2023, \aap, 677, A105,
  \dodoi{10.1051/0004-6361/202346754}

\bibitem[{{Dickinson} {et~al.}(2024){Dickinson}, {Milisavljevic}, {Garretson},
  {Dessart}, {Margutti}, {Chornock}, {Subrayan}, {Hillier}, {Golub}, {Li},
  {Logsdon}, {Rajagopal}, {Ridgway}, {Smith}, \& {Cynamon}}]{Dickinson24}
{Dickinson}, D., {Milisavljevic}, D., {Garretson}, B., {et~al.} 2024, arXiv
  e-prints, arXiv:2412.14406, \dodoi{10.48550/arXiv.2412.14406}

\bibitem[{Faran {et~al.}(2014)Faran, Poznanski, Filippenko, Chornock, Foley,
  Ganeshalingam, Leonard, Li, Modjaz, Nakar, Serduke, \&
  Silverman}]{faran_photometric_2014}
Faran, T., Poznanski, D., Filippenko, A.~V., {et~al.} 2014, MNRAS, 442, 844,
  \dodoi{10.1093/mnras/stu955}

\bibitem[{{Filippenko}(1982)}]{Filippenko1982}
{Filippenko}, A.~V. 1982, \pasp, 94, 715, \dodoi{10.1086/131052}

\bibitem[{{Filippenko}(1989)}]{Filippenko89}
---. 1989, \aj, 97, 726, \dodoi{10.1086/115018}

\bibitem[{{Filippenko}(1991{\natexlab{a}})}]{Filippenko91a}
{Filippenko}, A.~V. 1991{\natexlab{a}}, in Supernovae, ed. S.~E. {Woosley}, 467

\bibitem[{{Filippenko}(1991{\natexlab{b}})}]{Filippenko91b}
{Filippenko}, A.~V. 1991{\natexlab{b}}, in European Southern Observatory
  Conference and Workshop Proceedings, Vol.~37, European Southern Observatory
  Conference and Workshop Proceedings, ed. I.~J. {Danziger} \& K.~{Kjaer}

\bibitem[{Filippenko(1997)}]{filippenko_optical_1997}
Filippenko, A.~V. 1997, Annual Review of A \& A, 35, 309,
  \dodoi{10.1146/annurev.astro.35.1.309}

\bibitem[{{Filippenko} {et~al.}(2001){Filippenko}, {Li}, {Treffers}, \&
  {Modjaz}}]{Filippenko2001}
{Filippenko}, A.~V., {Li}, W.~D., {Treffers}, R.~R., \& {Modjaz}, M. 2001, in
  Astronomical Society of the Pacific Conference Series, Vol. 246, IAU Colloq.
  183: Small Telescope Astronomy on Global Scales, ed. B.~{Paczynski}, W.-P.
  {Chen}, \& C.~{Lemme}, 121

\bibitem[{Filippenko {et~al.}(2023)Filippenko, Zheng, \&
  Yang}]{filippenko_filippenko_2023}
Filippenko, A.~V., Zheng, W., \& Yang, Y. 2023, Transient Name Server
  AstroNote, 123, 1.
\newblock \url{https://ui.adsabs.harvard.edu/abs/2023TNSAN.123....1F}

\bibitem[{{Freedman} {et~al.}(2001){Freedman}, {Madore}, {Gibson}, {Ferrarese},
  {Kelson}, {Sakai}, {Mould}, {Kennicutt}, {Ford}, {Graham}, {Huchra},
  {Hughes}, {Illingworth}, {Macri}, \& {Stetson}}]{Freedman2001}
{Freedman}, W.~L., {Madore}, B.~F., {Gibson}, B.~K., {et~al.} 2001, \apj, 553,
  47, \dodoi{10.1086/320638}

\bibitem[{{Fulton} {et~al.}(2023){Fulton}, {Nicholl}, {Smith}, {Srivastav},
  {Young}, {McCollum}, {Moore}, {Sim}, {Weston}, {Sheng}, {Shingles}, {Sommer},
  {Aamer}, {Smartt}, {Stevance}, {Rhodes}, {Andersson}, {Denneau}, {Tonry},
  {Weiland}, {Lawrence}, {Siverd}, {Erasmus}, {Koorts}, {Anderson}, {Jordan},
  {Suc}, {Rest}, {Chen}, \& {Stubbs}}]{Fulton2023}
{Fulton}, M.~D., {Nicholl}, M., {Smith}, K.~W., {et~al.} 2023, Transient Name
  Server AstroNote, 124, 1

\bibitem[{{Gal-Yam} {et~al.}(2014){Gal-Yam}, {Arcavi}, {Ofek}, {Ben-Ami},
  {Cenko}, {Kasliwal}, {Cao}, {Yaron}, {Tal}, {Silverman}, {Horesh}, {De Cia},
  {Taddia}, {Sollerman}, {Perley}, {Vreeswijk}, {Kulkarni}, {Nugent},
  {Filippenko}, \& {Wheeler}}]{Gal-Yam14}
{Gal-Yam}, A., {Arcavi}, I., {Ofek}, E.~O., {et~al.} 2014, \nat, 509, 471,
  \dodoi{10.1038/nature13304}

\bibitem[{{Galbany} {et~al.}(2016){Galbany}, {Hamuy}, {Phillips}, {Suntzeff},
  {Maza}, {de Jaeger}, {Moraga}, {Gonz{\'a}lez-Gait{\'a}n}, {Krisciunas},
  {Morrell}, {Thomas-Osip}, {Krzeminski}, {Gonz{\'a}lez}, {Antezana},
  {Wishnjewski}, {McCarthy}, {Anderson}, {Guti{\'e}rrez}, {Stritzinger},
  {Folatelli}, {Anguita}, {Galaz}, {Green}, {Impey}, {Kim}, {Kirhakos},
  {Malkan}, {Mulchaey}, {Phillips}, {Pizzella}, {Prosser}, {Schmidt},
  {Schommer}, {Sherry}, {Strolger}, {Wells}, \& {Williger}}]{galbany16}
{Galbany}, L., {Hamuy}, M., {Phillips}, M.~M., {et~al.} 2016, \aj, 151, 33,
  \dodoi{10.3847/0004-6256/151/2/33}

\bibitem[{Gutiérrez {et~al.}(2017)Gutiérrez, Anderson, Hamuy, Morrell,
  González-Gaitan, Stritzinger, Phillips, Galbany, Folatelli, Dessart,
  Contreras, Valle, Freedman, Hsiao, Krisciunas, Madore, Maza, Suntzeff,
  Prieto, González, Cappellaro, Navarrete, Pizzella, Ruiz, Smith, \&
  Turatto}]{gutierrez_type_2017}
Gutiérrez, C.~P., Anderson, J.~P., Hamuy, M., {et~al.} 2017, ApJ, 850, 89,
  \dodoi{10.3847/1538-4357/aa8f52}

\bibitem[{{Hamann}(2023)}]{Hamann2023}
{Hamann}, N. 2023, Transient Name Server AstroNote, 127, 1

\bibitem[{{Hamuy} \& {Pinto}(2002)}]{hamuy02}
{Hamuy}, M., \& {Pinto}, P.~A. 2002, \apjl, 566, L63, \dodoi{10.1086/339676}

\bibitem[{Hamuy {et~al.}(2001)Hamuy, Pinto, Maza, Suntzeff, Phillips, Eastman,
  Smith, Corbally, Burstein, Li, Ivanov, Moro-Martin, Strolger, Souza, Anjos,
  Green, Pickering, Gonzalez, Antezana, Wischnjewsky, Galaz, Roth, Persson, \&
  Schommer}]{hamuy_distance_2001}
Hamuy, M., Pinto, P.~A., Maza, J., {et~al.} 2001, ApJ, 558, 615,
  \dodoi{10.1086/322450}

\bibitem[{{Henden} {et~al.}(2012){Henden}, {Krajci}, \& {Munari}}]{Henden2012}
{Henden}, A., {Krajci}, T., \& {Munari}, U. 2012, Information Bulletin on
  Variable Stars, 6024, 1

\bibitem[{{Hillier} \& {Dessart}(2012)}]{Hillier12}
{Hillier}, D.~J., \& {Dessart}, L. 2012, \mnras, 424, 252,
  \dodoi{10.1111/j.1365-2966.2012.21192.x}

\bibitem[{{Hillier} \& {Dessart}(2019)}]{Hillier19}
---. 2019, \aap, 631, A8, \dodoi{10.1051/0004-6361/201935100}

\bibitem[{{Hiramatsu} {et~al.}(2021){Hiramatsu}, {Howell}, {Moriya},
  {Goldberg}, {Hosseinzadeh}, {Arcavi}, {Anderson}, {Guti{\'e}rrez}, {Burke},
  {McCully}, {Valenti}, {Galbany}, {Fang}, {Maeda}, {Folatelli}, {Hsiao},
  {Morrell}, {Phillips}, {Stritzinger}, {Suntzeff}, {Gromadzki}, {Maguire},
  {M{\"u}ller-Bravo}, \& {Young}}]{Hiramatsu2021}
{Hiramatsu}, D., {Howell}, D.~A., {Moriya}, T.~J., {et~al.} 2021, \apj, 913,
  55, \dodoi{10.3847/1538-4357/abf6d6}

\bibitem[{{Hiramatsu} {et~al.}(2023){Hiramatsu}, {Tsuna}, {Berger}, {Itagaki},
  {Goldberg}, {Gomez}, {Kishalay}, {Hosseinzadeh}, {Bostroem}, {Brown},
  {Arcavi}, {Bieryla}, {Blanchard}, {Esquerdo}, {Farah}, {Howell}, {Matsumoto},
  {McCully}, {Newsome}, {Gonzalez}, {Pellegrino}, {Rhee}, {Terreran},
  {Vink{\'o}}, \& {Wheeler}}]{Hiramatsu2023}
{Hiramatsu}, D., {Tsuna}, D., {Berger}, E., {et~al.} 2023, \apjl, 955, L8,
  \dodoi{10.3847/2041-8213/acf299}

\bibitem[{{Hosseinzadeh} {et~al.}(2023){Hosseinzadeh}, {Farah}, {Shrestha},
  {Sand}, {Dong}, {Brown}, {Bostroem}, {Valenti}, {Jha}, {Andrews}, {Arcavi},
  {Haislip}, {Hiramatsu}, {Hoang}, {Howell}, {Janzen}, {Jencson}, {Kouprianov},
  {Lundquist}, {McCully}, {Meza Retamal}, {Modjaz}, {Newsome}, {Padilla
  Gonzalez}, {Pearson}, {Pellegrino}, {Ravi}, {Reichart}, {Smith}, {Terreran},
  \& {Vink{\'o}}}]{Hosseinzadeh2023}
{Hosseinzadeh}, G., {Farah}, J., {Shrestha}, M., {et~al.} 2023, \apjl, 953,
  L16, \dodoi{10.3847/2041-8213/ace4c4}

\bibitem[{{Itagaki}(2023)}]{2023TNSTR1158....1I}
{Itagaki}, K. 2023, Transient Name Server Discovery Report, 2023-1158, 1

\bibitem[{{Jacobson-Gal{\'a}n} {et~al.}(2022){Jacobson-Gal{\'a}n}, {Dessart},
  {Jones}, {Margutti}, {Coppejans}, {Dimitriadis}, {Foley}, {Kilpatrick},
  {Matthews}, {Rest}, {Terreran}, {Aleo}, {Auchettl}, {Blanchard}, {Coulter},
  {Davis}, {de Boer}, {DeMarchi}, {Drout}, {Earl}, {Gagliano}, {Gall},
  {Hjorth}, {Huber}, {Ibik}, {Milisavljevic}, {Pan}, {Rest}, {Ridden-Harper},
  {Rojas-Bravo}, {Siebert}, {Smith}, {Taggart}, {Tinyanont}, {Wang}, \&
  {Zenati}}]{Jacobson22}
{Jacobson-Gal{\'a}n}, W.~V., {Dessart}, L., {Jones}, D.~O., {et~al.} 2022,
  \apj, 924, 15, \dodoi{10.3847/1538-4357/ac3f3a}

\bibitem[{{Jacobson-Gal{\'a}n} {et~al.}(2023){Jacobson-Gal{\'a}n}, {Dessart},
  {Margutti}, {Chornock}, {Foley}, {Kilpatrick}, {Jones}, {Taggart}, {Angus},
  {Bhattacharjee}, {Braff}, {Brethauer}, {Burgasser}, {Cao}, {Carlile},
  {Chambers}, {Coulter}, {Dominguez-Ruiz}, {Dickinson}, {de Boer}, {Gagliano},
  {Gall}, {Gao}, {Gates}, {Gomez}, {Guolo}, {Halford}, {Hjorth}, {Huber},
  {Johnson}, {Karpoor}, {Laskar}, {LeBaron}, {Li}, {Lin}, {Loch}, {Lynam},
  {Magnier}, {Maloney}, {Matthews}, {McDonald}, {Miao}, {Milisavljevic}, {Pan},
  {Pradyumna}, {Ransome}, {Rees}, {Rest}, {Rojas-Bravo}, {Sandford},
  {Ascencio}, {Sanjaripour}, {Savino}, {Sears}, {Sharei}, {Smartt}, {Softich},
  {Theissen}, {Tinyanont}, {Tohfa}, {Villar}, {Wang}, {Wainscoat},
  {Westerling}, {Wiston}, {Wozniak}, {Yadavalli}, \&
  {Zenati}}]{Jacobson-Galan23}
{Jacobson-Gal{\'a}n}, W.~V., {Dessart}, L., {Margutti}, R., {et~al.} 2023,
  \apjl, 954, L42, \dodoi{10.3847/2041-8213/acf2ec}

\bibitem[{{Jacobson-Gal{\'a}n}
  {et~al.}(2024{\natexlab{a}}){Jacobson-Gal{\'a}n}, {Dessart}, {Davis},
  {Kilpatrick}, {Margutti}, {Foley}, {Chornock}, {Terreran}, {Hiramatsu},
  {Newsome}, {Padilla Gonzalez}, {Pellegrino}, {Howell}, {Filippenko},
  {Anderson}, {Angus}, {Auchettl}, {Bostroem}, {Brink}, {Cartier}, {Coulter},
  {de Boer}, {Drout}, {Earl}, {Ertini}, {Farah}, {Farias}, {Gall}, {Gao},
  {Gerlach}, {Guo}, {Haynie}, {Hosseinzadeh}, {Ibik}, {Jha}, {Jones},
  {Langeroodi}, {LeBaron}, {Magnier}, {Piro}, {Raimundo}, {Rest}, {Rest},
  {Rich}, {Rojas-Bravo}, {Sears}, {Taggart}, {Villar}, {Wainscoat}, {Wang},
  {Wasserman}, {Yan}, {Yang}, {Zhang}, \& {Zheng}}]{Jacobson24}
{Jacobson-Gal{\'a}n}, W.~V., {Dessart}, L., {Davis}, K.~W., {et~al.}
  2024{\natexlab{a}}, \apj, 970, 189, \dodoi{10.3847/1538-4357/ad4a2a}

\bibitem[{{Jacobson-Gal{\'a}n}
  {et~al.}(2024{\natexlab{b}}){Jacobson-Gal{\'a}n}, {Dessart}, {Davis},
  {Kilpatrick}, {Margutti}, {Foley}, {Chornock}, {Terreran}, {Hiramatsu},
  {Newsome}, {Padilla Gonzalez}, {Pellegrino}, {Howell}, {Filippenko},
  {Anderson}, {Angus}, {Auchettl}, {Bostroem}, {Brink}, {Cartier}, {Coulter},
  {de Boer}, {Drout}, {Earl}, {Ertini}, {Farah}, {Farias}, {Gall}, {Gao},
  {Gerlach}, {Guo}, {Haynie}, {Hosseinzadeh}, {Ibik}, {Jha}, {Jones},
  {Langeroodi}, {LeBaron}, {Magnier}, {Piro}, {Raimundo}, {Rest}, {Rest},
  {Rich}, {Rojas-Bravo}, {Sears}, {Taggart}, {Villar}, {Wainscoat}, {Wang},
  {Wasserman}, {Yan}, {Yang}, {Zhang}, \& {Zheng}}]{Jacobson-Galan24}
---. 2024{\natexlab{b}}, \apj, 970, 189, \dodoi{10.3847/1538-4357/ad4a2a}

\bibitem[{{Jencson} {et~al.}(2023){Jencson}, {Pearson}, {Beasor}, {Lau},
  {Andrews}, {Bostroem}, {Dong}, {Engesser}, {Gomez}, {Guolo}, {Hoang},
  {Hosseinzadeh}, {Jha}, {Karambelkar}, {Kasliwal}, {Lundquist}, {Meza
  Retamal}, {Rest}, {Sand}, {Shahbandeh}, {Shrestha}, {Smith}, {Strader},
  {Valenti}, {Wang}, \& {Zenati}}]{Jencson23}
{Jencson}, J.~E., {Pearson}, J., {Beasor}, E.~R., {et~al.} 2023, \apjl, 952,
  L30, \dodoi{10.3847/2041-8213/ace618}

\bibitem[{{Kilpatrick} {et~al.}(2023){Kilpatrick}, {Foley},
  {Jacobson-Gal{\'a}n}, {Piro}, {Smartt}, {Drout}, {Gagliano}, {Gall},
  {Hjorth}, {Jones}, {Mandel}, {Margutti}, {Ramirez-Ruiz}, {Ransome}, {Villar},
  {Coulter}, {Gao}, {Matthews}, {Taggart}, \& {Zenati}}]{Kilpatrick23}
{Kilpatrick}, C.~D., {Foley}, R.~J., {Jacobson-Gal{\'a}n}, W.~V., {et~al.}
  2023, \apjl, 952, L23, \dodoi{10.3847/2041-8213/ace4ca}

\bibitem[{Kirshner \& Kwan(1974)}]{kirshner_distances_1974}
Kirshner, R.~P., \& Kwan, J. 1974, ApJ, 193, 27, \dodoi{10.1086/153123}

\bibitem[{{Koltenbah}(2023)}]{Koltenbah2023}
{Koltenbah}, B. 2023, Transient Name Server AstroNote, 144, 1

\bibitem[{{Landolt}(1992)}]{Landolt92}
{Landolt}, A.~U. 1992, \aj, 104, 340, \dodoi{10.1086/116242}

\bibitem[{Landsman(1993)}]{landsman_idl_1993}
Landsman, W.~B. 1993, 52, 246.
\newblock \url{https://ui.adsabs.harvard.edu/abs/1993ASPC...52..246L}

\bibitem[{{Leonard} {et~al.}(2000){Leonard}, {Filippenko}, {Barth}, \&
  {Matheson}}]{Leonard00}
{Leonard}, D.~C., {Filippenko}, A.~V., {Barth}, A.~J., \& {Matheson}, T. 2000,
  \apj, 536, 239, \dodoi{10.1086/308910}

\bibitem[{{Li} {et~al.}(2024){Li}, {Hu}, {Li}, {Yang}, {Wang}, {Yan}, {Hu},
  {Zhang}, {Mao}, {Riise}, {Gao}, {Sun}, {Liu}, {Xiong}, {Wang}, {Mo},
  {Iskandar}, {Xi}, {Xiang}, {Wang}, {Sun}, {Zhang}, {Chen}, {Lin}, {Guo},
  {Liu}, {Cai}, {Zhou}, {Zhao}, {Chen}, {Zheng}, {Li}, {Zhang}, {Xu}, {Lyu},
  {Castro-Tirado}, {Chufarin}, {Potapov}, {Ionov}, {Korotkiy}, {Nazarov},
  {Sokolovsky}, {Hamann}, \& {Herman}}]{Li23}
{Li}, G., {Hu}, M., {Li}, W., {et~al.} 2024, \nat, 627, 754,
  \dodoi{10.1038/s41586-023-06843-6}

\bibitem[{{Li} {et~al.}(2003){Li}, {Filippenko}, {Chornock}, \& {Jha}}]{Li2003}
{Li}, W., {Filippenko}, A.~V., {Chornock}, R., \& {Jha}, S. 2003, \pasp, 115,
  844, \dodoi{10.1086/376432}

\bibitem[{{Limeburner}(2023)}]{Limeburner2023}
{Limeburner}, S. 2023, Transient Name Server AstroNote, 128, 1

\bibitem[{Mao {et~al.}(2023)Mao, Zhang, Cai, Chen, Chen, Gao, Li, Lyu, Qin,
  Sun, Xu, Zhang, Zhang, Zhao, Zheng, Zhou, \& Ye}]{mao_onset_2023}
Mao, Y., Zhang, M., Cai, G., {et~al.} 2023, TNS, 130, 1.
\newblock \url{https://ui.adsabs.harvard.edu/abs/2023TNSAN.130....1M}

\bibitem[{Miller \& Stone(1994)}]{miller_stone_1994}
Miller, J., \& Stone, R. 1994, The Kast Double Spectrograph

\bibitem[{{Morozova} {et~al.}(2015){Morozova}, {Piro}, {Renzo}, {Ott},
  {Clausen}, {Couch}, {Ellis}, \& {Roberts}}]{Morozova15}
{Morozova}, V., {Piro}, A.~L., {Renzo}, M., {et~al.} 2015, \apj, 814, 63,
  \dodoi{10.1088/0004-637X/814/1/63}

\bibitem[{{Niu} {et~al.}(2023){Niu}, {Sun}, {Maund}, {Zhang}, {Zhao}, \&
  {Liu}}]{Niu23}
{Niu}, Z., {Sun}, N.-C., {Maund}, J.~R., {et~al.} 2023, \apjl, 955, L15,
  \dodoi{10.3847/2041-8213/acf4e3}

\bibitem[{{Ofek} {et~al.}(2013){Ofek}, {Lin}, {Kouveliotou}, {Younes},
  {G{\"o}{\v{g}}{\"u}{\c{s}}}, {Kasliwal}, \& {Cao}}]{Ofek13}
{Ofek}, E.~O., {Lin}, L., {Kouveliotou}, C., {et~al.} 2013, \apj, 768, 47,
  \dodoi{10.1088/0004-637X/768/1/47}

\bibitem[{{Perley} {et~al.}(2023){Perley}, {Gal-Yam}, {Irani}, \&
  {Zimmerman}}]{perley23}
{Perley}, D.~A., {Gal-Yam}, A., {Irani}, I., \& {Zimmerman}, E. 2023, Transient
  Name Server AstroNote, 119, 1

\bibitem[{{Perley} \& {Irani}(2023)}]{Perley2023}
{Perley}, D.~A., \& {Irani}, I. 2023, Transient Name Server AstroNote, 120, 1

\bibitem[{{Pledger} \& {Shara}(2023)}]{Pledger23}
{Pledger}, J.~L., \& {Shara}, M.~M. 2023, \apjl, 953, L14,
  \dodoi{10.3847/2041-8213/ace88b}

\bibitem[{{Qin} {et~al.}(2024){Qin}, {Zhang}, {Bloom}, {Sollerman},
  {Zimmerman}, {Irani}, {Schulze}, {Gal-Yam}, {Kasliwal}, {Coughlin}, {Perley},
  {Fremling}, \& {Kulkarni}}]{Qin23}
{Qin}, Y.-J., {Zhang}, K., {Bloom}, J., {et~al.} 2024, \mnras, 534, 271,
  \dodoi{10.1093/mnras/stae2012}

\bibitem[{{Ransome} {et~al.}(2021){Ransome}, {Habergham-Mawson}, {Darnley},
  {James}, {Filippenko}, \& {Schlegel}}]{Ransome21}
{Ransome}, C.~L., {Habergham-Mawson}, S.~M., {Darnley}, M.~J., {et~al.} 2021,
  \mnras, 506, 4715, \dodoi{10.1093/mnras/stab1938}

\bibitem[{Riess {et~al.}(2022)Riess, Yuan, Macri, Scolnic, Brout, Casertano,
  Jones, Murakami, Anand, Breuval, Brink, Filippenko, Hoffmann, Jha, Kenworthy,
  Mackenty, Stahl, \& Zheng}]{riess_comprehensive_2022}
Riess, A.~G., Yuan, W., Macri, L.~M., {et~al.} 2022, ApJ Letters, 934, L7,
  \dodoi{10.3847/2041-8213/ac5c5b}

\bibitem[{{Riess} {et~al.}(2022){Riess}, {Yuan}, {Macri}, {Scolnic}, {Brout},
  {Casertano}, {Jones}, {Murakami}, {Anand}, {Breuval}, {Brink}, {Filippenko},
  {Hoffmann}, {Jha}, {D'arcy Kenworthy}, {Mackenty}, {Stahl}, \&
  {Zheng}}]{Riess2022}
{Riess}, A.~G., {Yuan}, W., {Macri}, L.~M., {et~al.} 2022, \apjl, 934, L7,
  \dodoi{10.3847/2041-8213/ac5c5b}

\bibitem[{{Rizzi} {et~al.}(2007){Rizzi}, {Tully}, {Makarov}, {Makarova},
  {Dolphin}, {Sakai}, \& {Shaya}}]{rizzi07}
{Rizzi}, L., {Tully}, R.~B., {Makarov}, D., {et~al.} 2007, \apj, 661, 815,
  \dodoi{10.1086/516566}

\bibitem[{{Saha} {et~al.}(2006){Saha}, {Thim}, {Tammann}, {Reindl}, \&
  {Sandage}}]{Saha2006}
{Saha}, A., {Thim}, F., {Tammann}, G.~A., {Reindl}, B., \& {Sandage}, A. 2006,
  \apjs, 165, 108, \dodoi{10.1086/503800}

\bibitem[{Sanders {et~al.}(2015)Sanders, Soderberg, Gezari, Betancourt,
  Chornock, Berger, Foley, Challis, Drout, Kirshner, Lunnan, Marion, Margutti,
  McKinnon, Milisavljevic, Narayan, Rest, Kankare, Mattila, Smartt, Huber,
  Burgett, Draper, Hodapp, Kaiser, Kudritzki, Magnier, Metcalfe, Morgan, Price,
  Tonry, Wainscoat, \& Waters}]{sanders_toward_2015}
Sanders, N.~E., Soderberg, A.~M., Gezari, S., {et~al.} 2015, ApJ, 799, 208,
  \dodoi{10.1088/0004-637X/799/2/208}

\bibitem[{Schlafly \& Finkbeiner(2011)}]{schlafly_measuring_2011}
Schlafly, E.~F., \& Finkbeiner, D.~P. 2011, ApJ, 737, 103,
  \dodoi{10.1088/0004-637X/737/2/103}

\bibitem[{{Schlegel}(1990)}]{Schlegel90}
{Schlegel}, E.~M. 1990, \mnras, 244, 269

\bibitem[{{Sgro} {et~al.}(2023){Sgro}, {Esposito}, {Blaclard}, {Gomez},
  {Marchis}, {Filippenko}, {Peluso}, {Lawrence}, {Verveen}, {Wagner}, {Nardi},
  {Wiart}, {Mirwald}, {Christensen}, {Eramia}, {Parker}, {Guillet}, {Kim},
  {Logan}, {Kyba}, {Toulmin}, {Vantaggiato}, {Adhis}, {Gary}, {Goodey},
  {Dickinson}, {Koster}, {Martin}, {Bonilla}, {Chung}, {Miny}, {Mortecrette},
  {Saibi}, {Gagnon}, {Simard}, {Vacon}, {Simard}, {Dreise}, {Funakoshi},
  {Vacon}, {Yaniz}, {Le Tarnec}, {Laugier}, {Siders}, {Sweitzer}, {Dvoracek},
  {Archer}, {Deitz}, {Bradley}, {Fukui}, {Sibbernsen}, {Borrot}, {Cross},
  {Heider}, {Yamaguchi}, {Hirsch}, {Leroux}, {Billiani}, {Lorber}, {Smallen},
  {Shimizu}, {Nishimura}, {Ryno}, {Cunningham}, {Gagnon}, {Primm}, {Rushton},
  {Sibbernsen}, {Mitchell}, {Yoblonsky}, {Leroux}, {Clerget}, {Stojanovi{\'c}},
  {Unique}, {Huth}, {Ang}, {Santoni}, {Foster}, {Poggiali}, {Xu}, {Kukita},
  {{\v{S}}{\'c}epanovi{\'c}}, {Saibi}, {Will}, {Latour}, {Haythornthwaite},
  {Cadieux}, {M{\"u}ller}, {Chung}, {Watanabe}, \& {Arnaud}}]{Sgro23}
{Sgro}, L.~A., {Esposito}, T.~M., {Blaclard}, G., {et~al.} 2023, Research Notes
  of the American Astronomical Society, 7, 141,
  \dodoi{10.3847/2515-5172/ace41f}

\bibitem[{{Shappee} \& {Stanek}(2011)}]{Shappee2011}
{Shappee}, B.~J., \& {Stanek}, K.~Z. 2011, \apj, 733, 124,
  \dodoi{10.1088/0004-637X/733/2/124}

\bibitem[{Silverman {et~al.}(2012)Silverman, Foley, Filippenko, Ganeshalingam,
  Barth, Chornock, Griffith, Kong, Lee, Leonard, Matheson, Miller, Steele,
  Barris, Bloom, Cobb, Coil, Desroches, Gates, Ho, Jha, Kandrashoff, Li,
  Mandel, Modjaz, Moore, Mostardi, Papenkova, Park, Perley, Poznanski, Reuter,
  Scala, Serduke, Shields, Swift, Tonry, Van~Dyk, Wang, \&
  Wong}]{silverman_berkeley_2012}
Silverman, J.~M., Foley, R.~J., Filippenko, A.~V., {et~al.} 2012, MNRAS, 425,
  1789, \dodoi{10.1111/j.1365-2966.2012.21270.x}

\bibitem[{{Smartt}(2009)}]{smartt09}
{Smartt}, S.~J. 2009, \araa, 47, 63,
  \dodoi{10.1146/annurev-astro-082708-101737}

\bibitem[{{Smartt}(2015)}]{smartt15}
---. 2015, \pasa, 32, e016, \dodoi{10.1017/pasa.2015.17}

\bibitem[{{Smith} {et~al.}(2023){Smith}, {Pearson}, {Sand}, {Ilyin},
  {Bostroem}, {Hosseinzadeh}, \& {Shrestha}}]{Smith23}
{Smith}, N., {Pearson}, J., {Sand}, D.~J., {et~al.} 2023, \apj, 956, 46,
  \dodoi{10.3847/1538-4357/acf366}

\bibitem[{{Soraisam} {et~al.}(2023){Soraisam}, {Szalai}, {Van Dyk}, {Andrews},
  {Srinivasan}, {Chun}, {Matheson}, {Scicluna}, \&
  {Vasquez-Torres}}]{Soraisam23}
{Soraisam}, M.~D., {Szalai}, T., {Van Dyk}, S.~D., {et~al.} 2023, \apj, 957,
  64, \dodoi{10.3847/1538-4357/acef22}

\bibitem[{{Stahl} {et~al.}(2019){Stahl}, {Zheng}, {de Jaeger}, {Filippenko},
  {Bigley}, {Blanchard}, {Blanchard}, {Brink}, {Cargill}, {Casper}, {Channa},
  {Choi}, {Choksi}, {Chu}, {Clubb}, {Cohen}, {Ellison}, {Falcon}, {Fazeli},
  {Fuller}, {Ganeshalingam}, {Gates}, {Gould}, {Halevi}, {Hayakawa},
  {Hestenes}, {Jeffers}, {Joubert}, {Kandrashoff}, {Kim}, {Kim}, {Kislak},
  {Kleiser}, {Kong}, {de Kouchkovsky}, {Krishnan}, {Kumar}, {Leja}, {Leonard},
  {Li}, {Li}, {Lu}, {Mason}, {Molloy}, {Pina}, {Rex}, {Ross}, {Stegman},
  {Tang}, {Thrasher}, {Wang}, {Wilkins}, {Yuk}, {Yunus}, \&
  {Zhang}}]{Stahl2019}
{Stahl}, B.~E., {Zheng}, W., {de Jaeger}, T., {et~al.} 2019, \mnras, 490, 3882,
  \dodoi{10.1093/mnras/stz2742}

\bibitem[{{Stathakis} \& {Sadler}(1991)}]{Stathakis91}
{Stathakis}, R.~A., \& {Sadler}, E.~M. 1991, \mnras, 250, 786,
  \dodoi{10.1093/mnras/250.4.786}

\bibitem[{{Stetson}(1987)}]{Stetson1987}
{Stetson}, P.~B. 1987, \pasp, 99, 191, \dodoi{10.1086/131977}

\bibitem[{{Sukhbold} {et~al.}(2016){Sukhbold}, {Ertl}, {Woosley}, {Brown}, \&
  {Janka}}]{Sukhbold16}
{Sukhbold}, T., {Ertl}, T., {Woosley}, S.~E., {Brown}, J.~M., \& {Janka}, H.~T.
  2016, \apj, 821, 38, \dodoi{10.3847/0004-637X/821/1/38}

\bibitem[{{Tartaglia} {et~al.}(2021){Tartaglia}, {Sand}, {Groh}, {Valenti},
  {Wyatt}, {Bostroem}, {Brown}, {Yang}, {Burke}, {Chen}, {Davis},
  {F{\"o}rster}, {Galbany}, {Haislip}, {Hiramatsu}, {Hosseinzadeh}, {Howell},
  {Hsiao}, {Jha}, {Kouprianov}, {Kuncarayakti}, {Lyman}, {McCully}, {Phillips},
  {Rau}, {Reichart}, {Shahbandeh}, \& {Strader}}]{Tartaglia21}
{Tartaglia}, L., {Sand}, D.~J., {Groh}, J.~H., {et~al.} 2021, \apj, 907, 52,
  \dodoi{10.3847/1538-4357/abca8a}

\bibitem[{{Teja} {et~al.}(2023){Teja}, {Singh}, {Basu}, {Anupama}, {Sahu},
  {Dutta}, {Swain}, {Nakaoka}, {Pathak}, {Bhalerao}, {Barway}, {Kumar},
  {A.~J.}, {Imazawa}, {Kumar}, \& {Kawabata}}]{Teja2023}
{Teja}, R.~S., {Singh}, A., {Basu}, J., {et~al.} 2023, \apjl, 954, L12,
  \dodoi{10.3847/2041-8213/acef20}

\bibitem[{{Terreran} {et~al.}(2016){Terreran}, {Jerkstrand}, {Benetti},
  {Smartt}, {Ochner}, {Tomasella}, {Howell}, {Morales-Garoffolo},
  {Harutyunyan}, {Kankare}, {Arcavi}, {Cappellaro}, {Elias-Rosa},
  {Hosseinzadeh}, {Kangas}, {Pastorello}, {Tartaglia}, {Turatto}, {Valenti},
  {Wiggins}, \& {Yuan}}]{Terreran16}
{Terreran}, G., {Jerkstrand}, A., {Benetti}, S., {et~al.} 2016, \mnras, 462,
  137, \dodoi{10.1093/mnras/stw1591}

\bibitem[{{Terreran} {et~al.}(2022){Terreran}, {Jacobson-Gal{\'a}n}, {Groh},
  {Margutti}, {Coppejans}, {Dimitriadis}, {Kilpatrick}, {Matthews}, {Siebert},
  {Angus}, {Brink}, {Filippenko}, {Foley}, {Jones}, {Tinyanont}, {Gall},
  {Pfister}, {Zenati}, {Ansari}, {Auchettl}, {El-Badry}, {Magnier}, \&
  {Zheng}}]{Terreran22}
{Terreran}, G., {Jacobson-Gal{\'a}n}, W.~V., {Groh}, J.~H., {et~al.} 2022,
  \apj, 926, 20, \dodoi{10.3847/1538-4357/ac3820}

\bibitem[{{Tody}(1986)}]{Tody1986}
{Tody}, D. 1986, in Society of Photo-Optical Instrumentation Engineers (SPIE)
  Conference Series, Vol. 627, Instrumentation in astronomy VI, ed. D.~L.
  {Crawford}, 733, \dodoi{10.1117/12.968154}

\bibitem[{Valenti {et~al.}(2016)Valenti, Howell, Stritzinger, Graham,
  Hosseinzadeh, Arcavi, Bildsten, Jerkstrand, McCully, Pastorello, Piro, Sand,
  Smartt, Terreran, Baltay, Benetti, Brown, Filippenko, Fraser, Rabinowitz,
  Sullivan, \& Yuan}]{valenti_diversity_2016}
Valenti, S., Howell, D.~A., Stritzinger, M.~D., {et~al.} 2016, MNRAS, 459,
  3939, \dodoi{10.1093/mnras/stw870}

\bibitem[{{Van Dyk}(2017)}]{VanDyk2017}
{Van Dyk}, S.~D. 2017, Philosophical Transactions of the Royal Society of
  London Series A, 375, 20160277, \dodoi{10.1098/rsta.2016.0277}

\bibitem[{{Van Dyk} {et~al.}(2003){Van Dyk}, {Li}, \& {Filippenko}}]{vandyk03}
{Van Dyk}, S.~D., {Li}, W., \& {Filippenko}, A.~V. 2003, \pasp, 115, 1289,
  \dodoi{10.1086/378308}

\bibitem[{{Van Dyk} {et~al.}(2000){Van Dyk}, {Peng}, {King}, {Filippenko},
  {Treffers}, {Li}, \& {Richmond}}]{VanDyk00}
{Van Dyk}, S.~D., {Peng}, C.~Y., {King}, J.~Y., {et~al.} 2000, \pasp, 112,
  1532, \dodoi{10.1086/317727}

\bibitem[{Van~Dyk {et~al.}(2019)Van~Dyk, Zheng, Maund, Brink, Srinivasan,
  Andrews, Smith, Leonard, Morozova, Filippenko, Conner, Milisavljevic, Jaeger,
  Long, Isaacson, Crossfield, Kosiarek, Howard, Fox, Kelly, Piro, Littlefair,
  Dhillon, Wilson, Butterley, Yunus, Channa, Jeffers, Falcon, Ross, Hestenes,
  Stegman, Zhang, \& Kumar}]{dyk_type_2019}
Van~Dyk, S.~D., Zheng, W., Maund, J.~R., {et~al.} 2019, ApJ, 875, 136,
  \dodoi{10.3847/1538-4357/ab1136}

\bibitem[{{Van Dyk} {et~al.}(2024){Van Dyk}, {Srinivasan}, {Andrews},
  {Soraisam}, {Szalai}, {Howell}, {Isaacson}, {Matheson}, {Petigura},
  {Scicluna}, {Stephens}, {Van Zandt}, {Zheng}, {Chun}, \&
  {Fillippenko}}]{VanDyk2023}
{Van Dyk}, S.~D., {Srinivasan}, S., {Andrews}, J.~E., {et~al.} 2024, \apj, 968,
  27, \dodoi{10.3847/1538-4357/ad414b}

\bibitem[{{Vasylyev} {et~al.}(2023){Vasylyev}, {Yang}, {Filippenko}, {Patra},
  {Brink}, {Wang}, {Chornock}, {Margutti}, {Gates}, {Burgasser}, {Karpoor},
  {LeBaron}, {Softich}, {Theissen}, {Wiston}, \& {Zheng}}]{Vasylyev23}
{Vasylyev}, S.~S., {Yang}, Y., {Filippenko}, A.~V., {et~al.} 2023, \apjl, 955,
  L37, \dodoi{10.3847/2041-8213/acf1a3}

\bibitem[{{Vink{\'o}} {et~al.}(2012){Vink{\'o}}, {S{\'a}rneczky}, {Tak{\'a}ts},
  {Marion}, {Heged{\"u}s}, {B{\'\i}r{\'o}}, {Borkovits}, {Szegedi-Elek},
  {Farkas}, {Klagyivik}, {Kiss}, {Kov{\'a}cs}, {P{\'a}l}, {Szak{\'a}ts},
  {Szalai}, {Szalai}, {Szatm{\'a}ry}, {Szing}, {Vida}, \&
  {Wheeler}}]{Vinko2012}
{Vink{\'o}}, J., {S{\'a}rneczky}, K., {Tak{\'a}ts}, K., {et~al.} 2012, \aap,
  546, A12, \dodoi{10.1051/0004-6361/201220043}

\bibitem[{{Vogl}(2020)}]{voglphd}
{Vogl}, C. 2020, PhD thesis, Technical University of Munich, Germany

\bibitem[{{Xiang} {et~al.}(2024){Xiang}, {Mo}, {Wang}, {Wang}, {Zhang}, {Lin},
  \& {Wang}}]{Xiang23}
{Xiang}, D., {Mo}, J., {Wang}, L., {et~al.} 2024, Science China Physics,
  Mechanics, and Astronomy, 67, 219514, \dodoi{10.1007/s11433-023-2267-0}

\bibitem[{{Yamanaka} {et~al.}(2023){Yamanaka}, {Fujii}, \&
  {Nagayama}}]{Yamanaka23}
{Yamanaka}, M., {Fujii}, M., \& {Nagayama}, T. 2023, \pasj, 75, L27,
  \dodoi{10.1093/pasj/psad051}

\bibitem[{{Yaron} {et~al.}(2023){Yaron}, {Bruch}, {Chen}, {Irani}, {Zimmerman},
  {Gal-Yam}, \& {Qin}}]{Yaron2023}
{Yaron}, O., {Bruch}, R., {Chen}, P., {et~al.} 2023, Transient Name Server
  AstroNote, 133, 1

\bibitem[{{Yaron} {et~al.}(2017){Yaron}, {Perley}, {Gal-Yam}, {Groh}, {Horesh},
  {Ofek}, {Kulkarni}, {Sollerman}, {Fransson}, {Rubin}, {Szabo}, {Sapir},
  {Taddia}, {Cenko}, {Valenti}, {Arcavi}, {Howell}, {Kasliwal}, {Vreeswijk},
  {Khazov}, {Fox}, {Cao}, {Gnat}, {Kelly}, {Nugent}, {Filippenko}, {Laher},
  {Wozniak}, {Lee}, {Rebbapragada}, {Maguire}, {Sullivan}, \&
  {Soumagnac}}]{Yaron17}
{Yaron}, O., {Perley}, D.~A., {Gal-Yam}, A., {et~al.} 2017, Nature Physics, 13,
  510, \dodoi{10.1038/nphys4025}

\bibitem[{{Zhang} {et~al.}(2023{\natexlab{a}}){Zhang}, {Lin}, {Wang}, {Zhao},
  {Li}, {Liu}, {Yan}, {Xiang}, {Wang}, \& {Bai}}]{zhangj2023}
{Zhang}, J., {Lin}, H., {Wang}, X., {et~al.} 2023{\natexlab{a}}, Science
  Bulletin, 68, 2548, \dodoi{10.1016/j.scib.2023.09.015}

\bibitem[{{Zhang} {et~al.}(2023{\natexlab{b}}){Zhang}, {Kennedy},
  {Oostermeyer}, {Bloom}, \& {Perley}}]{Zhang2023}
{Zhang}, K., {Kennedy}, D., {Oostermeyer}, B., {Bloom}, J., \& {Perley}, D.~A.
  2023{\natexlab{b}}, Transient Name Server AstroNote, 125, 1

\bibitem[{{Zimmerman} {et~al.}(2024){Zimmerman}, {Irani}, {Chen}, {Gal-Yam},
  {Schulze}, {Perley}, {Sollerman}, {Filippenko}, {Shenar}, {Yaron}, {Shahaf},
  {Bruch}, {Ofek}, {De Cia}, {Brink}, {Yang}, {Vasylyev}, {Ben Ami}, {Aubert},
  {Badash}, {Bloom}, {Brown}, {De}, {Dimitriadis}, {Fransson}, {Fremling},
  {Hinds}, {Horesh}, {Johansson}, {Kasliwal}, {Kulkarni}, {Kushnir}, {Martin},
  {Matuzewski}, {McGurk}, {Miller}, {Morag}, {Neil}, {Nugent}, {Post},
  {Prusinski}, {Qin}, {Raichoor}, {Riddle}, {Rowe}, {Rusholme}, {Sfaradi},
  {Sjoberg}, {Soumagnac}, {Stein}, {Strotjohann}, {Terwel}, {Wasserman},
  {Wise}, {Wold}, {Yan}, \& {Zhang}}]{Zimmerman24}
{Zimmerman}, E.~A., {Irani}, I., {Chen}, P., {et~al.} 2024, \nat, 627, 759,
  \dodoi{10.1038/s41586-024-07116-6}

\end{thebibliography}

\listofchanges
\end{document}